\documentclass[floats,floatfix,showpacs,amssymb,prd,twocolumn,superscriptaddress,nofootinbib]{revtex4-1}

\usepackage{graphicx,epsf, epsfig, amssymb}% Include figure files
\usepackage{bm}
\usepackage{longtable}
\usepackage[usenames,dvipsnames]{xcolor}
\usepackage{color}
\usepackage[breaklinks]{hyperref}
\usepackage{amsfonts,amsmath,amssymb,mathrsfs}
\usepackage{multirow}
\usepackage{enumerate}

\def\be{\begin{equation}}
\def\ee{\end{equation}}
\def\beq{\begin{eqnarray}}
\def\eeq{\end{eqnarray}}
\def\ben{\begin{enumerate}}
\def\een{\end{enumerate}}
\def\bi{\begin{itemize}}
\def\ei{\end{itemize}}
\def\f{\frac}

\newcommand{\nn}{\nonumber}

\def\etal{{\it et al.}} % Followed by a ~ and the rest of the text

\begin{document}

\title{Nonspinning black hole-neutron star mergers: \\ A model for the
  amplitude of gravitational waveforms}

\author{Francesco Pannarale}
\email{francesco.pannarale@aei.mpg.de}
\affiliation{ Max-Planck-Institut f{\"u}r Gravitationsphysik, Albert
  Einstein Institut, Potsdam 14476, Germany}

\author{Emanuele Berti}
\email{berti@phy.olemiss.edu}
\affiliation{Department of Physics and Astronomy, The University of
  Mississippi, University, Mississippi 38677, USA}

\author{Koutarou Kyutoku} 
\email{kyutoku@uwm.edu}
\affiliation{Department of Physics, University of Wisconsin-Milwaukee,
  P.O.Box 413, Milwaukee, Wisconsin 53201, USA}

\author{Masaru Shibata}
\email{mshibata@yukawa.kyoto-u.ac.jp}
\affiliation{Yukawa Institute for Theoretical Physics, Kyoto
  University, Kyoto 606-8502, Japan}


\date{\today}

\begin{abstract}
  Black hole-neutron star binary mergers display a much richer
  phenomenology than black hole-black hole mergers, even in the
  relatively simple case --- considered in this paper --- in which
  both the black hole and the neutron star are nonspinning. When the
  neutron star is tidally disrupted, the gravitational wave emission
  is radically different from the black hole-black hole case and it
  can be broadly classified in two groups, depending on the spatial
  extent of the disrupted material. We present a phenomenological
  model for the gravitational waveform amplitude in the frequency
  domain that encompasses the three possible outcomes of the merger:
  no tidal disruption, ``mild,'' and ``strong'' tidal disruption. The
  model is calibrated to general relativistic numerical simulations
  using piecewise polytropic neutron star equations of state. It
  should prove useful to extract information on the nuclear equation
  of state from future gravitational-wave observations, and also to
  obtain more accurate estimates of black hole-neutron star merger
  event rates in second- and third-generation interferometric
  gravitational-wave detectors. We plan to extend and improve the
  model as longer and more accurate gravitational waveforms become
  available, and we will make it publicly available online as a
  \textsc{Mathematica} package. We also present in the Appendix
  analytical fits of the projected KAGRA noise spectral density, which
  should be useful in data analysis applications.
\end{abstract}

\maketitle

%%%%%%%%%%%%%%%%%%%%%%%%%%%%%%%%%%%%%%%%%%%%%%%%%%%%%%%%%%%%%%%%%%%%%%%%%%%%%%%
\section{Introduction}
%%%%%%%%%%%%%%%%%%%%%%%%%%%%%%%%%%%%%%%%%%%%%%%%%%%%%%%%%%%%%%%%%%%%%%%%%%%%%%%
Numerical relativity has made impressive strides in the past couple of
decades. Several research groups can now simulate the late inspiral
and merger of compact binaries, which are the main target for second-
and third-generation gravitational-wave (GW) detectors, such as
Advanced LIGO/Virgo \cite{2010CQGra..27h4006H}, KAGRA
\cite{Somiya:2011np,KAGRAnoise}, LIGO-India \cite{indigo} and the
Einstein Telescope \cite{2010CQGra..27s4002P}. These systems are
composed of either black holes (BHs) or neutron stars (NSs), and they
belong to three families: BH-BH, NS-NS, and BH-NS binaries. In
general, the calculations are resource intensive and time consuming,
so fully numerical simulations covering many cycles and spanning the
whole parameter space of these binaries --- including masses, spins,
nuclear equation of state (EOS), etc.~--- are still beyond the reach
of present-day computers.

For this reason, semianalytical waveform models are necessary to
bridge the gap between the early inspiral (where the binary dynamics
can be treated via perturbative methods) and the merger
phase. Analytical models covering the full inspiral and merger have
several applications. First and foremost, they can be used to reduce
the computational cost of building GW detection templates. The
community has engaged in large-scale efforts, such as the
NINJA/NINJA-2 and NRAR collaborations
\cite{Aylott:2009ya,Ajith:2012az,NRAR2013}, to optimize the task of
injecting waveforms in detector data and of constructing
matched-filtering templates. Semianalytical merger models are also
valuable to improve event rate estimates, which are currently based on
rather rough approximations to the actual binary waveforms
\cite{Abadie:2010cf,Dominik:2012kk}: see
e.g.~\cite{Marassi:2011si,Zhu:2012xw} for preliminary efforts in this
direction. 
Last but not least (and more ambitiously), semianalytical models
incorporating the characteristic signatures predicted by numerical
merger simulations may prove useful to constrain the properties of
compact binaries. For example, Read
\etal~\cite{Read:2009yp,Read:2013zra} analyzed general relativistic
simulations of NS-NS mergers to show that EOS information can be
extracted (at least in principle) from departures from the
point-particle limit of the gravitational waveform produced during the
late inspiral. Bauswein \etal~\cite{Bauswein:2012ya} used a
three-dimensional relativistic smoothed-particle hydrodynamics code to
demonstrate a correlation between the NS radii and the peak frequency
of the postmerger GW signal from NS-NS mergers. Finally, Lackey
\etal~\cite{Lackey:2011vz,Lackey:2013axa} recently studied GW
constraints on the NS EOS for BH-NS binaries using a Fisher matrix
analysis of ``hybrid waveforms'' obtained by combining either
post-Newtonian (PN) or effective-one-body (EOB) models with numerical
waveforms. 

With few exceptions (see e.g. \cite{Zhu:2012xw}), event rate estimates
for advanced GW detectors using population synthesis calculations and
models of GW backgrounds from compact binaries rely on very simple
approximations for the gravitational waveforms
\cite{2010CQGra..27q3001A,2010ApJ...715L.138B,Belczynski:2012cx,Dominik:2012kk,2013arXiv1308.1546D}. An
implementation of more complex models, such as the one developed here,
could potentially allow us to use GW observations to better understand
the astrophysical formation scenarios leading to compact binary
formation, and perhaps to constrain the nuclear EOS. For BH-NS
binaries, this may be achieved through comparisons between the
observed GW cutoff frequencies and those predicted by the models.

It is fair to state that semianalytical models and numerical
simulations are most advanced for BH-BH systems. Phenomenological
waveform models spanning inspiral, merger, and ringdown (IMR) were
initially proposed by Ajith \etal~for nonspinning binaries
\cite{Ajith:2007qp,Ajith:2007kx,Ajith:2007xh}, and later extended to
spinning, nonprecessing binaries \cite{Ajith:2009bn}. The original
(nonspinning) and improved (spinning) versions of this model are
sometimes referred to as ``PhenomA'' \cite{Ajith:2007kx} and
``PhenomB'' \cite{Ajith:2009bn}, or alternatively as ``PhenV1'' and
``PhenV2'': see e.g.~\cite{Damour:2010zb}. An upgraded version of
these models was subsequently proposed by Santamar{\'{\i}}a
\etal~\cite{Santamaria:2010yb}: this last model is sometimes called
``PhenomC'' or ``PhenV3,'' and it has the important feature of
reducing by construction to the correct PN limit in the early
inspiral. Progress in tuning the EOB model to numerical BH-BH
simulations has also been remarkable: see
e.g.~\cite{Taracchini:2012ig,Damour:2012ky} for the latest
incarnations of these models, and \cite{Damour:2010zb,Ohme:2011zm} for
comparisons between waveforms of the EOB and PhenomX (X=A,\,B,\,C)
families.

Phenomenological BH-BH waveforms have been extensively used in
parameter estimation studies. PhenomX waveforms were used in data
analysis applications for both Earth-based \cite{Ajith:2009fz} and
space-based detectors
\cite{McWilliams:2009bg,McWilliams:2010eq,AmaroSeoane:2012km,AmaroSeoane:2012je}.
Systematic errors in parameter estimation are likely to be important,
but there are relatively few studies in this direction. For example,
PN waveforms were used to quantify errors in estimating full IMR
PhenomA waveform parameters in \cite{Bose:2012vb}, while systematic
errors of EOB models in the LISA context were investigated in
\cite{Littenberg:2012uj}. Parameter estimation accuracy and systematic
errors depend mostly on GW phasing, which relies on long and accurate
simulations. Such long and accurate simulations are particularly hard
to achieve for compact binaries containing NSs. For NS-NS or BH-NS
binaries (unlike BH-BH binaries) the outcome of the merger depends on
several physical assumptions (e.g.~on the nuclear EOS, magnetic
fields, neutrino emission, and so on) that are currently poorly
constrained by laboratory experiments and astrophysical
observations. General relativistic simulations of NS-NS mergers,
however, have been studied for a long time, and they are now long and
accurate enough to be compared with analytical models (see
\cite{FaberRasiolrr-2012-8} for a review on the current status of
studies of coalescing binary NSs). These studies hold the promise to
constrain the EOS of matter at supranuclear densities, e.g.~via the
measurement of tidal deformation parameters: see
e.g.~\cite{Read:2009yp,Read:2013zra,Baiotti:2011am,Bernuzzi:2011aq,Bernuzzi:2012ci,Hotokezaka:2013mm}
for recent work in this area, with particular attention to
gravitational waveform accuracy.

For BH-NS systems, present simulations are comparatively short.
Difficulties arise because BH-NS binaries are expected to have
relatively large mass ratio, which causes complications for both
analytical and numerical approaches (see
\cite{ShibataTaniguchilrr-2011-6} for a review on the current status
of studies of coalescing BH-NS systems). For typical values of the
BH-NS mass ratio, the convergence of the PN approximation is expected
to be slower than in the NS-NS case (see e.g.~\cite{Berti:2007cd} for
a systematic study in the context of initial data). On the other hand,
numerical codes must track very different dynamical time scales,
making simulations heavily resource intensive (see
e.g.~\cite{Kyutoku:2011vz,Foucart:2012vn,Foucart:2013} for
investigations on the impact of the BH spin, the NS EOS, and realistic
mass ratios on the gravitational wave emission). Recent investigations
\cite{Nitz:2013mxa,Harry:2013tca} studied the impact of precession and
of different PN approximants on the detection of BH-NS binary
inspirals. The state of the art for gravitational waveform modeling
includes an attempt to incorporate higher harmonics in the inspiral
signal \cite{Cho:2012ed}, and recent work (involving some of us) on
the construction of hybrid waveforms to measure EOS parameters from
BH-NS mergers \cite{Lackey:2013axa}.

An important caveat in building phenomenological models is that, with
few recent exceptions \cite{Foucart:2013}, BH-NS simulations are
generally too short to guarantee accurate phasing estimates in the
whole parameter space.\footnote{For example, Hannam \etal
  \cite{Hannam2010} studied the minumum number of numerical waveform
  cycles that are necessary to ensure an accurate phase and amplitude
  modelling in the case of BH-BH binaries: this number grows with the
  binary mass ratio, i.e.~towards a more BH-NS like scenario. Issues
  in building hybrid (PN/EOB-numerical) GWs for BH-NS and NS-NS
  binaries are discussed in \cite{Lackey:2013axa} and
  \cite{Read:2013zra}, respectively.} Therefore, in this paper we
adopt a conservative approach and focus on the analytical modeling of
the GW {\em amplitude} in the frequency domain. For consistency and
continuity with previous studies, we find it convenient to build our
GW amplitude model on the foundations of the BH-BH PhenomC model
\cite{Santamaria:2010yb}. Additional parameters needed to reproduce
the complex phenomenology of BH-NS systems are introduced and tuned
using the simulations reported in
\cite{Kyutoku:2010zd,Lackey:2011vz}. Our main interest here is to
reproduce the high-frequency behavior of the GW spectrum reported in
these simulations, where most of the interesting EOS-related
phenomenology takes place. We will show that our model is useful to
improve signal-to-noise-ratio (SNR) estimates for BH-NS systems (which
depend only on the GW amplitude in the frequency domain) and to obtain
estimates of the cutoff frequency of the merger signal -- a
potentially measurable quantity -- in different physical scenarios.
The model can (and will) be improved as longer, more accurate
simulations become available. We note that the knowledge of the
amplitude alone allows one to estimate the best possible SNR, and that
this best possible SNR may be realized in searches only if the phase
of the waveform is known too.

The plan of the paper is as follows. In Sec.~\ref{sec:PhenomGWfs} we
anticipate our main results and provide a concise recipe to implement
our model. Section~\ref{sec:NRdata} reviews the numerical simulations
used to calibrate and verify the model, and
Sec.~\ref{sec:PhenoMixedGWfs} illustrates the logic we followed to
build the BH-NS model upon the BH-BH PhenomC model. In
Sec.~\ref{sec:results} we compare our model of GW spectra with all
binaries for which we have numerical data, in order to validate it and
test its accuracy. In Sec.~\ref{sec:SNRs} we compare SNRs obtained for
BH-NS binaries when using our model, the PhenomC model, and the
commonly employed restricted PN approximation. We use different noise
curves and conclude that, while the use of numerical waveforms can
induce SNR corrections as large as $\sim 10\%$ with respect to the
``standard'' restricted PN approximation, BH-BH phenomenological
models are accurate enough to compute the SNR even when used to model
(nonspinning) BH-NS mergers. In Sec.~\ref{sec:fCut} we use our model
to compute the cutoff frequency of the merger signal in the three
different physical scenarios (no tidal disruption, mild tidal
disruption, and strong tidal disruption). These results could find
application in theoretical calculations of tidal disruption, which in
turn can be used to assess the detectability of EOS effects in BH-NS
mergers. Appendix \ref{app:factors} clarifies the relation between
different conventions on the waveform amplitude used in the
literature. Finally, Appendix \ref{app:KAGRA} provides (to our
knowledge, for the first time) analytical fits to different proposed
KAGRA noise curves \cite{KAGRAnoise}. Throughout the paper, unless
specified otherwise, we use geometrical units ($G=c=1$).

%%%%%%%%%%%%%%%%%%%%%%%%%%%%%%%%%%%%%%%%%%%%%%%%%%%%%%%%%%%%%%%%%%%%%%%%%%%%%%
\section{Executive summary of the inspiral-merger-ringdown
  model}\label{sec:PhenomGWfs}
%%%%%%%%%%%%%%%%%%%%%%%%%%%%%%%%%%%%%%%%%%%%%%%%%%%%%%%%%%%%%%%%%%%%%%%%%%%%%%
We begin this paper by summarizing our model for the frequency domain
GW amplitude. This section is essentially a step-by-step recipe to
facilitate the implementation (and possibly improvement) of the model
in SNR calculations and data analysis codes. We begin by reviewing the
BH-BH PhenomC construction, and then list the modifications of PhenomC
that are necessary in order to reproduce the more complex
phenomenology of nonspinning BH-NS mergers. For the purpose of
computing SNRs, one may use fewer parameters than those used in our
model (Section \ref{sec:SNRs}). However our goal here is to reproduce
the rich phenomenology of BH-NS mergers and to predict the cutoff
frequency. In this paper we introduce the minimum number of parameters
that we found to be necessary for this purpose. Note that we could
have followed the PhenomC construction step by step, thus reabsorbing
all of our additional parameters in the fit of the PhenomC model. We
chose, however, not to follow this approach, and to show our ``BH-NS
corrections'' explicitly.

%%%%%%%%%%%%%%%%%%%%%%%%%%%%%%%%%%%%%%%%%%%%%%%%%%%%%%%%%%%%%%%%%%%%%%%%%%%%%%
\subsection{The black hole-black hole merger model}
%%%%%%%%%%%%%%%%%%%%%%%%%%%%%%%%%%%%%%%%%%%%%%%%%%%%%%%%%%%%%%%%%%%%%%%%%%%%%%
The BH-BH PhenomC waveforms of \cite{Santamaria:2010yb} are built in
the frequency domain, which is particularly convenient for SNR
calculations. In this section, unless otherwise noted and in
accordance with the conventions of \cite{Santamaria:2010yb}, all
frequencies are to be intended as multiplied by the sum $m_0=M_1+M_2$
of the two initial BH masses (in other words, we are using units in
which $m_0=1$). The amplitude $\tilde{A}_\text{Phen}(f)$ of the
frequency-domain signal
$\tilde{h}_\text{Phen}(f)=\tilde{A}_\text{Phen}(f)e^{i\Phi_\text{Phen}}$
is split in a sum of two terms:
\beq
\label{eq:PhenomAmp}
\tilde{A}_\text{Phen}(f) &= \tilde{A}_\text{PM}(f)w_{f_0,d}^-(f) +
\tilde{A}_\text{RD}(f)w_{f_0,d}^+(f)\,.
\eeq
Here $\tilde{A}_\text{PM}(f)$ models the inspiral and premerger
amplitude, and in turn it is decomposed in a sum of the form
\be
\tilde{A}_\text{PM}(f)=\tilde{A}_\text{PN}(f)+\gamma_1f^{5/6}\,,
\ee
where the first term is a PN inspiral contribution (the coefficients
of which are collected in the Appendix of \cite{Santamaria:2010yb}),
and the second term is intended to model premerger (strong-field)
modifications to the PN inspiral. The amplitude $\gamma_1$ of this
second contribution is fitted to BH-BH hybrid waveform data.

The second term of Eq.\,(\ref{eq:PhenomAmp}),
\be
\tilde{A}_\text{RD}(f)=\delta_1\mathcal{L}(f,f_\text{RD}(\chi_\text{f},m_0),
\delta_2\mathcal{Q}(\chi_\text{f}))f^{-7/6}\,,
\ee
is the ringdown amplitude, modeled with a Lorentzian
$\mathcal{L}(f,f_0,\sigma)\equiv \sigma^2/[(f-f_0)^2+\sigma^2/4]$. The
ringdown amplitude, $\delta_1$, is fitted to BH-BH hybrid waveform
data, and so is $\delta_2$, which accounts for inaccuracies in the fit
(taken from \cite{Rezzolla-etal-2007b}) used to calculate the remnant
BH spin parameter $\chi_\text{f}$. The remnant spin, $\chi_\text{f}$,
is used to determine the $l=m=2$, $n=0$ (later $220$, for brevity)
quasinormal mode (QNM) ringdown frequency, $f_\text{RD}$, and quality
factor, $\mathcal{Q}$, of the BH remnant, using the fitting formula
provided in \cite{BertiCardosoWill}. This ringdown frequency
$f_\text{RD}$ also depends on the mass of the BH remnant, which is
approximated to $m_0$. Finally, $w_{{f_0,d}}^\pm(f)$ are the windowing
functions
\beq
w_{{f_0,d}}^\pm(f)\equiv \f{1}{2}\left[ 1\pm
  \tanh\left(\f{4(f-f_0)}{d}\right)\right]\,,
\eeq
where the PhenomC model for BH-BH binaries sets $f_0 =0.98f_\text{RD}$
and $d=0.015$.

Summarizing, we have an IMR amplitude model for BH-BH gravitational
waveforms in the frequency-domain which contains: (1) a PN
contribution from the inspiral, (2) a premerger term with amplitude
$\gamma_1$ fitted to BH-BH hybrid waveforms, and (3) a ringdown term
given by a Lorentzian with overall amplitude set by the coefficient
$\delta_1$ and a second fitting coefficient $\delta_2$ (which accounts
for errors in the fit to $\chi_\text{f}$). All of these coefficients
are determined empirically by comparison with BH-BH hybrid waveforms.

%%%%%%%%%%%%%%%%%%%%%%%%%%%%%%%%%%%%%%%%%%%%%%%%%%%%%%%%%%%%%%%%%%%%%%%%%%%%%%
\subsection{The black hole-neutron star model:\\The algorithm, step by
  step}
\label{sec:PhenoMixedAlgorithm}
%%%%%%%%%%%%%%%%%%%%%%%%%%%%%%%%%%%%%%%%%%%%%%%%%%%%%%%%%%%%%%%%%%%%%%%%%%%%%%
In order to generalize the PhenomC model \cite{Santamaria:2010yb}, we
start by writing the frequency-domain GW amplitude of BH-NS mergers in
a similar way, i.e.~as a sum of three terms:
\begin{eqnarray}
\label{eq:PhenoMixed2}
\tilde{A}_\text{Phen}(f) &=&
\tilde{A}_\text{PN}(f)w_{\epsilon_\text{ins}\tilde{f}_0,d+\sigma_\text{tide}}^- \nn\\
&+& 1.25\gamma_1f^{5/6}w_{\tilde{f}_0,d+\sigma_\text{tide}}^- \nn\\
&+& \tilde{\mathcal{A}}_\text{RD}(f)w_{\tilde{f}_0,d+\sigma_\text{tide}}^+\,.
\end{eqnarray}
Here the PN inspiral contribution $\tilde{A}_\text{PN}(f)$, the
premerger amplitude $\gamma_1$ and the parameter $d=0.015$ are
identical to those used in \cite{Santamaria:2010yb}. The ringdown
amplitude is
\beq
\label{eq:RingdownAmp}
\tilde{\mathcal{A}}_\text{RD}(f)=\epsilon_\text{tide}\delta_1\mathcal{L}(f,
f_\text{RD}(\chi_\text{f},M_\text{f}),
\alpha\delta_2\mathcal{Q}(\chi_\text{f}))f^{-7/6}\,,~~
\eeq
where the coefficients $\delta_1$ and $\delta_2$ are again calculated
according to the fits reported in \cite{Santamaria:2010yb}. However,
we also introduce the correction parameters $\epsilon_\text{ins}$,
$\sigma_\text{tide}$, $\epsilon_\text{tide}$, and $\alpha$, and
replace $f_0$ with $\tilde{f}_0$. The first three parameters and
$\tilde{f}_0$ tend to $1$, $0$, $1$, and $f_0$ (respectively) in the
BH-BH limit, so that one recovers the binary BH model of
\cite{Santamaria:2010yb}. The parameter $\alpha$ and the factor of
$1.25$ appearing in Eq.\,(\ref{eq:PhenoMixed2}) are in apparent
discrepancy with the BH-BH limit, and we discuss them at length
below. In short, we believe that longer and more accurate comparisons
of BH-NS and BH-BH waveforms generated by different codes are required
to clarify the role of these parameters.

Given a BH-NS system formed by a nonspinning BH with gravitational
mass $M_\text{BH}$ and a nonspinning NS with gravitational mass
$M_\text{NS}$, baryonic mass $M_\text{b,NS}$, radius $R_\text{NS}$,
and compactness $\mathcal{C}=M_\text{NS}/R_\text{NS}$ (where all of
these quantities refer to BHs or NSs {\em in isolation}), the recipe
to use Eq.\,(\ref{eq:PhenoMixed2}) is the following:
\ben
\item Determine the coefficients $\gamma_1$, $\delta_1$, and
  $\delta_2$ using the BH-BH fits provided in
  \cite{Santamaria:2010yb}.
\item Determine the correction $\alpha$ to $\delta_2$ according to the
  fit in Eq.\,(\ref{eq:fit1}) below, where
  $\nu=M_\text{BH}M_\text{NS}/(M_\text{BH}+M_\text{NS})^2$ denotes the
  symmetric mass ratio.
\item Solve the equation 
  \begin{align}
    \label{eq:xi-tide}
    \frac{M_\text{NS}\xi_\text{tide}^3}{M_\text{BH}} =
    \frac{3[\xi_\text{tide}-2(M_\text{BH}/R_\text{NS})]}{\xi_\text{tide}-3(M_\text{BH}/R_\text{NS})}\,
  \end{align}
  for $\xi_\text{tide}$, a coefficient providing relativistic
  corrections to the standard Newtonian estimate of the orbital radius
  at \emph{mass shedding}: cf. Eq.\,(7) of \cite{Foucart:2012nc},
  which in turn builds upon classic results by
  \cite{Fishbone1973}. Then use this quantity to calculate the mass of
  the torus remaining around the NS at late times, $M_\text{b,torus}$,
  according to the fitting formula \cite{Foucart:2012nc}
  \begin{align}
    \label{eq:mtorus-fit}
    \frac{M_\text{b,torus}}{M_\text{b,NS}}=0.296\xi_\text{tide}(1-2\mathcal{C})-0.171\frac{r_\text{ISCO}}{R_\text{NS}}\,,
  \end{align}
  where $r_\text{ISCO}$ is the radius of the innermost stable circular
  orbit (ISCO) of the initial BH: see \cite{Bardeen:1972fi} and
  Eqs.\,(\ref{eqs:rISCO}) below. Notice the dependence of
  Eqs.\,(\ref{eq:xi-tide}) and (\ref{eq:mtorus-fit}) on the NS EOS
  (via the NS radius, its compactness, and its baryonic mass).
\item Calculate the spin parameter $\chi_\text{f}$ and gravitational
  mass $M_\text{f}$ of the BH remnant, following the model of
  \cite{Pannarale2012}. This requires solving numerically for
  $\chi_\text{f}$ the closed expression
  \begin{widetext}
  \begin{align}
    \chi_\text{f} = \frac{l_z(\bar{r}_\text{ISCO,f},\chi_\text{f})M_\text{BH}\{f(\nu)M_\text{b,NS}+[1-f(\nu)]M_\text{NS}-M_\text{b,torus}\}}{[m_0\left\{1-[1-e(\bar{r}_\text{ISCO,i},0)]\nu\right\}-e(\bar{r}_\text{ISCO,f},\chi_\text{f})M_\text{b,torus}]^2}\,,\nn\\
  \end{align}
  \end{widetext}
  where $M=M_\text{BH}+M_\text{NS}$,
  \begin{align}
    \label{eq:fbridge}
    f(\nu) = \left\{
      \begin{array}{ll}
        0 & \nu \leq 0.16 \\
        \frac{1}{2}\Big[1-\cos\Big(\frac{\pi(\nu - 0.16)}{2/9-0.16}\Big)\Big] & 0.16<\nu<2/9 \\
        1 & \nu\geq 2/9 \\
      \end{array}
    \right.\nn\\
  \end{align}
  and $l_z(\bar{r},\chi)$ and $e(\bar{r},\chi)$ denote the angular
  momentum and energy per unit mass of a point particle on a circular
  orbit of radius $\bar{r}$ around a Kerr BH of unit mass and
  dimensionless spin parameter $\chi$. They are given by the
  expressions
  \begin{align}
    \label{eq:e}
    e(\bar{r},\chi) &= \frac{\bar{r}^2-2\bar{r}\pm
      \chi\sqrt{\bar{r}}}{\bar{r}(\bar{r}^2-3\bar{r}\pm
      2\chi\sqrt{\bar{r}})^{1/2}}\,,\\
    \label{eq:lz}
    l_z(\bar{r},\chi) &= \pm \frac{\bar{r}^2\mp
      2\chi\sqrt{\bar{r}}+\chi^2}{\sqrt{\bar{r}}(\bar{r}^2-3\bar{r}\pm
      2\chi\sqrt{\bar{r}})^{1/2}}\,,
  \end{align}
  whereas the ISCO radii are calculated according to
  \begin{align}
    \bar{r}_\text{ISCO} &= [3+Z_2\mp\sqrt{(3-Z_1)(3+Z_1+2Z_2)}]\,,\nn\\
    Z_1 &= 1 +
    (1-\chi^2)^{1/3}\left[(1+\chi)^{1/3}+(1-\chi)^{1/3}\right]\,,\nn\\
    \label{eqs:rISCO}
    Z_2 &= \sqrt{3\chi^2+Z_1^2}\,.
  \end{align}
  The upper/lower sign holds for prograde/retrograde orbits
  \cite{Bardeen:1972fi}. Once $\chi_\text{f}$ is determined,
  $M_\text{f}$ follows from
  \begin{align}
    \label{eq:model-Mf}
    M_\text{f} &=
    m_0\left\{1-[1-e(\bar{r}_\text{ISCO,i},0)]\nu\right\}\nn\\
    &-M_\text{b,torus}e(\bar{r}_\text{ISCO,f},\chi_\text{f})\,.
  \end{align}
\item Compute the GW reference frequency for the \emph{onset} of the
  NS tidal disruption:
  \begin{align}
    \label{eq:ftide}
    f_\text{tide}=\pm\f{1}{\pi(\chi_\text{f}M_\text{f}+\sqrt{\tilde{r}_\text{tide}^3/M_\text{f}})}\,,
  \end{align}
  where the orbital radius at the onset of tidal disruption is
  \begin{align}\label{tidaldisr}
    \tilde{r}_\text{tide}=\xi_\text{tide}R_\text{NS}(1-2\mathcal{C})\,.
  \end{align}
\item Use $\chi_\text{f}$ and $M_\text{f}$ to determine the $220$
  ringdown frequency $f_\text{RD}$ and quality factor $\mathcal{Q}$ of
  the BH remnant, that is, following the fits of
  \cite{BertiCardosoWill}, calculate
  \begin{align}
    f_\text{RD} &= [1.5251 - 1.1568(1-\chi_\text{f})^{0.1292}]/(2\pi M_\text{f})\,,\\
    \mathcal{Q} &= 0.700 + 1.4187(1-\chi_\text{f})^{0.4990}\,.
   \end{align}
\item Set 
  \be
  \tilde{f}_0=\min [f_\text{tide}, \tilde f_\text{RD}]\,,
  \ee
  where $\tilde f_\text{RD} \equiv 0.99\times 0.98f_\text{RD}$.
\item[8a.] If $\tilde{f}_0=\tilde f_\text{RD}$, then the merger is
  ``nondisruptive,'' and the NS matter accretion is coherent until the
  plunge, so the merger and the pure inspiral contributions to
  Eq.\,(\ref{eq:PhenoMixed1}) end at the same frequency,
  i.e.~$\epsilon_\text{ins}=1$. $\epsilon_\text{tide}$ and
  $\sigma_\text{tide}$ are instead determined according to the fits of
  Sec.~\ref{sec:ND}.
\item[8b.] If $\tilde{f}_0=f_\text{tide}$ and $M_\text{b,torus}>0$,
  then the merger is ``disruptive,'' the NS material is scattered
  around, and the ringdown contribution to Eq.\,(\ref{eq:PhenoMixed1})
  vanishes, i.e.~$\epsilon_\text{tide}=0$. $\epsilon_\text{ins}\neq 1$
  and $\sigma_\text{tide}\neq 0$ are determined according to the fits
  of Sec.~\ref{sec:D}.
\item[8c.] If $\tilde{f}_0=f_\text{tide}$ and $M_\text{b,torus}=0$,
  then the merger is ``mildly disruptive.'' The parameter
  $\epsilon_\text{ins}$ is found as for the disruptive cases, while
  $\epsilon_\text{tide}$ is determined as for the nondisruptive cases,
  i.e.~the shutoff of the signal has an intermediate behavior between
  a tidal disruption shutoff and a QNM ringdown. $\sigma_\text{tide}$
  takes the value $0.041$.
\een

In Sec.~\ref{sec:NRdata} below we will review the numerical
simulations used to calibrate this model. Then, in
Sec.~\ref{sec:PhenoMixedGWfs} we will clarify how physical intuition
on tidal disruption was used to build the model itself.

%%%%%%%%%%%%%%%%%%%%%%%%%%%%%%%%%%%%%%%%%%%%%%%%%%%%%%%%%%%%%%%%%%%%%%%%%%%%%%%
\section{The Numerical Simulations}\label{sec:NRdata}
%%%%%%%%%%%%%%%%%%%%%%%%%%%%%%%%%%%%%%%%%%%%%%%%%%%%%%%%%%%%%%%%%%%%%%%%%%%%%%%
The numerical simulations used to calibrate our model adopt piecewise
polytropic EOSs, which are meant to reproduce nuclear-theory based
EOSs with a small number of polytropic constants $\kappa_i$ and
exponents $\Gamma_i$ \cite{Read:2008iy}:
\be
P(\rho)=\kappa_i \rho^{\Gamma_i}\quad {\rm for}\quad \rho_{i-1}\leq
\rho <\rho_i \quad (i=1\,,\dots\,,n)\,.
\ee
Since the pressure is required to be continuous, i.e.
\be
\kappa_i \rho_i^{\Gamma_i}=\kappa_{i+1} \rho_i^{\Gamma_{i+1}}\,,
\ee
the EOS is completely specified once we assign $\kappa_1$, $\Gamma_i$
and $\rho_i$ ($i=1\,\dots\,,n$). More specifically, we consider a
two-region piecewise polytrope\footnote{Strictly speaking, some of our
  piecewise polytropic EOSs (Bss, Bs, HBss, HBs, and Hss) do not
  support the recent observations of NSs with mass $M_{\rm NS}\geq
  2\,M_\odot$ \cite{Demorest:2010bx,Antoniadis:2013pzd}, but this does
  not necessarily mean that these EOS models are not realistic. The
  reason is that regions of very high density are not relevant for the
  BH-NS binaries studied here, where the NS typically has mass
  $M<1.4\,M_\odot$; therefore we can conceivably modify the
  high-density EOS in order to satisfy observational constraints on
  the maximum mass without altering our conclusions on gravitational
  waveforms from BH-NS binaries.}. We set the ``crustal'' polytropic
parameters to be $\Gamma_1=1.35692395$ and
$\kappa_1/c^2=3.99873692\times 10^{-8} ({\rm g}/{\rm
  cm}^3)^{1-\Gamma_1}$, and we vary $\Gamma_2$. Instead of specifying
$\rho_1$ we assign the pressure $P_{\rm fidu}$ at the fiducial density
$\rho_{\rm fidu}=10^{14.7}~{\rm g}/{\rm cm}^3$, because this parameter
is correlated with the NS radius. The relations
\be
P_{\rm fidu}=\kappa_2 \rho_{\rm fidu}^{\Gamma_2}\,, \quad
\kappa_1 \rho_1^{\Gamma_1}=\kappa_2 \rho_1^{\Gamma_2}\,,
\ee
then determine the values of $\kappa_2$ and $\rho_1$. 

The numerical runs used to build and test our model are collected in
Tables \ref{tab:NRruns2model} and \ref{tab:NRruns2tests}. The
beginning of each run name identifies the NS EOS, following the naming
scheme for piecewise polytropes introduced in
\cite{Read:2008iy,Read:2009yp,Kyutoku:2010zd}. In particular, 2H,
1.5H, 1.25H, H, HB, and B denote EOSs with the same value of the core
polytropic exponent $\Gamma_2 = 3.0$, but decreasing values of the
pressure at the fiducial density $\log
P_\text{fidu}=\{34.9,\,34.7,\,34.6,\,34.5,\,34.4,\,34.3\}$, so that 2H
is the stiffest and B is the softest EOS in the group; an appended
``l,'' ``s,'' or ``ss'' means that the EOS has the same $\log
P_\text{fidu}$ but a different value of $\Gamma_2$ (``l'' stands for
larger and ``s'' for ``smaller'' core polytropic exponent, so
$\Gamma_2 = 3.3,\,2.7,\,2.4$ for an EOS with appended ``l,'' ``s,'' or
``ss,'' respectively). The rest of the name encodes the NS mass (e.g.,
M$135$ means that $M_\text{NS}=1.35M_\odot$) and the binary mass ratio
$Q\equiv M_\text{BH}/M_\text{NS}$.

%TTTTTTTTTTTTTTTTTTTTTTTTTTTTTTTTTTTTTTTTTTTTTTTTTTTTTTTTTTTTTTTTTTTTTTTTTTTTTT
\begin{table}[!t]
  \caption{\label{tab:NRruns2model} Physical parameters of the
    numerical-relativity simulations used to develop the waveform
    model \cite{Kyutoku:2010zd,Lackey:2011vz}. The pressure at the
    fiducial density $10^{14.7}$g/cm$^3$ is in dyne/cm$^3$, whereas
    the NS mass is in solar masses.}
  \begin{tabular}{|l||cccc|cc|}
    \colrule
    \colrule
    Run label & $\Gamma_2$ & $\log P_\text{fidu}$ & $M_\text{NS}$ & $\mathcal{C}$ & $Q$ & $\chi_\text{i}$\\
    \colrule
    {\ttfamily B-M135-Q5}       & $3.0$ & $34.3$ & $1.35$ & $0.1819$ & $5$ & $0$ \\
    {\ttfamily H-M135-Q5}       & $3.0$ & $34.5$ & $1.35$ & $0.1624$ & $5$ & $0$ \\
    {\ttfamily 2H-M135-Q5}      & $3.0$ & $34.9$ & $1.35$ & $0.1309$ & $5$ & $0$ \\
    {\ttfamily B-M135-Q4}       & $3.0$ & $34.3$ & $1.35$ & $0.1819$ & $4$ & $0$ \\
    {\ttfamily H-M135-Q4}       & $3.0$ & $34.5$ & $1.35$ & $0.1624$ & $4$ & $0$ \\
    {\ttfamily 2H-M135-Q4}      & $3.0$ & $34.9$ & $1.35$ & $0.1309$ & $4$ & $0$ \\
    {\ttfamily B-M135-Q3}       & $3.0$ & $34.3$ & $1.35$ & $0.1819$ & $3$ & $0$ \\
    {\ttfamily HB-M135-Q3}      & $3.0$ & $34.4$ & $1.35$ & $0.1718$ & $3$ & $0$ \\
    {\ttfamily H-M135-Q3}       & $3.0$ & $34.5$ & $1.35$ & $0.1624$ & $3$ & $0$ \\
    {\ttfamily 2H-M135-Q3}      & $3.0$ & $34.9$ & $1.35$ & $0.1309$ & $3$ & $0$ \\
    {\ttfamily B-M135-Q2}       & $3.0$ & $34.3$ & $1.35$ & $0.1819$ & $2$ & $0$ \\
    {\ttfamily HB-M135-Q2}      & $3.0$ & $34.4$ & $1.35$ & $0.1718$ & $2$ & $0$ \\
    {\ttfamily H-M135-Q2}       & $3.0$ & $34.5$ & $1.35$ & $0.1624$ & $2$ & $0$ \\
    {\ttfamily 2H-M135-Q2}      & $3.0$ & $34.9$ & $1.35$ & $0.1309$ & $2$ & $0$ \\
    \colrule
    \colrule
  \end{tabular}
\end{table}
%TTTTTTTTTTTTTTTTTTTTTTTTTTTTTTTTTTTTTTTTTTTTTTTTTTTTTTTTTTTTTTTTTTTTTTTTTTTTTT

%TTTTTTTTTTTTTTTTTTTTTTTTTTTTTTTTTTTTTTTTTTTTTTTTTTTTTTTTTTTTTTTTTTTTTTTTTTTTTT
\begin{table}[!t]
  \caption{\label{tab:NRruns2tests} Physical parameters of the numerical-relativity simulations used to test the waveform model and assess its validity beyond the runs used to build it (see Table \ref{tab:NRruns2model} for additional information). The three groups of runs allow us to test the model for different values of $M_\text{NS}$, $\log P_\text{fidu}$ (for $\Gamma_2=3.0$), and $\Gamma_2$, respectively.}
  \begin{tabular}{|l||cccc|cc|}
    \colrule
    \colrule
    Run label & $\Gamma_2$ & $\log P_\text{fidu}$ & $M_\text{NS}$ & $\mathcal{C}$ & $Q$ & $\chi_\text{i}$\\
    \colrule
    {\ttfamily B-M12-Q2}        & $3.0$ & $34.3$ & $1.20$ & $0.1614$ & $2$ & $0$ \\
    {\ttfamily HB-M12-Q2}       & $3.0$ & $34.4$ & $1.20$ & $0.1527$ & $2$ & $0$ \\
    {\ttfamily H-M12-Q2}        & $3.0$ & $34.5$ & $1.20$ & $0.1447$ & $2$ & $0$ \\
    {\ttfamily 2H-M12-Q2}       & $3.0$ & $34.9$ & $1.20$ & $0.1172$ & $2$ & $0$ \\
    \colrule
    {\ttfamily 1.5H-M135-Q5}    & $3.0$ & $34.7$ & $1.35$ & $0.1456$ & $5$ & $0$ \\
    {\ttfamily 1.5H-M135-Q4}    & $3.0$ & $34.7$ & $1.35$ & $0.1456$ & $4$ & $0$ \\
    {\ttfamily 1.5H-M135-Q3}    & $3.0$ & $34.7$ & $1.35$ & $0.1456$ & $3$ & $0$ \\
    {\ttfamily 1.5H-M135-Q2}    & $3.0$ & $34.7$ & $1.35$ & $0.1456$ & $2$ & $0$ \\
    {\ttfamily 1.25H-M135-Q2}   & $3.0$ & $34.6$ & $1.35$ & $0.1537$ & $2$ & $0$ \\
    \colrule
    {\ttfamily Bl-M135-Q2}      & $3.3$ & $34.3$ & $1.35$ & $0.1798$ & $2$ & $0$ \\
    {\ttfamily HBl-M135-Q2}     & $3.3$ & $34.4$ & $1.35$ & $0.1719$ & $2$ & $0$ \\
    {\ttfamily Hl-M135-Q2}      & $3.3$ & $34.5$ & $1.35$ & $0.1638$ & $2$ & $0$ \\
    {\ttfamily 1.25Hl-M135-Q2}  & $3.3$ & $34.6$ & $1.35$ & $0.1565$ & $2$ & $0$ \\
    {\ttfamily 1.5Hl-M135-Q2}   & $3.3$ & $34.7$ & $1.35$ & $0.1497$ & $2$ & $0$ \\
    {\ttfamily Bs-M135-Q2}      & $2.7$ & $34.3$ & $1.35$ & $0.1856$ & $2$ & $0$ \\
    {\ttfamily HBs-M135-Q2}     & $2.7$ & $34.4$ & $1.35$ & $0.1723$ & $2$ & $0$ \\
    {\ttfamily Hs-M135-Q2}      & $2.7$ & $34.5$ & $1.35$ & $0.1605$ & $2$ & $0$ \\
    {\ttfamily 1.25Hs-M135-Q2}  & $2.7$ & $34.6$ & $1.35$ & $0.1497$ & $2$ & $0$ \\
    {\ttfamily 1.5Hs-M135-Q2}   & $2.7$ & $34.7$ & $1.35$ & $0.1399$ & $2$ & $0$ \\
    {\ttfamily Bss-M135-Q2}     & $2.4$ & $34.3$ & $1.35$ & $0.1941$ & $2$ & $0$ \\
    {\ttfamily HBss-M135-Q2}    & $2.4$ & $34.4$ & $1.35$ & $0.1741$ & $2$ & $0$ \\
    {\ttfamily Hss-M135-Q2}     & $2.4$ & $34.5$ & $1.35$ & $0.1577$ & $2$ & $0$ \\
    {\ttfamily 1.25Hss-M135-Q2} & $2.4$ & $34.6$ & $1.35$ & $0.1435$ & $2$ & $0$ \\
    {\ttfamily 1.5Hss-M135-Q2}  & $2.4$ & $34.7$ & $1.35$ & $0.1312$ & $2$ & $0$ \\
    \colrule
    \colrule
  \end{tabular}
\end{table}
%TTTTTTTTTTTTTTTTTTTTTTTTTTTTTTTTTTTTTTTTTTTTTTTTTTTTTTTTTTTTTTTTTTTTTTTTTTTTTT

Our simulations are such that the cases with $\Gamma_2=3.0$ span mass
ratios $Q=\{2,3,4,5\}$, whereas the cases with
$\Gamma_2=\{2.4,2.7,3.3\}$ all have $Q=2$. If we were to use all the
data we would run the risk of being biased by core stiffness effects
for low BH masses. In this first paper, we therefore decided to build
the IMR GW amplitude model on the runs with $\Gamma_2=3.0$ EOSs, which
span all the available values of $Q$, and to leave the $Q=2$,
$M_\text{NS}=1.35M_\odot$, $\Gamma_2\neq 3$ ones (last part of Table
\ref{tab:NRruns2tests}) as test cases to verify the validity of the
model against variations of the NS core polytropic exponent. Further,
we select a subset of $\Gamma_2=3.0$ simulations as additional test
cases. These are nine simulations: four with $M_\text{NS}=1.2M_\odot$
(first block in Table \ref{tab:NRruns2tests}), and five with EOS
$1.5$H or $1.25$H (second block in Table \ref{tab:NRruns2tests}). The
reasoning behind this procedure is that once the model is built, we
may test its validity against binaries with $M_\text{NS}\neq
1.35M_\odot$ and a subset of EOSs with $\Gamma_2=3.0$ that were not
used to calibrate it. The $\Gamma_2=3.0$, $1.5$H runs have the
additional benefit of allowing us to test the model over different
values of $Q$.

Once again, when more runs for $\Gamma_2\neq 3.0$ and $Q\neq 2$ will
be available, we plan to generalize the model so that the dependence
on $\Gamma_2$ somehow appears explicitly. This may require using the
NS Love numbers (as opposed to the NS compactness) in our expressions
and fits used to determine the GW spectrum
(cf. \cite{Lackey:2013axa}).

%%%%%%%%%%%%%%%%%%%%%%%%%%%%%%%%%%%%%%%%%%%%%%%%%%%%%%%%%%%%%%%%%%%%%%%%%%%%%%%
\section{Modeling Nonspinning Black Hole-Neutron Star
  Waveforms}\label{sec:PhenoMixedGWfs}
%%%%%%%%%%%%%%%%%%%%%%%%%%%%%%%%%%%%%%%%%%%%%%%%%%%%%%%%%%%%%%%%%%%%%%%%%%%%%%%
As discussed in Sec.~\ref{sec:PhenomGWfs}, the phenomenological
coefficients $\gamma_1$, $\delta_1$, and $\delta_2$ appearing in the
PhenomC model of \cite{Santamaria:2010yb} are fitted to BH-BH hybrid
waveforms. It is natural to expect that these coefficients should be
somehow corrected in a BH-NS binary model. A second aspect one may
{\em a priori} envisage to modify, in the spirit of the PhenomC model,
is the connection between the inspiral and premerger phases, which in
the BH-BH case are ``turned off'' together using a unique windowing
function $w_{f_0,d}^-$. In the PhenomC model the $\gamma_1f^{5/6}$
term of Eq.\,(\ref{eq:PhenomAmp}) is supposed to represent the merger
contribution to the amplitude, so we should somehow separate it from
$\tilde{A}_\text{PN}(f)$ in the case of mixed binaries: the NS can be
elongated and disrupted prior to merger, so the behavior of a BH-NS
premerger can differ significantly from the BH-BH case. These
considerations lead us to the following generalization of the
amplitude model in Eq.\,(\ref{eq:PhenomAmp}):
\beq
\tilde{A}_\text{Phen}(f) &=& \tilde{A}_\text{PN}(f)w_{f_1,d_1}^- \nn\\
&+& \kappa\gamma_1f^{5/6}w_{f_2,d_3}^- \nn\\
&+& \tilde{\mathcal{A}}_\text{RD}(f)w_{f_3,d_3}^+\,,
\eeq
where we explicitly wrote down three different frequencies and widths
for the windowing functions, and we introduced a correction $\kappa$
to the BH-BH coefficient $\gamma_1$. For ease of comparison, we choose
to model the PN contribution $\tilde{A}_\text{PN}(f)$ using the same
PN approximation for the inspiral as in \cite{Santamaria:2010yb},
i.e.~we neglect tidal effects on the GW amplitude of the inspiral (see
e.g.~\cite{Baiotti:2010xh,Kyutoku:2010zd,Bernuzzi:2011aq} for recent
studies of the influence of tidal effects in the late inspiral). The
ringdown amplitude, $\tilde{\mathcal{A}}_\text{RD}$, is now modified
as explained in Eq.\,(\ref{eq:RingdownAmp}). Once more, the functional
form is similar to the $\tilde{A}_\text{RD}$ used in the model of
\cite{Santamaria:2010yb}, but the BH-BH fitted quantities $\delta_1$
and $\delta_2$ are corrected by two new ``fudge factors''
$\epsilon_\text{tide}$ and $\alpha$, respectively. The parameter
$\epsilon_\text{tide}$ plays a physical role: it tends to unity when
tidal effects are irrelevant and the merger has a BH-BH-like behavior;
vice versa, it tends to zero when tidal effects take over (so that the
ringdown of the BH remnant is not strongly excited). The reason for
introducing $\alpha$ is, instead, that BH-BH PhenomC waveforms were
constructed against binaries the large majority of which has $Q<4$
(i.e., $\nu > 4/25 = 0.16$), whereas the BH-NS binaries considered here
may also have larger mass ratios: therefore $\alpha$ effectively
corrects the fit for $\delta_2$ of the PhenomC waveforms, which is
biased towards low mass ratios. Additionally, $\alpha$ takes care of
the fact that in our model we allow the mass of the BH remnant in a
BH-NS merger to be different from $M_\text{BH}+M_\text{NS}$ (at
variance with \cite{Santamaria:2010yb}); it also compensates for using
the BH-NS model of \cite{Pannarale2012}, rather than the BH-BH model
of \cite{Rezzolla-etal-2007b}, in our prediction of the final spin
$\chi_\text{f}$.

When the BH-NS coalescence is nondisruptive, this generalization
should reduce to Eq.\,(\ref{eq:PhenomAmp}). Furthermore, the three IMR
contributions should be connected smoothly. In order to satisfy these
constraints we set
\beq
d_i = d + \sigma_\text{tide}\,,
\eeq
where $d=0.015$ is the BH-BH windowing width and $\sigma_\text{tide}$
is a tidal (or finite-size) correction to $d$, which must tend to zero
as NS tidal distortions become weaker. Next, and again for smoothness
and continuity reasons, we set $f_2=f_3$. For physical reasons, we
also impose $f_1\leq f_2$: this means that the ending frequency of the
``pure'' inspiral contribution should never be larger than the ending
frequency of the merger contribution. We expect that $f_1\rightarrow
f_2$ as the coalescence becomes more and more BH-BH-like and tidal
effects become more and more negligible, so we write
$f_1=\epsilon_\text{ins}f_2$, with $0<\epsilon_\text{ins}\leq 1$.

All in all, we have 
\beq
\label{eq:PhenoMixed1}
\tilde{A}_\text{Phen}(f) &=&
\tilde{A}_\text{PN}(f)w_{\epsilon_\text{ins}\tilde{f}_0,d+\sigma_\text{tide}}^- \nn\\
&+& \kappa\gamma_1f^{5/6}w_{\tilde{f}_0,d+\sigma_\text{tide}}^- \nn\\
&+&
\tilde{\mathcal{A}}_\text{RD}(f)w_{\tilde{f}_0,d+\sigma_\text{tide}}^+\,,
\eeq
where we made the notational change $f_2\to \tilde{f}_0$, hinting to
the fact that this frequency should be close to the $220$ QNM ringdown
frequency of the BH remnant as tidal effects become smaller.

Equation (\ref{eq:PhenoMixed1}) is our general framework, and from now
on we specify the details of our model. Let us begin by examining the
BH remnant. In the case of the BH-BH phenomenological GW model in
Eq.\,(\ref{eq:PhenomAmp}), the fit of \cite{Rezzolla-etal-2007b} is
used to predict the spin parameter $\chi_\text{f}$ of the BH remnant,
and its mass $M_\text{f}$ is assumed to be equal to $m_0$. This
enables one to calculate $f_\text{RD}$ and $\mathcal{Q}$, which enter
the ringdown amplitude model through $\tilde{A}_\text{RD}$. This
approach cannot reproduce all data for the BH remnant of BH-NS binary
mergers accurately enough, as it is based on results of BH-BH merger
simulations. We may use the model of \cite{Pannarale2012} to predict
$\chi_\text{f}$ and $M_\text{f}$ more accurately. The approach of
\cite{Pannarale2012} relies on the prediction of \cite{Foucart:2012nc}
for the mass $M_\text{b,torus}$ of the torus remnant (possibly)
produced by a BH-NS coalescence. Reference \cite{Foucart:2012nc} also
provides a prediction for the orbital radius at the onset of tidal
disruption, i.e.~at mass shedding. This is given in
Eq.\,(\ref{tidaldisr}), where $\xi_\text{tide}$ is the solution of
Eq.\,(\ref{eq:xi-tide}). The prediction for $\tilde{r}_\text{tide}$
relies on several approximations (e.g.~it is coordinate dependent),
but it may be usefully exploited in combination with the predictions
for $\chi_\text{f}$ and $M_\text{f}$ to define a GW reference
frequency as in Eq.\,(\ref{eq:ftide}). This is consistent with the
calculation of $M_\text{b,torus}$, and tends to infinity as
$\mathcal{C}\rightarrow 1/2$ (that is, in the BH limit). In addition
to writing out Eq.\,(\ref{eq:PhenoMixed1}), we have thus chosen how to
determine $\chi_\text{f}$ and $M_\text{f}$ (and hence how to calculate
the $f_\text{RD}$ and $\mathcal{Q}$ entering
$\tilde{\mathcal{A}}_\text{RD}$), and introduced the useful frequency
$f_\text{tide}$.

We then proceed by finding the values of $\kappa$, $\alpha$,
$\epsilon_\text{tide}$, $\sigma_\text{tide}$, $\epsilon_\text{ins}$,
and $\tilde{f}_0$ which best reproduce the high-frequency behavior of
our numerical GW amplitude.  As a final step, we must look for
correlations between the values we found for $\kappa$, $\alpha$,
$\epsilon_\text{tide}$, $\sigma_\text{tide}$, $\epsilon_\text{ins}$,
and $\tilde{f}_0$, on one hand, and binary and remnant parameters, on
the other.

In this study we focus on high frequencies for two reasons: (1) we are
interested in the regime where the BH-NS phenomenology departs from
the BH-BH case, and EOS-dependent effects emerge at high frequencies;
(2) current simulations do not allow us to accurately handle the
inspiral regime (for example, we do not make attempts to reduce the
residual orbital eccentricity). In this regard, it is useful to remark
that the influence of resolution on frequency-domain GW spectra was
investigated in Fig.~15 of \cite{Kyutoku:2010zd}. That study suggests
that systematic errors due to resolution should be subdominant
compared with effects due to (i) the finite length of the simulations,
and (ii) the residual eccentricity of the initial data. These
important aspects should be addressed with future, longer, and more
accurate numerical simulations, that will presumably reduce the
differences between the numerical data and the analytical PN-based
description of the inspiral regime.

%FFFFFFFFFFFFFFFFFFFFFFFFFFFFFFFFFFFFFFFFFFFFFFFFFFFFFFFFFFFFFFFFFFFFFFFFFFFFF
\begin{figure}[!t]
\includegraphics[scale=0.35,clip=true]{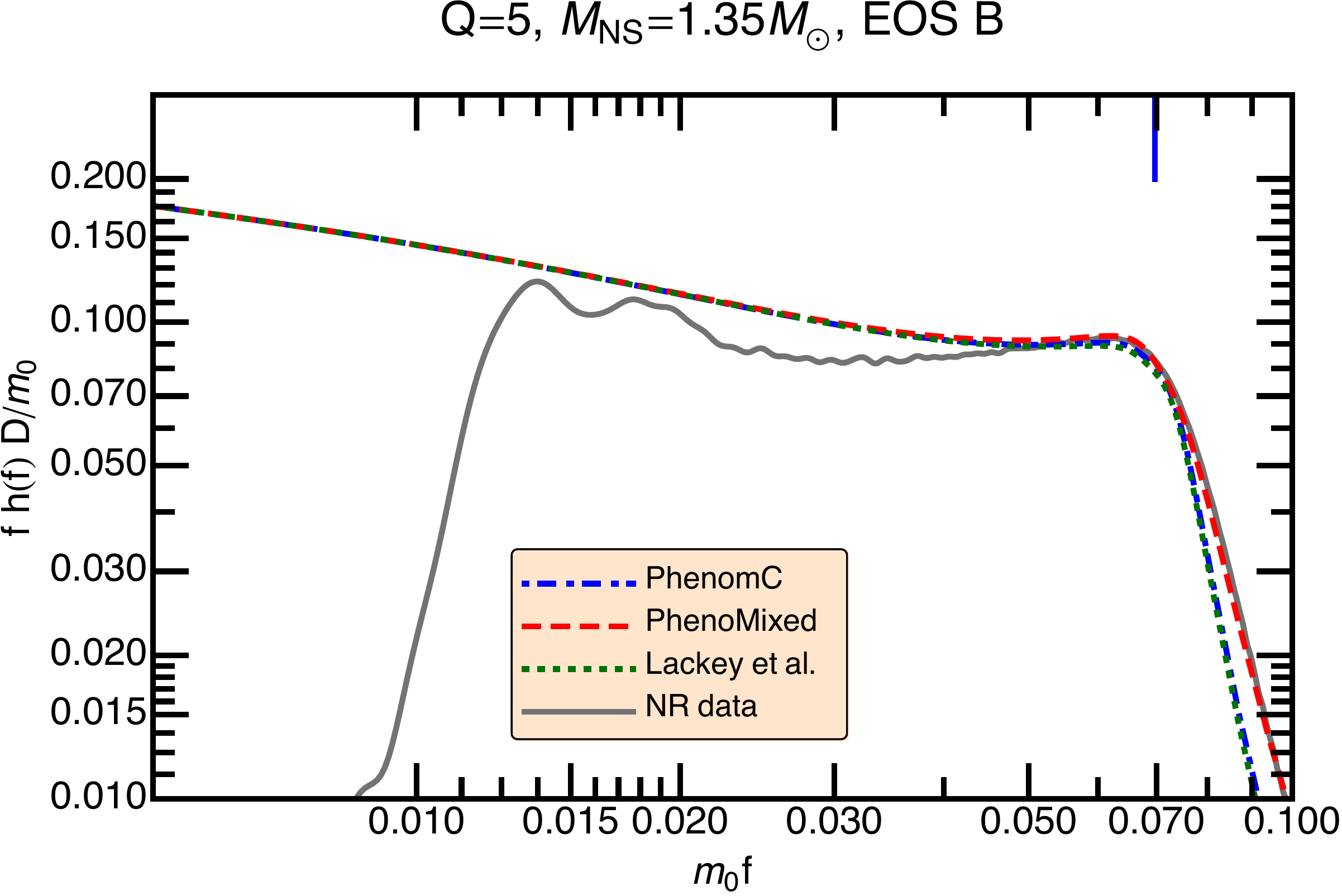}
\caption{GW amplitude of a BH-NS binary merger with a $1.35M_\odot$ NS
  with EOS B, a nonspinning BH, and a mass ratio $Q=5$ (Table
  \ref{tab:NRruns2model}, {\ttfamily B-M135-Q5}). The GW strain of the
  NR simulation (continuous grey) is compared to the BH-BH
  phenomenological model of \cite{Santamaria:2010yb} (dot-dashed
  blue), to the BH-NS model of \cite{Lackey:2013axa} (dotted green),
  and to our phenomenological model (dashed red). The location of
  $m_0\tilde{f}_\text{RD}$ is shown by a short, blue, vertical line in
  the top of the graph; the mass-shedding frequency $m_0
  f_\text{tide}$ is higher than $0.1$. \label{Fig:MNS135Q5EOSB}}
\end{figure}
%FFFFFFFFFFFFFFFFFFFFFFFFFFFFFFFFFFFFFFFFFFFFFFFFFFFFFFFFFFFFFFFFFFFFFFFFFFFFF

%%%%%%%%%%%%%%%%%%%%%%%%%%%%%%%%%%%%%%%%%%%%%%%%%%%%%%%%%%%%%%%%%%%%%%%%%%%%%%%
\subsection{Black hole-black hole like mergers}
%%%%%%%%%%%%%%%%%%%%%%%%%%%%%%%%%%%%%%%%%%%%%%%%%%%%%%%%%%%%%%%%%%%%%%%%%%%%%%%
Let us first consider a ``BH-BH-like'' mixed binary merger with the
softest EOS, namely the $Q=5$ case with EOS B and
$M_\text{NS}=1.35M_\odot$ ({\ttfamily B-M135-Q5} in Table
\ref{tab:NRruns2model}). No torus is produced in this merger, and
$f_\text{tide}>f_\text{RD}$. In Fig.~\ref{Fig:MNS135Q5EOSB} we show
the numerical data for the GW spectrum (grey curve), the prediction of
the PhenomC model of \cite{Santamaria:2010yb} (blue, dot-dashed
curve), the prediction of the BH-NS model of Lackey and collaborators
\cite{Lackey:2013axa} (green, dotted curve), and the prediction of
Eq.\,(\ref{eq:PhenoMixed1}) with $\kappa$, $\alpha$,
$\epsilon_\text{tide}$, $\sigma_\text{tide}$, $\epsilon_\text{ins}$,
and $\tilde{f}_0$ tuned to mimic the high-frequency behavior of the
numerical data (red, dashed curve). Throughout this paper, in the GW
strain plots, we find it convenient to use the dimensionless frequency
$m_0f$, in place of $f$. As a reference for quick conversions between
the two, the following formula may be used:
\begin{align}
\label{eq:frequency-conversion}
f=2030\,\text{Hz}\times\frac{m_0f}{0.01}\frac{1}{\hat{M}_\text{NS}(1+Q)}\,,
\end{align}
where $\hat{M}_\text{NS}$ is the NS mass in solar mass units. For
example, for a system with mass ratio $Q=5$ and a NS of $1.35M_\odot$,
$m_0f=0.1$ corresponds to $\sim 2500\,$Hz.

Note that the initial match between the numerical data and the
analytic spectra is not obtained ``artificially'' by rescaling the
data to achieve the matching. It is, instead, obtained mathematically
by a careful comparison of numerical and analytical conventions on the
waveform amplitude (see Appendix \ref{app:factors}). As mentioned
previously, the deviation from the matching is then due to residual
eccentricity in the numerical simulation (see e.g.~Fig.~5 in
\cite{Ajith2008}) and to the sudden onset of the GW emission: both
features can be cured, the former numerically and the latter
analytically, but addressing these aspects is beyond the scope of this
paper. Setting $\tilde{f}_0=\tilde f_\text{RD}\equiv 0.99\times 0.98
f_\text{RD}$ provides a better high-frequency matching than the BH-BH
PhenomC prescription of using $0.98 f_\text{RD}$. As anticipated, this
small difference is not surprising, as the ringdown frequency is
following from a different model for the properties of the BH
remnant. This prescription for $\tilde{f}_0$ works for all
nondisruptive coalescences. We set $\kappa=1.25$ to achieve a better
matching of the knee in the waveform spectrum. This same rescaling of
$\gamma_1$ works for all other waveforms used to build our model. With
the data currently available, the nature of this correction is
unclear: it could be a ``universal'' correction, a consequence of
residual eccentricity in the data, an artifact of trying to match GWs
that are not the BH-BH hybrid ones with $\gamma_1$ values obtained
from the BH-BH hybrid waveforms themselves, or it could have some
other origin. It is reassuring, though, that the rescaling is
unique. Perhaps longer, more accurate waveforms will lead to the
conclusion that $\gamma_1$ should not be rescaled when passing from
BH-BH to BH-NS mergers. We set $\sigma_\text{tide}=0$,
$\epsilon_\text{ins}=1$, and $\epsilon_\text{tide}=1$, so that $d$
(the shutoff frequency of the PN inspiral contribution) and $\delta_1$
were not corrected: these values match our expectations, given the
nondisruptive nature of this specific coalescence. Finally, the
``fudge factor'' correcting $\delta_2$ was set to $\alpha=1.35$.

%%%%%%%%%%%%%%%%%%%%%%%%%%%%%%%%%%%%%%%%%%%%%%%%%%%%%%%%%%%%%%%%%%%%%%%%%%%%%%%
\subsection{Stiffening the EOS: Tidal effects}\label{sec:classification}
%%%%%%%%%%%%%%%%%%%%%%%%%%%%%%%%%%%%%%%%%%%%%%%%%%%%%%%%%%%%%%%%%%%%%%%%%%%%%%%

%FFFFFFFFFFFFFFFFFFFFFFFFFFFFFFFFFFFFFFFFFFFFFFFFFFFFFFFFFFFFFFFFFFFFFFFFFFFFF
\begin{figure}[b]
\includegraphics[scale=0.35,clip=true]{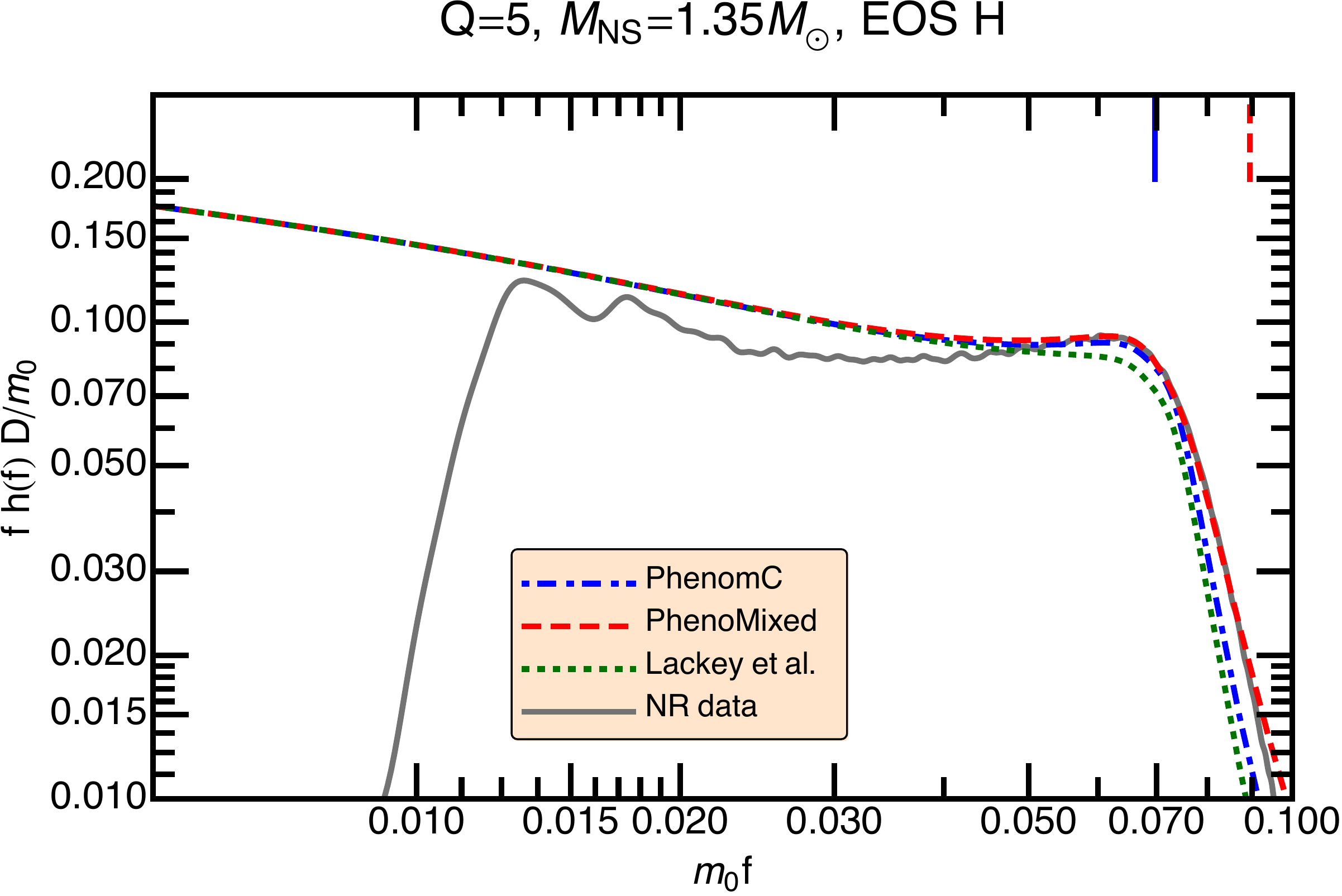}
\caption{BH-NS merger case {\ttfamily H-M135-Q5} (see Table
  \ref{tab:NRruns2model}). The GW strain of the NR simulation
  (continuous grey) is compared to the BH-BH phenomenological model of
  \cite{Santamaria:2010yb} (dot-dashed blue), to the BH-NS model of
  \cite{Lackey:2013axa} (dotted green), and to our BH-NS
  phenomenological model (dashed red). The location of the QNM
  frequency of the BH remnant is shown by the short, vertical, blue
  line in the top of the graph; the frequency at the onset of
  disruption frequency $m_0 f_\text{tide}$ is indicated by the short,
  vertical, dashed, red line. \label{Fig:MNS135Q5EOSH}}
\end{figure}
%FFFFFFFFFFFFFFFFFFFFFFFFFFFFFFFFFFFFFFFFFFFFFFFFFFFFFFFFFFFFFFFFFFFFFFFFFFFFF

In Fig.~\ref{Fig:MNS135Q5EOSH} we repeat the procedure for the $Q=5$,
$M_\text{NS}=1.35M_\odot$, EOS H binary ({\ttfamily H-M135-Q5} in
Table \ref{tab:NRruns2model}). We set once again $\kappa=1.25$ (note
that this value of $\kappa$ seems to always give a good match,
independently of the EOS stiffness), $\epsilon_\text{ins}=1$,
$\epsilon_\text{tide}=1$, $\sigma_\text{tide}=0$, $\alpha=1.35$, and
$\tilde{f}_0=\tilde f_\text{RD}$. The frequency at the onset of tidal
disruption $f_\text{tide}$, marked by the short, vertical, dotted red
line, is now closer to $\tilde{f}_\text{RD}$, marked by the short,
vertical blue line. This means that the EOS stiffening gradually
increases the relevance of tidal effects. The behavior of this merger,
however, is still very BH-BH-like.

%FFFFFFFFFFFFFFFFFFFFFFFFFFFFFFFFFFFFFFFFFFFFFFFFFFFFFFFFFFFFFFFFFFFFFFFFFFFFF
\begin{figure}[!t]
  \includegraphics[scale=0.35,clip=true]{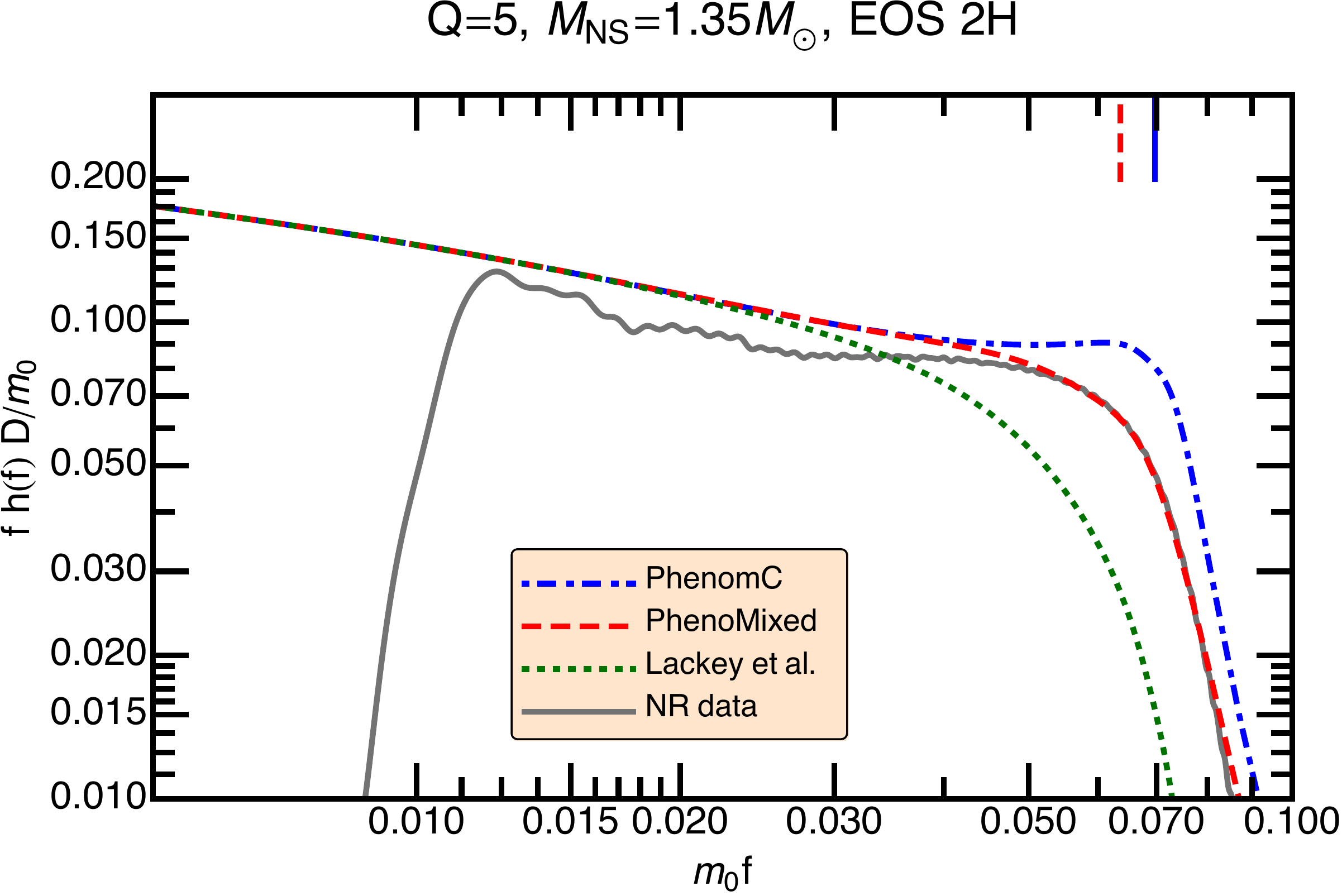}
  \caption{BH-NS merger {\ttfamily 2H-M135-Q5} (see Table
    \ref{tab:NRruns2model}). In this plot (and in the remainder of
    this paper) we use the same color and linestyle conventions as in
    Fig.~\ref{Fig:MNS135Q5EOSH}. \label{Fig:MNS135Q5EOS2H}}
\end{figure}
%FFFFFFFFFFFFFFFFFFFFFFFFFFFFFFFFFFFFFFFFFFFFFFFFFFFFFFFFFFFFFFFFFFFFFFFFFFFFF

By further stiffening the EOS, we eventually hit a disruptive merger,
for which $f_\text{tide}<f_\text{RD}$. This case is reported in
Fig.~\ref{Fig:MNS135Q5EOS2H}, where we consider data from the run
{\ttfamily 2H-M135-Q5} in Table \ref{tab:NRruns2model}. In this case,
setting $\epsilon_\text{ins}=1$ and $\alpha=1.35$, but
$\epsilon_\text{tide}=0.49$, $\sigma_\text{tide}=0.041$, and
$\tilde{f}_0=f_\text{tide}$ is required. As anticipated,
$\sigma_\text{tide}$ and $\epsilon_\text{tide}$ need to be greater
than $0$ and smaller than $1$, respectively, since this is a
disruptive merger. This run shows that we must require the following
prescription on the ``windowing'' frequency:
\beq
\label{eq:ourf0prescrition}
\tilde{f}_0=\min [f_\text{tide}, \tilde f_\text{RD}]\,.
\eeq
In the $\mathcal{C}\rightarrow 1/2$ limit, i.e.~in the nonspinning
BH-BH limit, $\tilde{f}_0\equiv \tilde f_\text{RD}=0.99\times
0.98f_\text{RD}$. This differs slightly from the $f_0=0.98f_\text{RD}$
used in \cite{Santamaria:2010yb}. The origin of this discrepancy is
(again) in the different methods used to calculate the BH remnant
parameters determining $f_\text{RD}$.

Incidentally, the plots (and in particular
Fig.~\ref{Fig:MNS135Q5EOS2H}) show that while the amplitude model
developed by Lackey \etal~\cite{Lackey:2013axa} is accurate enough for
BH-BH-like mergers, it becomes increasingly inaccurate when the NS EOS
is particularly stiff (the same conclusion seems to hold in all other
cases we have investigated). A possible origin of the discrepancy may
be the fact that the model of Lackey and collaborators was calibrated
also to the EOSs with $\Gamma_2=2.4$, $2.7$, and $3.3$ ($15$
altogether), in addition to the six $\Gamma_2=3.0$ EOSs, for a total
of $21$ EOSs. The 2H EOS is just one EOS out of $21$ in this
catalog. The price to pay for fitting more EOSs may thus be that
results for the 2H EOS are less accurate.

%FFFFFFFFFFFFFFFFFFFFFFFFFFFFFFFFFFFFFFFFFFFFFFFFFFFFFFFFFFFFFFFFFFFFFFFFFFFFF
\begin{figure}[!t]
  \includegraphics[scale=0.32,clip=true]{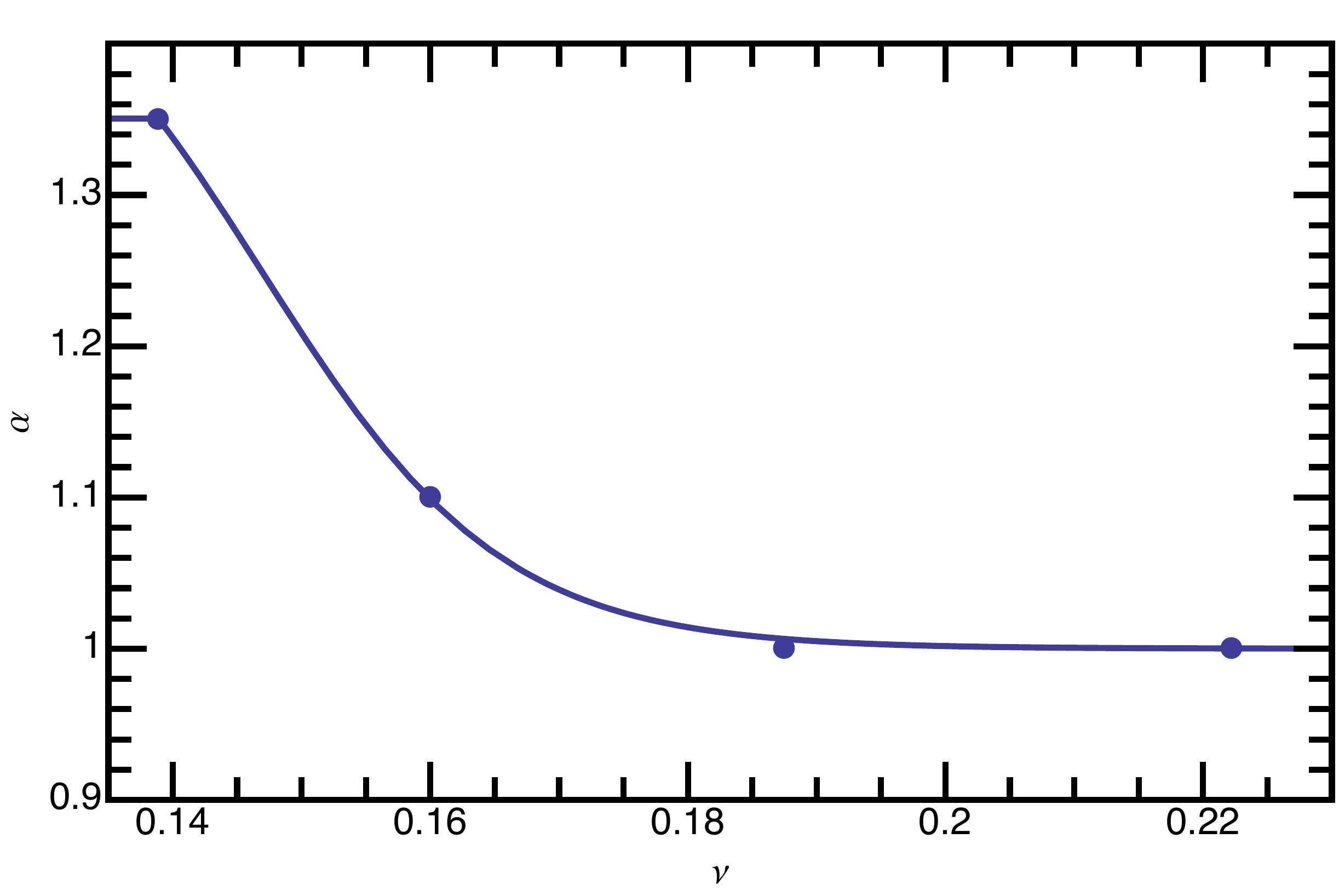}
  \caption{Values of $\alpha$ (the factor correcting $\delta_2$)
    versus $\nu$. The fitting function is of the form
    $w_{\nu_0,d_0}^-(\nu)+1$, with $\nu_0=0.146872$ and
    $d_0=0.0749456$, for $\nu\geq 5/36$, i.e.~for $Q\leq 5$. For $\nu<
    5/36$ we assume $\alpha$ to be constant and equal to
    $w_{\nu_0,d_0}^-(5/36)+1\simeq 1.35$.
    \label{Fig:Fit1}}
\end{figure}
%FFFFFFFFFFFFFFFFFFFFFFFFFFFFFFFFFFFFFFFFFFFFFFFFFFFFFFFFFFFFFFFFFFFFFFFFFFFFF

%FFFFFFFFFFFFFFFFFFFFFFFFFFFFFFFFFFFFFFFFFFFFFFFFFFFFFFFFFFFFFFFFFFFFFFFFFFFFF
\begin{figure*}[htb]
  \begin{tabular*}{\textwidth}{c@{\extracolsep{\fill}}c}
  \includegraphics[scale=0.32,clip=true]{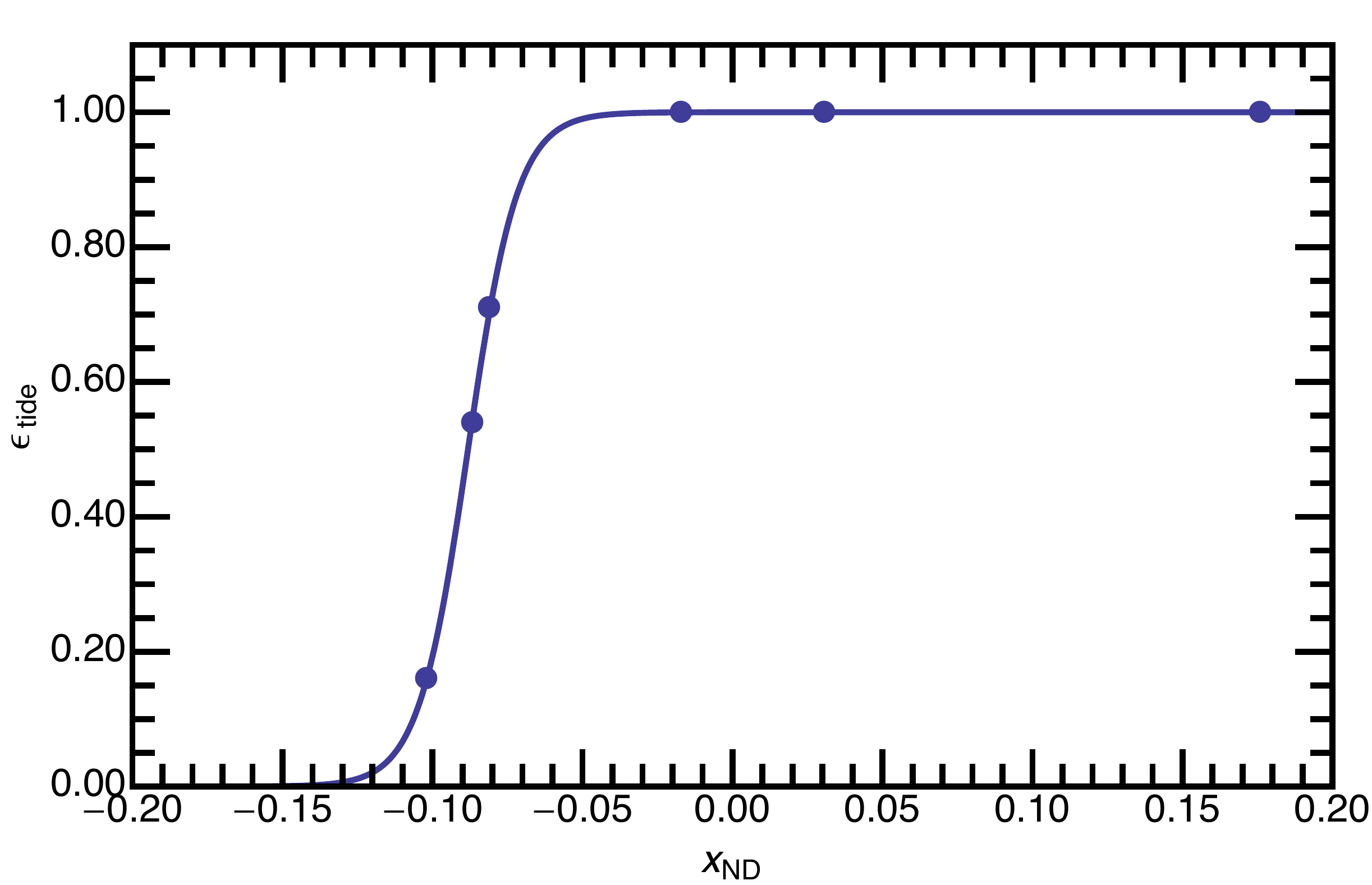}
  &
  \includegraphics[scale=0.32,clip=true]{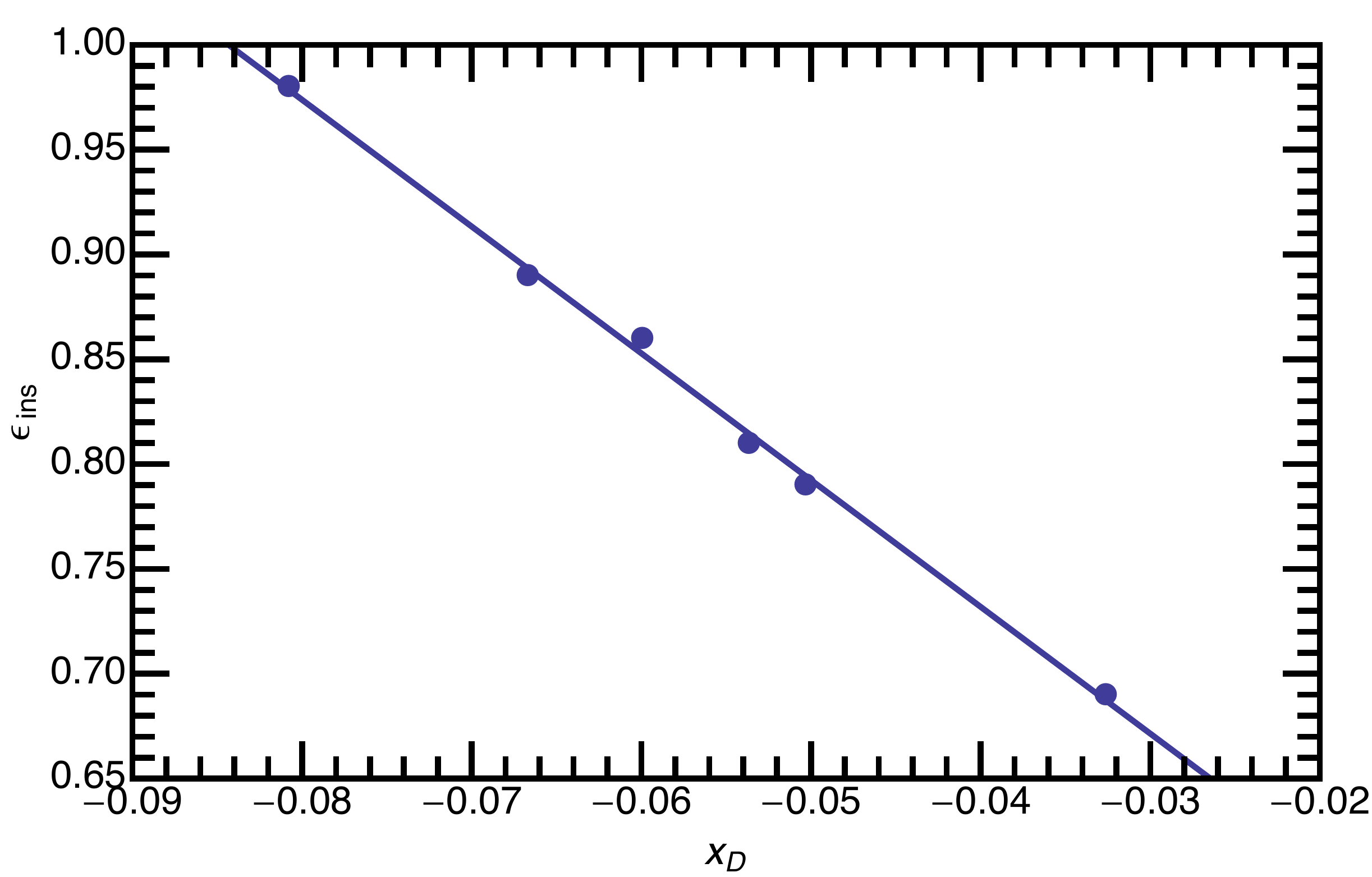}
  \\
  \includegraphics[scale=0.32,clip=true]{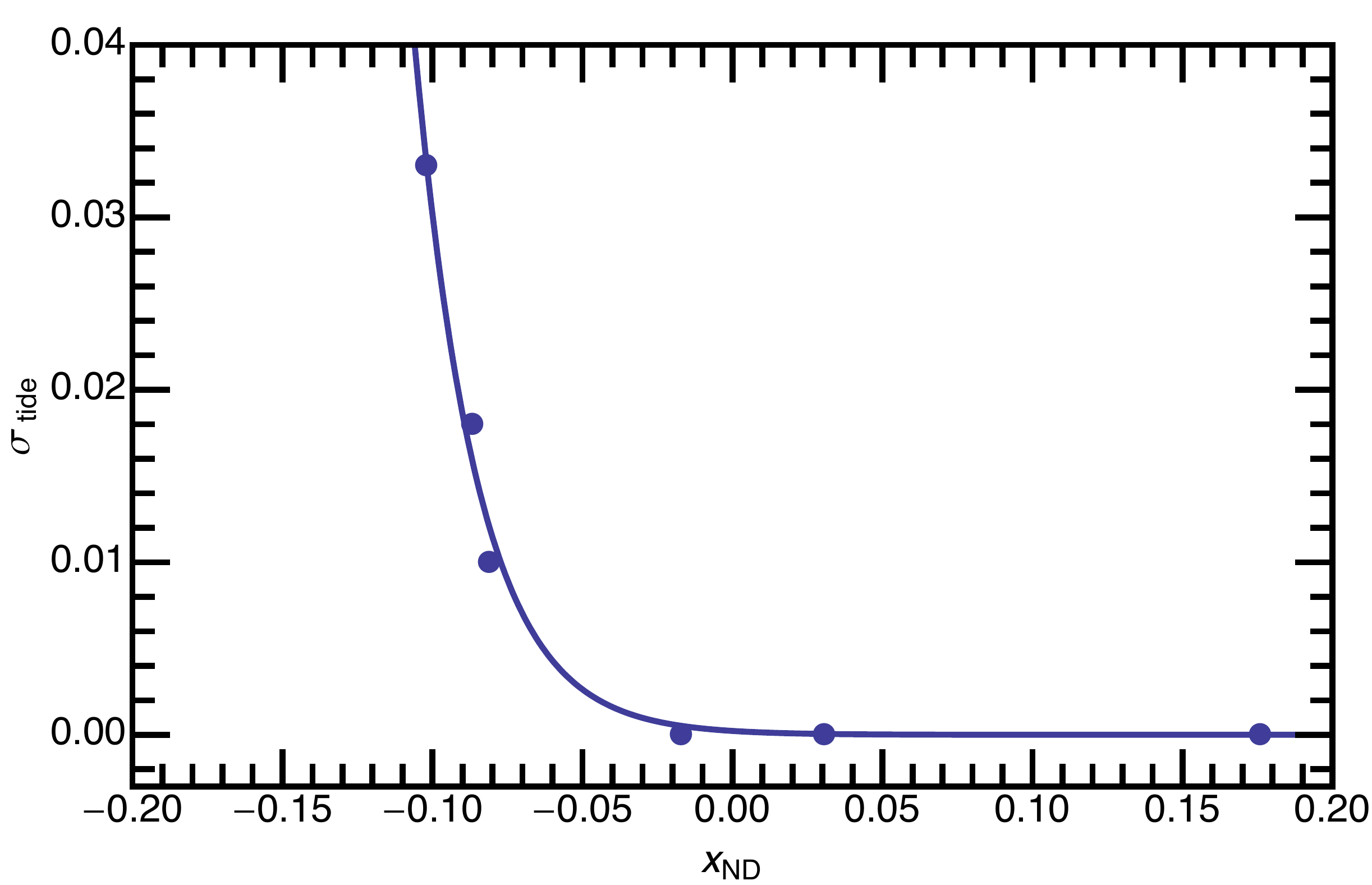}
  &  
  \includegraphics[scale=0.32,clip=true]{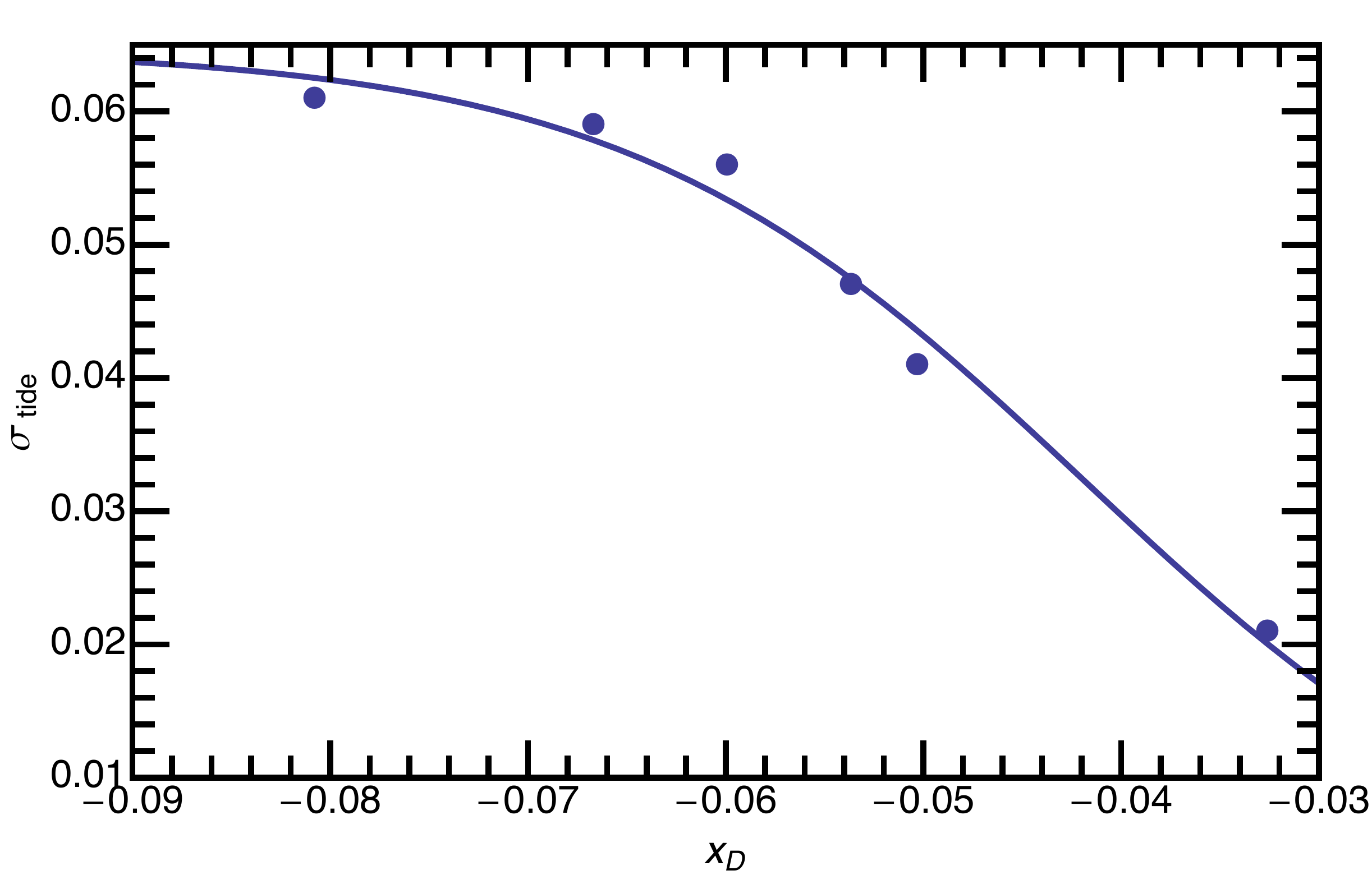}
  \end{tabular*}
  \caption{Correction parameters appearing in our phenomenological
    models for nondisruptive mergers (left panels) and disruptive
    mergers (right panels).
    Top left: correction factor $\epsilon_\text{tide}$ to
    $\delta_1$ versus $x_\text{ND}$, as defined in
    Eq.\,(\ref{eq:ND-combo}), for the six nondisruptive cases with
    $M_\text{NS}=1.35M_\odot$ and $\Gamma_2=3.0$. The data points are
    fitted by the function $2w_{x_0,d_0}^+(x_\text{ND})$, with
    $x_0=-0.088166$ and $d_0=0.066167$. 
    Top right: correction factor $\epsilon_\text{ins}$ to
    $\tilde{f}_0$ versus $x_\text{D}$, as defined in
    Eq.\,(\ref{eq:D-combo}), for the six disruptive cases with
    $M_\text{NS}=1.35M_\odot$ and $\Gamma_2=3.0$. The data points are
    fitted by the linear function $ax_\text{D}+b$, with $a=-6.04599$
    and $b=0.490086$. 
    Bottom left: the additive correction $\sigma_\text{tide}$ to
    $d$ versus $x_\text{ND}$, as defined in Eq.\,(\ref{eq:ND-combo}), for
    the six nondisruptive cases with $M_\text{NS}=1.35M_\odot$ and
    $\Gamma_2=3.0$. The data points are fitted by the function
    $2w_{x_0,d_0}^-(x_\text{ND})$, with $x_0=-0.170321$ and
    $d_0=0.162074$.
    Bottom right: the additive correction $\sigma_\text{tide}$ to
    $d$ versus the $x_\text{D}$, as defined in Eq.\,(\ref{eq:D-combo}),
    for the six nondisruptive cases with $M_\text{NS}=1.35M_\odot$
    and $\Gamma_2=3.0$. The data points are fitted by the function
    $Aw_{x_0,d_0}^-(x_\text{D})$, with $x_0=-0.0419235$,
    $d_0=0.0930419$, and $A=0.129459$. 
    \label{Fig:Fit2}}
\end{figure*}
%FFFFFFFFFFFFFFFFFFFFFFFFFFFFFFFFFFFFFFFFFFFFFFFFFFFFFFFFFFFFFFFFFFFFFFFFFFFFF

Finally, we must understand the behavior of $\alpha$,
$\epsilon_\text{ins}$, $\sigma_\text{tide}$, and
$\epsilon_\text{tide}$. By trying to reproduce the GW spectra of all
runs in Table \ref{tab:NRruns2model}, we found that $\alpha$ depends
on the binary mass ratio $Q=M_\text{BH}/M_\text{NS}$, or,
equivalently, on the symmetric mass ratio $\nu=Q/(1+Q)^2$:
cf. Fig.~\ref{Fig:Fit1}. Unfortunately, this dependence is poorly
constrained, as our runs span values of $Q$ from $2$ to $5$. We know,
however, that our model for $\chi_\text{f}$ (based on
\cite{Pannarale2012}) differs negligibly from the model for
$\chi_\text{f}$ used in \cite{Santamaria:2010yb} for $Q\geq5$
(i.e.~$\nu\leq 5/36$). We therefore decided to make the conservative
choice of fitting our data in Fig.~\ref{Fig:Fit1} with the function
\beq
\label{eq:fit1}
\alpha = w_{\nu_0,d_0}^-(\nu)+1
\eeq
for $\nu \geq 5/36$, while assuming $\alpha$ to be constant and equal
to its value at $Q=5$ (corresponding to $\nu=5/36$) for $Q\geq
5$. This yields $\nu_0=0.146872$ and $d_0=0.0749456$.

In order to understand the behavior of $\epsilon_\text{ins}$,
$\sigma_\text{tide}$, and $\epsilon_\text{tide}$, we found it useful
to divide the numerical runs in three groups (following the
classification suggested in \cite{Kyutoku:2011vz} for the outcome of a
BH-NS coalescence):
\begin{enumerate}[(i)]
\item mergers with $f_\text{tide}>\tilde f_\text{RD}$
  (``nondisruptive'');
\item mergers with $f_\text{tide}<\tilde f_\text{RD}$ and
  $M_\text{b,torus}=0$ (``mildly disruptive'');
\item mergers with $f_\text{tide}<\tilde f_\text{RD}$ and
  $M_\text{b,torus}>0$ (``strongly disruptive'').
\end{enumerate}
The first group shows a clear QNM excitation, the third group shows a
sharp high-frequency cutoff of tidal origin, and the behavior of the
second group is somewhere in between the other two. Out of the cases
used to build our model (as listed in Table \ref{tab:NRruns2model}) we
have $6$, $2$, and $6$ data sets for the first, second, and third
group, respectively, so that the second group is not very
populated. We expect to have more data (and to clarify the behavior of
this second group) in the near future, when we will consider waveforms
from merging binaries with nonzero initial spins.

%%%%%%%%%%%%%%%%%%%%%%%%%%%%%%%%%%%%%%%%%%%%%%%%%%%%%%%%%%%%%%%%%%%%%%%%%%%%%%
\subsubsection{Nondisruptive mergers}\label{sec:ND}
%%%%%%%%%%%%%%%%%%%%%%%%%%%%%%%%%%%%%%%%%%%%%%%%%%%%%%%%%%%%%%%%%%%%%%%%%%%%%%
For nondisruptive mergers tidal effects are weak, the NS matter moves
coherently and, thus, the merger and the inspiral contributions to the
GW spectrum ``fade out'' at the same frequency: this implies that
$\epsilon_\text{ins}=1$.

Let us now turn to $\epsilon_\text{tide}$, the factor correcting
$\delta_1$ in Eq.\,(\ref{eq:RingdownAmp}). As shown in the top-left
panel of Fig.~\ref{Fig:Fit2}, values of this parameter which allow our
model to reproduce the spectra of the six nondisruptive mergers in
Table \ref{tab:NRruns2model} have a regular behavior if plotted as a
function of the dimensionless quantity,
\beq
\label{eq:ND-combo}
x_\text{ND}\equiv \left(\f{f_\text{tide} - \tilde f_\text{RD}}{\tilde
    f_\text{RD}}\right)^2 - 0.6\mathcal{C}\,.
\eeq
This functional form captures the fact that a ``large'' NS suppresses
ringdown excitation via destructive interference
(cf.~\cite{Berti:2006hb} for a toy model illustrating this
phenomenon). The frequency at the onset of tidal disruption for a
large NS is closer to the QNM frequency of the BH remnant. A fit of
the form
\be
\epsilon_\text{tide}=2w_{x_0,d_0}^+(x_\text{ND})
\ee
yields $x_0=-0.0881657$ and $d_0=0.0661666$. The windowing function
choice is motivated by observing that the ringdown amplitude is
smoothly suppressed as tidal effects take over, i.e.~as the NS
disruption frequency approaches the QNM frequency of the BH remnant
from above.

A similar approach is used for $\sigma_\text{tide}$, which must vanish
as the coalescence becomes more and more BH-BH-like. On the other
hand, as tidal effects increase, $\sigma_\text{tide}$ must grow and
``smear'' the signal shutoff. In this case, we fitted the six data
points as follows:
\be
\sigma_\text{tide}=2w_{x_0,d_0}^-(x_\text{ND})\,,
\ee
obtaining $x_0=-0.170321$ and $d_0=0.162074$. This fit is shown along
with the data in the top-right panel of Fig.~\ref{Fig:Fit2}.

To summarize, for nondisruptive mergers we can set
\beq
\epsilon_\text{ins} &=& 1\,,\\
\label{eq:epstideND}
\epsilon_\text{tide} &=& 2w_{x_0,d_0}^+(x_\text{ND}) \quad \text{with}~
\left\{
  \begin{array}{l}
    x_0=-0.0881657\\
    d_0=0.0661666\,,\\
  \end{array}
\right.\\
\sigma_\text{tide} &=& 2w_{x_0,d_0}^-(x_\text{ND}) \quad \text{with}~
\left\{
  \begin{array}{l}
    x_0=-0.170321\\
    d_0=0.162074\,.
  \end{array}
\right.~~~~
\eeq
%

%%%%%%%%%%%%%%%%%%%%%%%%%%%%%%%%%%%%%%%%%%%%%%%%%%%%%%%%%%%%%%%%%%%%%%%%%%%%%%
\subsubsection{Disruptive mergers}\label{sec:D}
%%%%%%%%%%%%%%%%%%%%%%%%%%%%%%%%%%%%%%%%%%%%%%%%%%%%%%%%%%%%%%%%%%%%%%%%%%%%%%
Tidal effects must be taken into account in the phenomenology of
disruptive BH-NS mergers. The NS matter is scattered around and
accretes onto the BH incoherently, no ringdown of the BH remnant is
manifest in the GW spectrum, and therefore we have
$\epsilon_\text{tide}=0$.

As tidal effects grow stronger, the PN inspiral description must be
suppressed at smaller frequencies than the merger contribution. An
effective description of the end of the merger contribution may be
obtained by turning it off at $f_\text{tide}$
[cf.~Eq.\,(\ref{eq:ourf0prescrition})], so that
$\epsilon_\text{ins}\tilde{f}_0$ ends the inspiral contribution.

The values of $\epsilon_\text{ins}$ that allow us to reproduce the
spectra of the six disruptive mergers of Table \ref{tab:NRruns2model}
are well correlated with the following dimensionless quantity:
\beq
\label{eq:D-combo}
x_\text{D} \equiv \f{M_\text{b,torus}}{M_\text{b,NS}} +
2.23\mathcal{C} - 1.02\sqrt{\nu}.
\eeq
A good linear fit to the data, as shown in the bottom-left panel of
Fig.~\ref{Fig:Fit2}, is
\be
\epsilon_\text{ins}=0.490086-6.04599x_\text{D}\,.
\ee

The data for $\sigma_\text{tide}$ also show a correlation with
$x_\text{D}$. A good fit is
\be
\sigma_\text{tide}=A w^-_{x_0,d_0}(x_\text{D})
\ee
with $x_0=-0.0419235$, $d_0=0.0930419$, and $A=0.129459$.  As shown in
the bottom-right panel of Fig.~\ref{Fig:Fit2}, this fit is less robust
than the previous ones, and our present model for $\sigma_\text{tide}$
is quite likely to change as more data become available. However, in
the rest of the paper we will show that this has a minor effect on the
agreement between the BH-NS phenomenological waveforms and the
numerical data.

%FFFFFFFFFFFFFFFFFFFFFFFFFFFFFFFFFFFFFFFFFFFFFFFFFFFFFFFFFFFFFFFFFFFFFFFFFFFFF
\begin{figure*}[ht]
\begin{tabular*}{\textwidth}{c@{\extracolsep{\fill}}c}
\includegraphics[scale=0.35,clip=true]{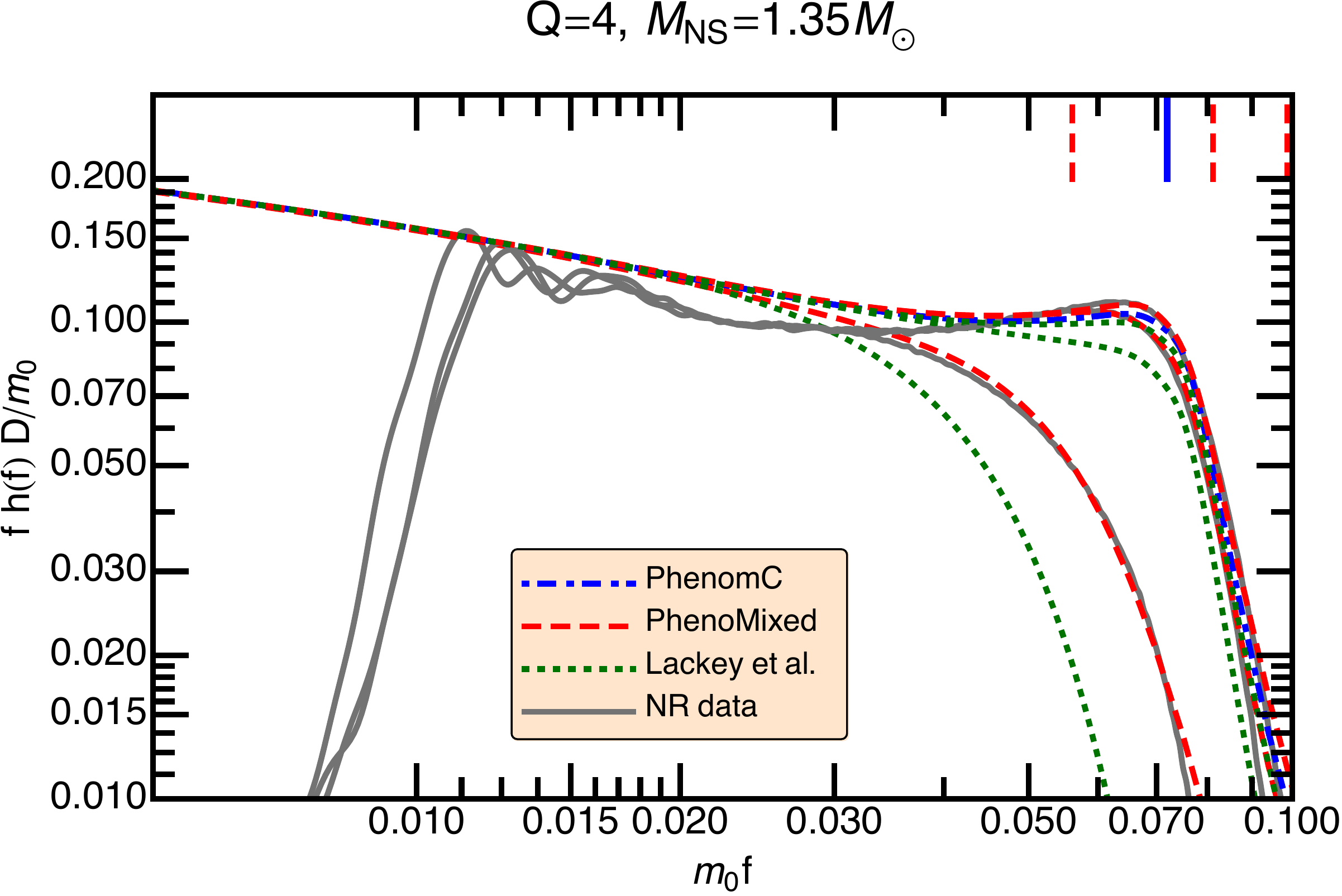}
&
\includegraphics[scale=0.35,clip=true]{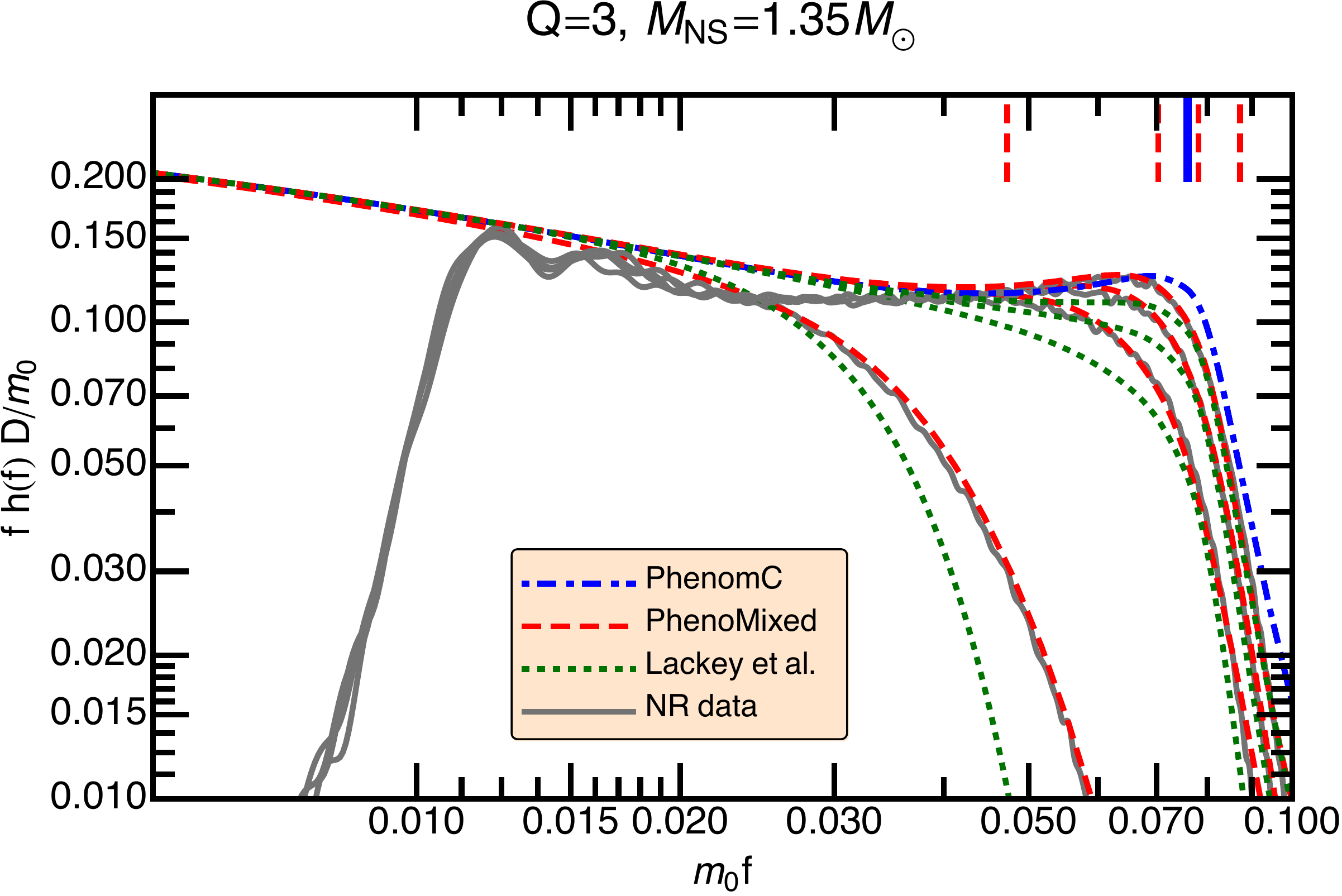}
\\
\end{tabular*}
\includegraphics[scale=0.35,clip=true]{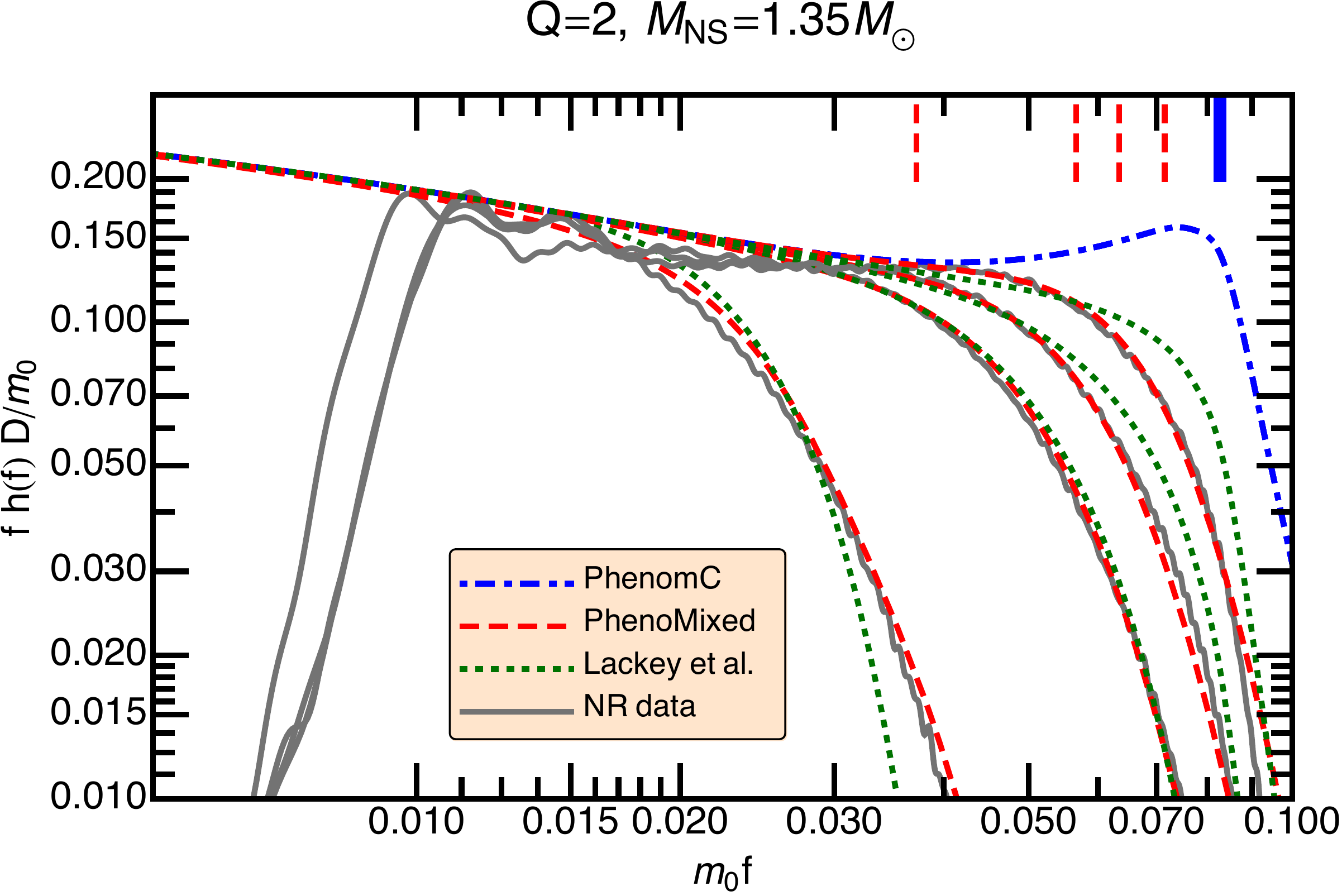}
\caption{The numerical simulations used to build the model (solid
  lines), discussed in Sec.~\ref{subs:build}, are compared to our
  phenomenological amplitude model (dashed lines), to the
  phenomenological amplitude model of Lackey
  \etal~\cite{Lackey:2013axa} (dotted lines), and to the BH-BH model of
  Santamar{\'{\i}}a \etal~\cite{Santamaria:2010yb} (dash-dotted line).
  Top left: Cases 
  {\ttfamily 2H-M135-Q4},
  {\ttfamily H-M135-Q4}, and
  {\ttfamily B-M135-Q4} (from left to right, so that the rightmost
  model is the closest to the BH-BH case).
  Top right: Cases 
  {\ttfamily 2H-M135-Q3}, 
  {\ttfamily H-M135-Q3}, 
  {\ttfamily HB-M135-Q3}, and
  {\ttfamily B-M135-Q3}.
  Bottom: Cases 
  {\ttfamily 2H-M135-Q2},
  {\ttfamily H-M135-Q2},
  {\ttfamily HB-M135-Q2}, and
  {\ttfamily B-M135-Q2}.
  \label{Fig:MNS135QX}}
\end{figure*}
%FFFFFFFFFFFFFFFFFFFFFFFFFFFFFFFFFFFFFFFFFFFFFFFFFFFFFFFFFFFFFFFFFFFFFFFFFFFFF

%FFFFFFFFFFFFFFFFFFFFFFFFFFFFFFFFFFFFFFFFFFFFFFFFFFFFFFFFFFFFFFFFFFFFFFFFFFFFF
\begin{figure*}[thb]
\begin{tabular*}{\textwidth}{c@{\extracolsep{\fill}}c}
\includegraphics[scale=0.35,clip=true]{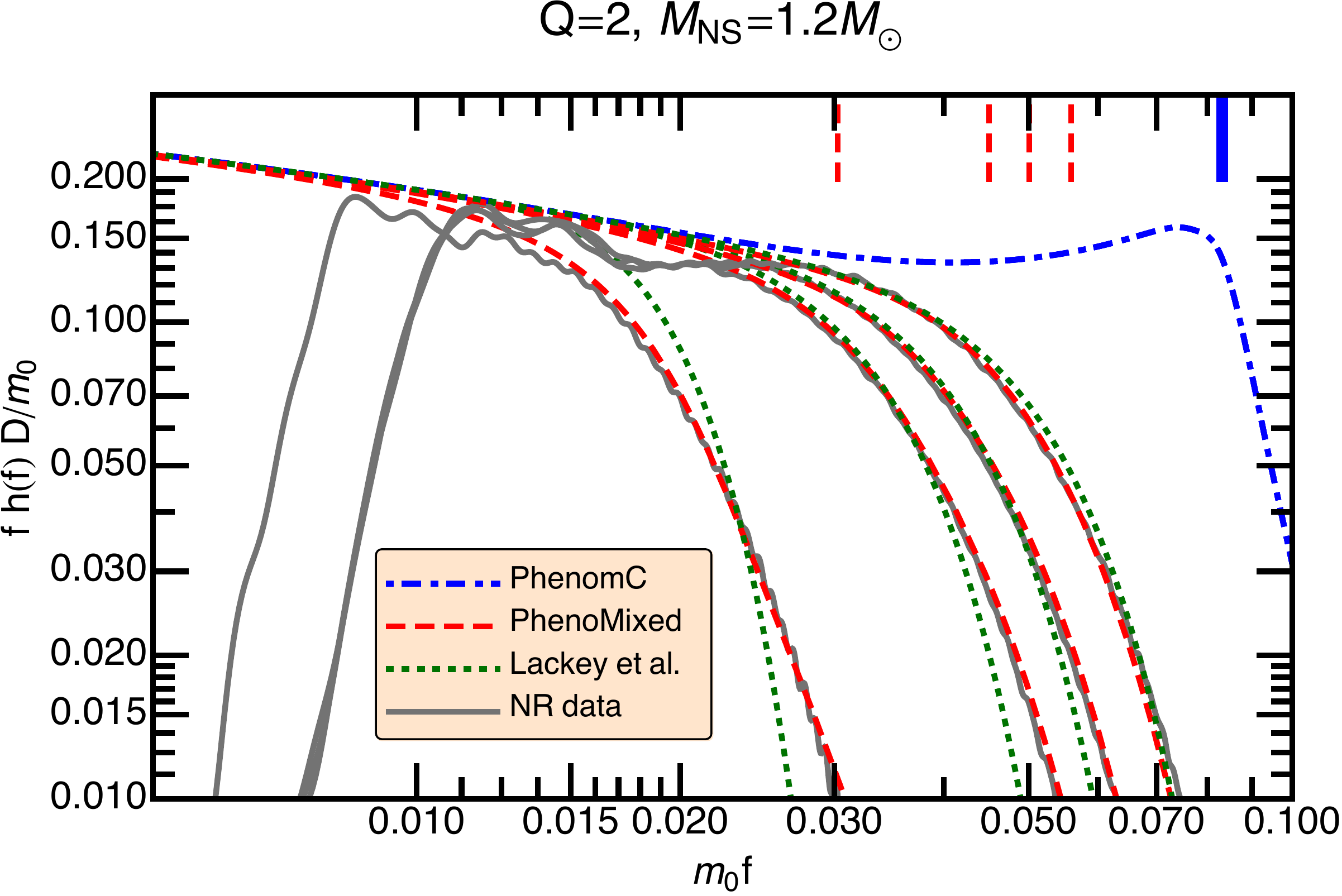}
&
\includegraphics[scale=0.35,clip=true]{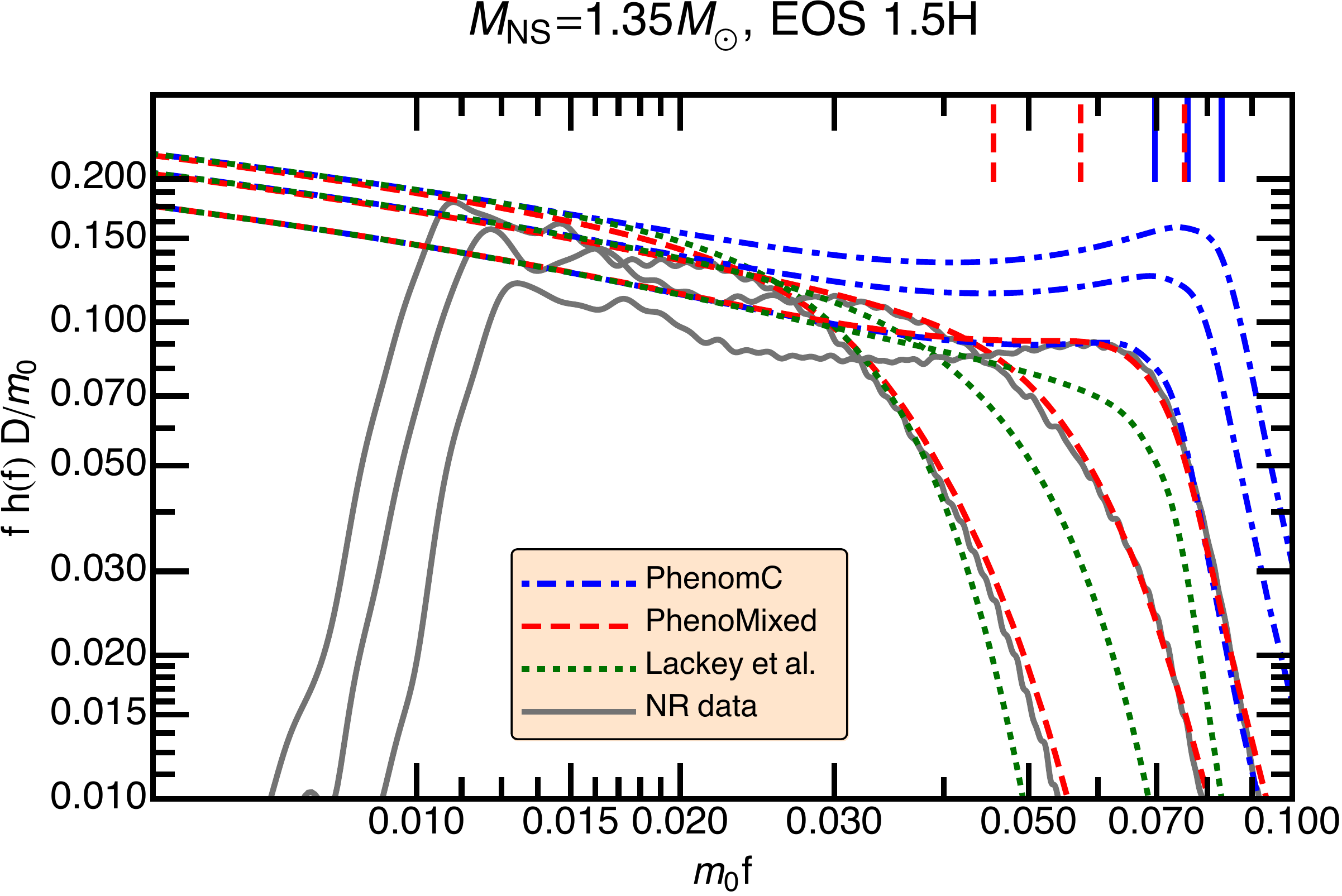}
\end{tabular*}
\caption{Left: First set of numerical simulations used to test our
  model (cases {\ttfamily 2H-M12-Q2}, {\ttfamily H-M12-Q2}, {\ttfamily
    HB-M12-Q2}, and {\ttfamily B-M12-Q2}), as discussed in
  Sec.~\ref{subs:test}. Right: Second set of numerical simulations
  used to test our model (cases {\ttfamily 1.5H-M135-Q2}, {\ttfamily
    1.5H-M135-Q3}, and {\ttfamily 1.5H-M135-Q5}), as discussed in
  Sec.~\ref{subs:test2}; we obtain similar results for the two remaining 
  cases (not shown in this plot to avoid cluttering),
  i.e.~{\ttfamily 1.5H-M135-Q4} and {\ttfamily 1.25H-M135-Q2}.
  \label{Fig:test}}
\end{figure*}
%FFFFFFFFFFFFFFFFFFFFFFFFFFFFFFFFFFFFFFFFFFFFFFFFFFFFFFFFFFFFFFFFFFFFFFFFFFFFF

%FFFFFFFFFFFFFFFFFFFFFFFFFFFFFFFFFFFFFFFFFFFFFFFFFFFFFFFFFFFFFFFFFFFFFFFFFFFFF
\begin{figure*}[thb]
\begin{tabular*}{\textwidth}{c@{\extracolsep{\fill}}c}
\includegraphics[scale=0.35,clip=true]{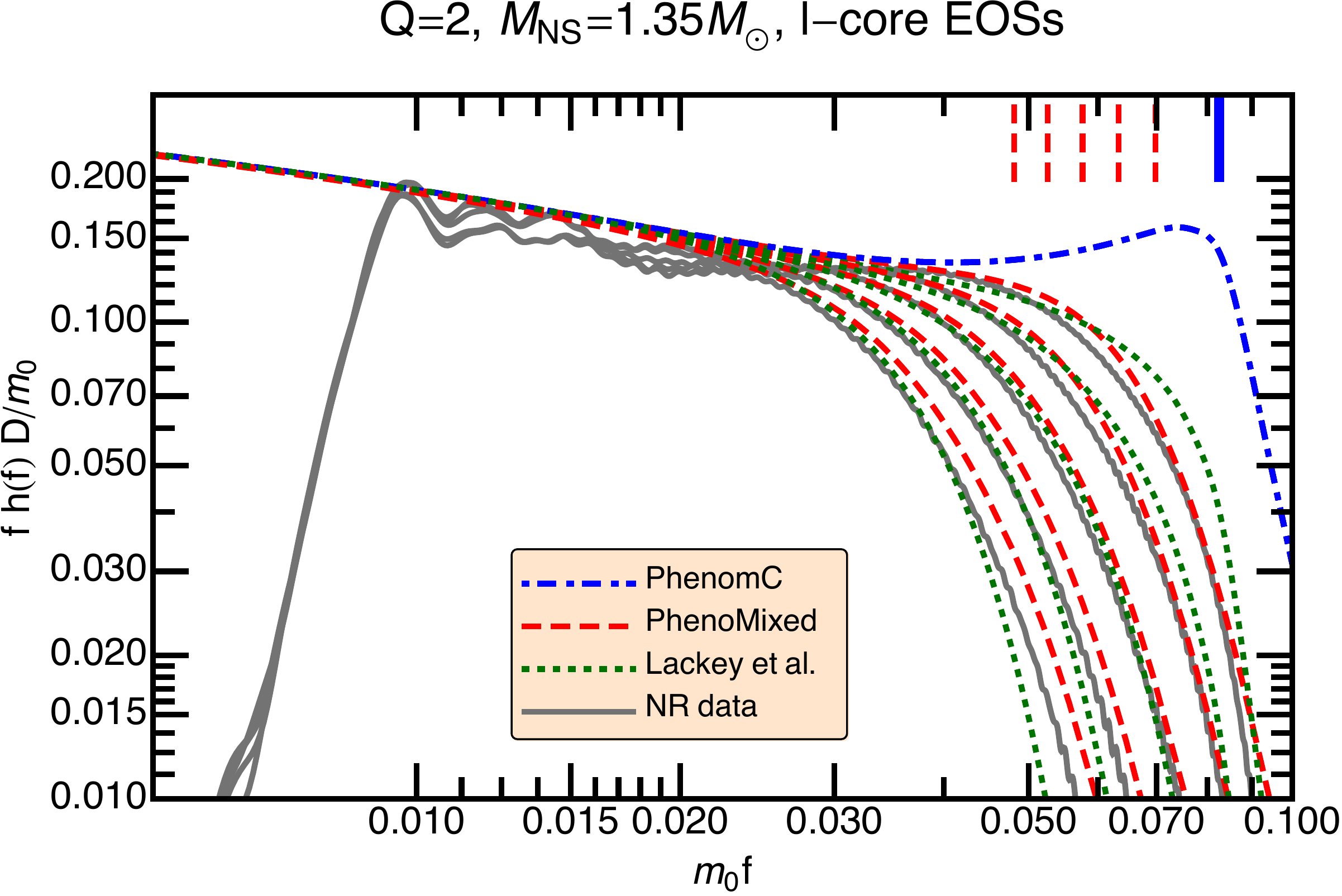}
&
\includegraphics[scale=0.35,clip=true]{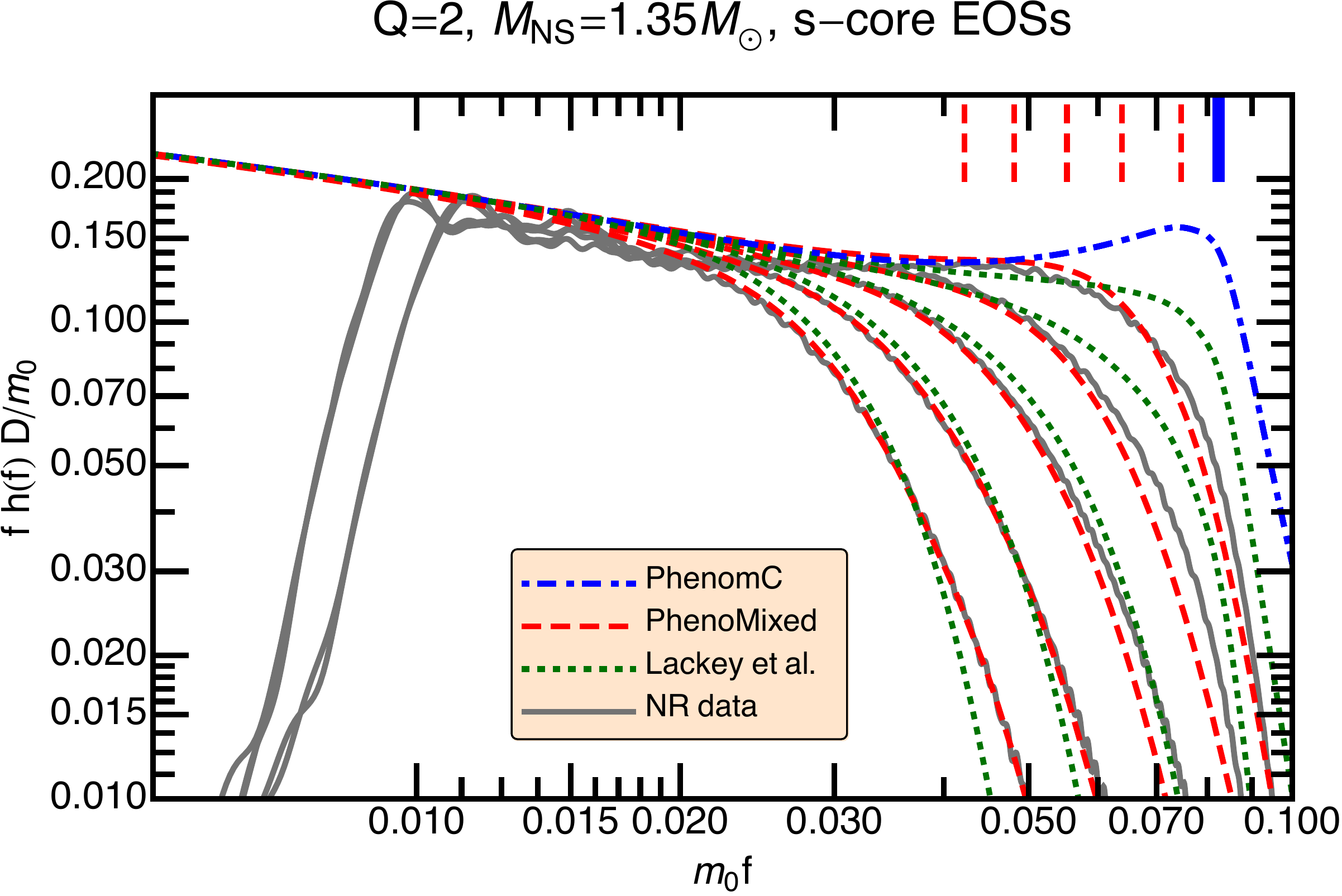}
\end{tabular*}
\includegraphics[scale=0.35,clip=true]{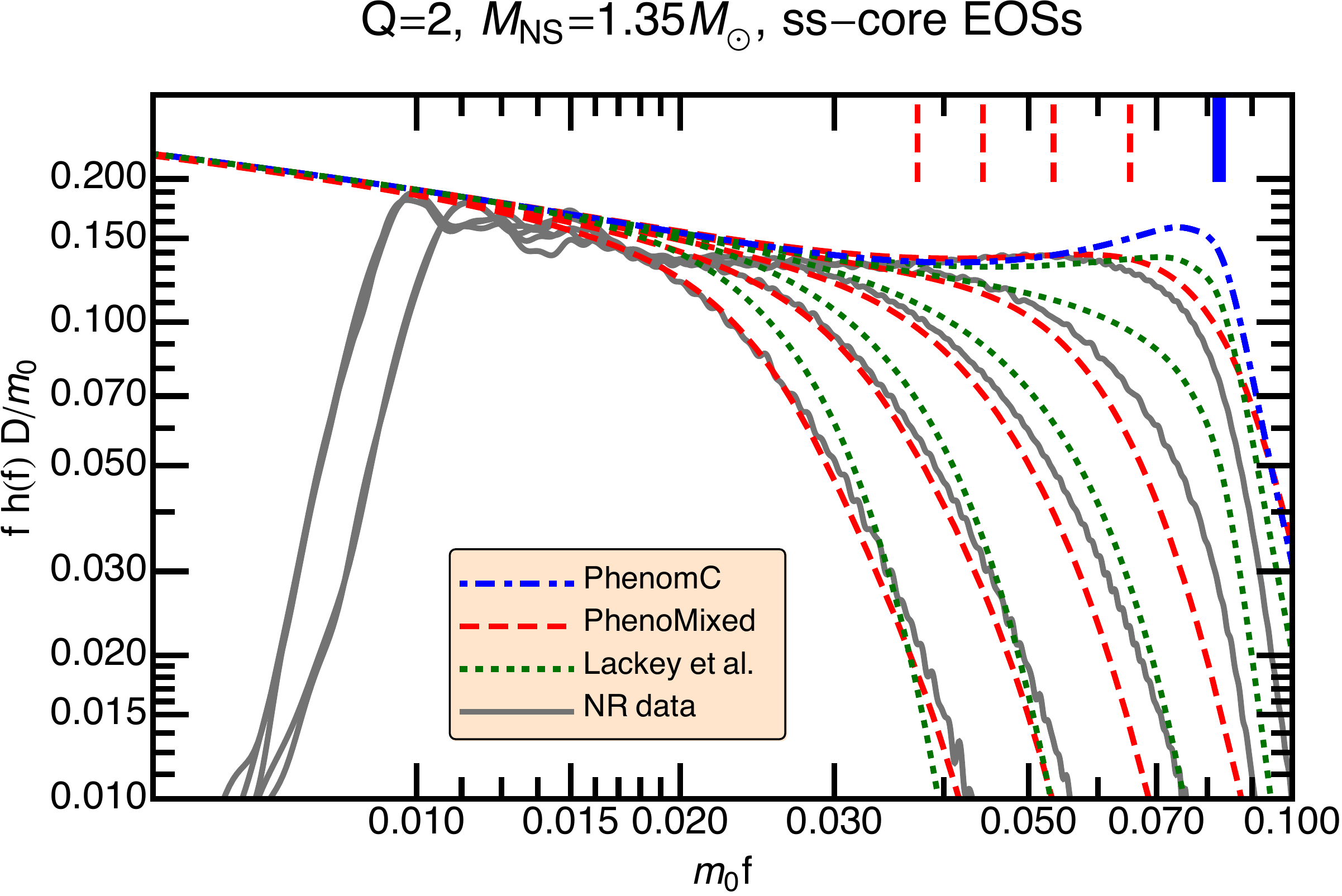}
\caption{Top left:
Cases 
{\ttfamily 1.5Hl-M135-Q2},
{\ttfamily 1.25Hl-M135-Q2},
{\ttfamily Hl-M135-Q2}, 
{\ttfamily HBl-M135-Q2}, and
{\ttfamily Bl-M135-Q2}.
Top right:
Cases 
{\ttfamily 1.5Hs-M135-Q2},
{\ttfamily 1.25Hs-M135-Q2},
{\ttfamily Hs-M135-Q2}, 
{\ttfamily HBs-M135-Q2}, and
{\ttfamily Bs-M135-Q2}.
Bottom: 
Cases 
{\ttfamily 1.5Hss-M135-Q2},
{\ttfamily 1.25Hss-M135-Q2},
{\ttfamily Hss-M135-Q2}, 
{\ttfamily HBss-M135-Q2}, and
{\ttfamily Bss-M135-Q2}.
  \label{Fig:MNS135Q2EOSx}}
\end{figure*}
%FFFFFFFFFFFFFFFFFFFFFFFFFFFFFFFFFFFFFFFFFFFFFFFFFFFFFFFFFFFFFFFFFFFFFFFFFFFFF

To summarize, disruptive mergers are well reproduced by setting
\beq
\epsilon_\text{tide} &=& 0\,,\\
\label{eq:epsinsD}
\epsilon_\text{ins} &=& ax_\text{D}+b \quad \text{with}~\left\{
  \begin{array}{l}
    a=-6.04599\\
    b=0.490086\,,\\
  \end{array}
\right.\\
\sigma_\text{tide} &=& Aw_{x_0,d_0}^-(x_\text{D}) \quad \text{with}
~\left\{
  \begin{array}{l}
    A=0.129459\\
    x_0=-0.0419235\\
    d_0=0.0930419\,.\\
  \end{array}
\right.~~~
\eeq
%

%%%%%%%%%%%%%%%%%%%%%%%%%%%%%%%%%%%%%%%%%%%%%%%%%%%%%%%%%%%%%%%%%%%%%%%%%%%%%%
\subsubsection{Mildly disruptive mergers}\label{sec:MD}
%%%%%%%%%%%%%%%%%%%%%%%%%%%%%%%%%%%%%%%%%%%%%%%%%%%%%%%%%%%%%%%%%%%%%%%%%%%%%%
Only two of the binaries in Table \ref{tab:NRruns2model} have
$M_\text{b,torus}=0$ and $f_\text{tide}<\tilde{f}_\text{RD}$, so that
this regime is relatively poorly constrained by simulations. We expect
the phenomenology in this case to be intermediate between the one of
disruptive and nondisruptive mergers. In this sense it is reassuring
to observe that:
\bi
\item[1)] $\epsilon_\text{ins}$ may be determined as prescribed for
  disruptive mergers in Eq.\,(\ref{eq:epsinsD}), and
\item[2)] $\epsilon_\text{tide}$ may be determined as prescribed for
  nondisruptive cases in Eq.\,(\ref{eq:epstideND}).
\ei
These observations confirm that the nature of mildly disruptive
coalescences is indeed somewhere in between the disruptive and
nondisruptive cases. The value $\sigma_\text{tide}=0.041$ works for
both mildly disruptive cases. We expect a better understanding of this
class of mergers to emerge from future studies of BH-NS binaries with
a spinning BH.

%%%%%%%%%%%%%%%%%%%%%%%%%%%%%%%%%%%%%%%%%%%%%%%%%%%%%%%%%%%%%%%%%%%%%%%%%%%%%%
\section{Model-Data Comparisons}\label{sec:results}
%%%%%%%%%%%%%%%%%%%%%%%%%%%%%%%%%%%%%%%%%%%%%%%%%%%%%%%%%%%%%%%%%%%%%%%%%%%%%%
We may now see the model at work. In this section we collect and
discuss the results for the GW spectra of all runs in Tables
\ref{tab:NRruns2model} and \ref{tab:NRruns2tests}. We follow the
convention adopted in Figs.~\ref{Fig:MNS135Q5EOSB},
\ref{Fig:MNS135Q5EOSH}, and \ref{Fig:MNS135Q5EOS2H} and show the
numerical data with a grey, continuous line, the prediction of our
model with a red, dashed line, the prediction of the PhenomC model
with a blue, dot-dashed line, and the prediction of the BH-NS model of
\cite{Lackey:2013axa} with a green, dotted line. In addition, the
locations of the frequencies $\tilde{f}_\text{RD}$ and $f_\text{tide}$
are marked by short, vertical lines (straight blue and dashed red,
respectively). We first show the remaining\footnote{Cases {\ttfamily
    B-M135-Q5}, {\ttfamily H-M135-Q5}, and {\ttfamily 2H-M135-Q5} were
  already discussed in Sec.~\ref{sec:PhenoMixedGWfs}.} GW spectra of
the binaries in Table \ref{tab:NRruns2model}, upon which our model is
built. We then test our model against the binaries in Table
\ref{tab:NRruns2tests}. We begin by looking at cases with $M\neq
1.35M_\odot$, i.e.~the four runs in the first block of Table
\ref{tab:NRruns2tests}; we then consider the five cases in the second
block of Table \ref{tab:NRruns2tests}, i.e.~those with a
$\Gamma_2=3.0$ core description, but that were not used when building
the model; finally, we look at the fifteen binaries in which the NS
core EOS has $\Gamma_2\neq 3.0$.

%%%%%%%%%%%%%%%%%%%%%%%%%%%%%%%%%%%%%%%%%%%%%%%%%%%%%%%%%%%%%%%%%%%%%%%%%%%%%%
\subsection{The $14$ cases used to build the model}
\label{subs:build}
%%%%%%%%%%%%%%%%%%%%%%%%%%%%%%%%%%%%%%%%%%%%%%%%%%%%%%%%%%%%%%%%%%%%%%%%%%%%%%
In Figs.~\ref{Fig:MNS135Q5EOSB}, \ref{Fig:MNS135Q5EOSH}, and
\ref{Fig:MNS135Q5EOS2H} we showed how the model of
Sec.~\ref{sec:PhenoMixedAlgorithm}, with hand-tuned parameters,
performs for a subset of the binaries it was built upon, namely the
three cases with mass ratio $Q=5$. The variations in the GW spectra
yielded by our model if we use the fits of the previous section for
the parameters, instead of hand-tuned values, are negligible.

In Fig.~\ref{Fig:MNS135QX} we illustrate the performance of our model
for the three binaries with $Q=4$, the four with $Q=3$, and the four
with $Q=2$, respectively (see Table \ref{tab:NRruns2model}). We apply
the full procedure outlined in Sec.~\ref{sec:PhenoMixedAlgorithm} and
explained in the previous section. Our model reproduces the
high-frequency phenomenology very well, both for nondisruptive and
disruptive mergers. In the case of disruptive mergers, in particular,
when there is little or no QNM ringdown excitation of the BH remnant,
we achieve a considerable improvement over the BH-BH phenomenological
waveforms of \cite{Santamaria:2010yb} and also over the model by
Lackey \etal~\cite{Lackey:2013axa}. Therefore our model should yield
more accurate results for practical applications, including
e.g.~calculations of cutoff frequencies and SNRs. During the inspiral
phase our model and the PhenomC model match, by construction, and we
expect longer and more accurate numerical waveforms to better match
the PN description of the inspiral GW amplitude, especially for mass
ratios closer to unity, where the convergence of the PN expansion
works best.

In conclusion, our model reproduces the high-frequency GW amplitude
phenomenology for the fourteen runs upon which it was
built. Furthermore, it performs much better than the BH-BH
phenomenological waveform model and better than the model of
\cite{Lackey:2013axa}, especially when mergers are disruptive.

%%%%%%%%%%%%%%%%%%%%%%%%%%%%%%%%%%%%%%%%%%%%%%%%%%%%%%%%%%%%%%%%%%%%%%%%%%%%%%
\subsection{The four test cases with $M_\text{NS}\neq 1.35M_\odot$}
\label{subs:test}
%%%%%%%%%%%%%%%%%%%%%%%%%%%%%%%%%%%%%%%%%%%%%%%%%%%%%%%%%%%%%%%%%%%%%%%%%%%%%%
So far, we considered only binaries with $M_\text{NS}=1.35M_\odot$. A
first useful series of tests we can provide for our model thus
involves binaries with a different NS mass. This may be done with the
aid of our numerical data for four coalescences in which the NS has a
mass of $1.2M_\odot$ (first group of cases in Table
\ref{tab:NRruns2tests}). We stress, once more, the fact that the data
from these runs were \emph{not} used to build the model.

The left panel of Fig.~\ref{Fig:test} shows how our phenomenological
model performs for these binaries. The tests are successful, in that
(1) they correctly capture the phenomenology of all four mergers, and
(2) they provide a more accurate description when compared to the
PhenomC or Lackey \etal~\cite{Lackey:2013axa} amplitude models. It
must be noted that all cases share the same, low mass ratio $Q=2$, and
that they are all disruptive mergers. In this sense, these tests may
be viewed as being still limited, but the runs we have represent the
current state of the art for nonspinning BH-NS mergers.

%%%%%%%%%%%%%%%%%%%%%%%%%%%%%%%%%%%%%%%%%%%%%%%%%%%%%%%%%%%%%%%%%%%%%%%%%%%%%%
\subsection{The five test cases with $\Gamma_2=3.0$}
\label{subs:test2}
%%%%%%%%%%%%%%%%%%%%%%%%%%%%%%%%%%%%%%%%%%%%%%%%%%%%%%%%%%%%%%%%%%%%%%%%%%%%%%
We now move to the second group of test cases in Table
\ref{tab:NRruns2tests}. These five sets of data share a $\Gamma_2=3.0$
EOS description for the NS core and were not used when building the
model. They are obtained by setting $\log P_\text{fidu}$ to its value
corresponding to the $1.5$H or $1.25$H EOS choices. Four data sets
spanning different values of the binary mass ratio are available for
the first choice, while a single data set for $Q=2$ is available for
the $1.25$H EOS. Three of our results for the $1.5$H EOS data are
reported in the right panel of Fig.~\ref{Fig:test}, where we
demonstrate that these BH-NS merger typologies are correctly
reproduced by our model. The remaining $1.5$H EOS and the single
$1.25$H EOS data sets are reproduced with similar accuracy and are not
shown to avoid cluttering in the plot. This ensures that our
formulation is universal for data obtained with two-component
piecewise polytropes having $\Gamma_2=3.0$ in the core.

%%%%%%%%%%%%%%%%%%%%%%%%%%%%%%%%%%%%%%%%%%%%%%%%%%%%%%%%%%%%%%%%%%%%%%%%%%%%%%
\subsection{The $15$ test cases with $\Gamma_2\neq 3$}
%%%%%%%%%%%%%%%%%%%%%%%%%%%%%%%%%%%%%%%%%%%%%%%%%%%%%%%%%%%%%%%%%%%%%%%%%%%%%%
The third and last set of BH-NS binary merger simulations in Table
\ref{tab:NRruns2tests} is relative to NS models in which the core EOS
description employs a polytropic exponent $\Gamma_2$ different from
$3.0$. All of these test cases are limited to mass ratio
$Q=2$. However, EOS-related effects on the binary dynamics, and hence
on the GW phenomenology, are enhanced by low values of the binary mass
ratio: with the exception of the astrophysically unlikely case of
binaries with $Q<2$, these data sets are therefore the most
challenging possible test beds.

We present some results for the $\Gamma_2\neq3.0$ tests in the three
panels of Fig.~\ref{Fig:MNS135Q2EOSx}, which refer to $\Gamma_2=3.3$,
$\Gamma_2=2.7$, and $\Gamma_2=2.4$, respectively (these are denoted as
{\ttfamily l}, {\ttfamily s}, and {\ttfamily ss} in Table
\ref{tab:NRruns2tests}). In each plot, we consider the results for the
$1.5$Hx, $1.25$Hx, Hx, HBx, and Bx EOS, where
$\rm{x}=\{\text{l},\text{s},\text{ss}\}$. Overall, for cases with
$\Gamma_2=\{3.3,2.7\}$ our model shows good agreement with the
numerical data. An excellent match is evident for the {\ttfamily
  1.25Hl-M135-Q2} and for {\ttfamily 1.5Hl-M135-Q2} data sets. With
$\Gamma_2=2.4$, we run into our three worst test outcomes. These occur
in the case of runs {\ttfamily Bss-M135-Q2}, {\ttfamily HBss-M135-Q2},
and {\ttfamily Hss-M135-Q2}, in order of decreasing agreement between
the numerical waveforms and our new phenomenological waveforms. Case
{\ttfamily Bss-M135-Q2}, in particular, is a somewhat critical
test. This happens because $f_\text{RD}\simeq f_\text{tide}$, which we
know to be the hardest regime to model. For test cases {\ttfamily
  1.25Hss-M135-Q2} and {\ttfamily 1.5Hss-M135-Q2}, on the other hand,
we achieve a very good match between the model and the data.

The $\Gamma_2=\{3.3, 2.7, 2.4\}$, $Q=2$ test cases thus tell us that
the model starts breaking down for systems in which the BH mass is
low, and the NS is very compact and has an exceptionally soft
core. This unfavorable region of the space of parameters is small, and
probably astrophysically marginal, since current observations are
gradually ruling out EOSs that predict a significant softening in the
core \cite{Ozel:2006km,Antoniadis:2013pzd,Demorest:2010bx}. Therefore
our test pool provides a solid confirmation of the validity of our
model.

%%%%%%%%%%%%%%%%%%%%%%%%%%%%%%%%%%%%%%%%%%%%%%%%%%%%%%%%%%%%%%%%%%%%%%%%%%%%%%
\section{Signal-to-Noise-Ratio Comparisons}\label{sec:SNRs}
%%%%%%%%%%%%%%%%%%%%%%%%%%%%%%%%%%%%%%%%%%%%%%%%%%%%%%%%%%%%%%%%%%%%%%%%%%%%%%
In this section we compare SNRs computed using our BH-NS amplitude
model, the BH-BH PhenomC model of \cite{Santamaria:2010yb}, and the
restricted PN model used in several classic papers on GW data analysis
\cite{Cutler:1994ys,Poisson:1995ef}. The SNR $\rho$ for a
frequency-domain signal $\tilde{h}(f)$ and a detector with noise power
spectral density $S_h(f)$ is defined as
\be\label{SNRdef}
\rho \equiv 4 \Re \int_{f_\text{start}}^{f_{\text{end}}} df
\frac{\tilde{h}(f)\tilde{h}^*(f)}{S_h(f)}\,,
\ee
where $S_h(f)$ is the noise power spectral density of the detector,
and $\tilde{h}(f)$ is defined in Eq.\,(\ref{GWstrain}) of Appendix
\ref{app:factors} -- i.e., it is a weighted average of the plus ($+$)
and cross ($\times$) polarization states. For any given binary system,
we define the starting frequency $f_\text{start}$ to be $10\,$Hz for
second-generation detectors, and $1\,$Hz for third-generation
detectors. Our convention on the ending frequency $f_\text{end}$ will
be discussed below.

In order to make our comparisons as universal as possible, we will
consider {\em ratios} of SNRs, so that possible overall factors coming
from distance, orientation, and inclination of the source cancel out.
We define the following quantities, which are useful to understand the
impact of modeling on detectability:
\be
\label{eq:assessRPN}
\epsilon_\text{RPN}\equiv 1-\f{\rho_\text{RPN}}{\rho_\text{BHNS}}\,,
\ee
\be
\label{eq:assessPhenomC}
\epsilon_\text{BHBH}\equiv 1-\f{\rho_\text{BHBH}}{\rho_\text{BHNS}}\,.
\ee
The first quantity ($\epsilon_\text{RPN}$) measures the SNR deviation
between a BH-NS waveform and a restricted PN (RPN) amplitude model ---
i.e., a zero-order amplitude expansion --- obtained for the same
masses of the binary constituents. Naturally, RPN waveforms (which are
supposed to be accurate for inspirals only) deviate significantly from
merger waveforms after the binary members cross the ISCO. Therefore,
in order to provide a fair comparison, in this case we will follow
much of the existing GW literature
(e.g.~\cite{Cutler:1994ys,Poisson:1995ef}) and truncate the SNR
calculation at an upper frequency $f_\text{end}$ that corresponds to
the conventional Schwarzschild ISCO $r=6m_0$ for a binary of total
mass $m_0$.

The second quantity ($\epsilon_\text{BHBH}$) measures the deviation
between a BH-NS waveform amplitude model and the corresponding BH-BH
waveform amplitude model for a nonspinning binary with the same
masses. In this comparison\footnote{We also performed a similar
  comparison between our amplitude model and that of Lackey
  \etal~\cite{Lackey:2013axa}, finding that
  $\epsilon_\text{Lackey}\equiv
  1-\f{\rho_\text{Lackey}}{\rho_\text{BHNS}}<0.01$ in all cases.
  Therefore the waveform model of \cite{Lackey:2013axa} is more than
  appropriate for SNR calculations in the case of nonspinning
  binaries.} it makes sense to consider the whole waveform, and
therefore we set $f_\text{end}=5000\,$Hz for all binaries.

%TTTTTTTTTTTTTTTTTTTTTTTTTTTTTTTTTTTTTTTTTTTTTTTTTTTTTTTTTTTTTTTTTTTTTTTTTTTTTT
\begin{table*}[!b]
  \caption{$1-\rho_\text{RPN}/\rho_\text{BHNS}$ [in round brackets: $1-\rho_\text{BHBH}/\rho_\text{BHNS}$] for several detectors. $\rho_\text{RPN}$ is the SNR calculated with the restricted PN model, while $\rho_\text{BHNS}$ and $\rho_\text{BHBH}$ are the SNRs obtained using our phenomenological BH-NS model and the PhenomC model. The numbers reported are percentages. \label{tab:SNRs}}
  \begin{tabular}{|l||rr|r|rrr|rr|}
    \colrule
    \colrule
    \multirow{2}{*}{Run Label} & \multicolumn{2}{|c|}{LIGO} &  \multicolumn{1}{|c|}{Virgo} & \multicolumn{3}{|c|}{KAGRA} & \multicolumn{2}{|c|}{ET} \\
    & \multicolumn{1}{|c}{Adv} & \multicolumn{1}{c|}{AdvZDHP} &  \multicolumn{1}{|c|}{Adv} & \multicolumn{1}{|c}{varBRSE} & \multicolumn{1}{c}{maxBRSE} & \multicolumn{1}{c|}{varDRSE} & \multicolumn{1}{|c}{B} & \multicolumn{1}{c|}{C} \\
    \colrule
    {\ttfamily EOSBQ5M135}      & $-9.6$ ($0.0 $) & $-6.5$ ($0.0 $) & $-6.7$ ($0.0 $) & $-7.0$ ($0.0 $) & $-7.2$ ($0.0 $) & $-8.1$ ($0.0 $) & $-6.5$ ($0.0 $) & $-4.8$ ($0.0 $) \\
    {\ttfamily EOSHQ5M135}      & $-9.6$ ($0.0 $) & $-6.5$ ($0.0 $) & $-6.7$ ($0.0 $) & $-7.0$ ($0.0 $) & $-7.2$ ($0.0 $) & $-8.1$ ($0.0 $) & $-6.5$ ($0.0 $) & $-4.8$ ($0.0 $) \\
    {\ttfamily EOS2HQ5M135}     & $-9.6$ ($0.0 $) & $-6.5$ ($0.0 $) & $-6.7$ ($0.0 $) & $-7.0$ ($0.0 $) & $-7.2$ ($0.0 $) & $-8.1$ ($0.0 $) & $-6.5$ ($0.0 $) & $-4.8$ ($0.0 $) \\
    {\ttfamily EOSBQ4M135}      & $-7.8$ ($0.0 $) & $-5.4$ ($0.0 $) & $-5.8$ ($0.0 $) & $-5.8$ ($0.0 $) & $-5.9$ ($0.0 $) & $-6.6$ ($0.0 $) & $-5.3$ ($0.0 $) & $-4.0$ ($0.0 $) \\
    {\ttfamily EOSHQ4M135}      & $-7.8$ ($0.0 $) & $-5.4$ ($0.0 $) & $-5.8$ ($0.0 $) & $-5.8$ ($0.0 $) & $-5.9$ ($0.0 $) & $-6.6$ ($0.0 $) & $-5.3$ ($0.0 $) & $-4.0$ ($0.0 $) \\
    {\ttfamily EOS2HQ4M135}     & $-8.4$ ($-0.5$) & $-5.9$ ($-0.5$) & $-6.3$ ($-0.6$) & $-6.3$ ($-0.5$) & $-6.4$ ($-0.5$) & $-7.2$ ($-0.5$) & $-5.9$ ($-0.5$) & $-4.5$ ($-0.5$) \\
    {\ttfamily EOSBQ3M135}      & $-6.1$ ($0.0 $) & $-4.3$ ($0.0 $) & $-4.8$ ($0.0 $) & $-4.6$ ($0.0 $) & $-4.6$ ($0.0 $) & $-5.2$ ($0.0 $) & $-4.2$ ($0.0 $) & $-3.2$ ($0.0 $) \\
    {\ttfamily EOSHBQ3M135}     & $-6.1$ ($0.0 $) & $-4.3$ ($0.0 $) & $-4.8$ ($0.0 $) & $-4.6$ ($0.0 $) & $-4.6$ ($0.0 $) & $-5.2$ ($0.0 $) & $-4.2$ ($0.0 $) & $-3.2$ ($0.0 $) \\
    {\ttfamily EOSHQ3M135}      & $-6.1$ ($0.0 $) & $-4.3$ ($0.0 $) & $-4.8$ ($0.0 $) & $-4.6$ ($0.0 $) & $-4.6$ ($0.0 $) & $-5.2$ ($0.0 $) & $-4.2$ ($0.0 $) & $-3.2$ ($0.0 $) \\
    {\ttfamily EOS2HQ3M135}     & $-7.1$ ($-1.0$) & $-5.2$ ($-0.9$) & $-5.9$ ($-1.0$) & $-5.5$ ($-0.9$) & $-5.5$ ($-0.9$) & $-6.1$ ($-0.9$) & $-5.1$ ($-0.9$) & $-4.0$ ($-0.8$) \\
    {\ttfamily EOSBQ2M135}      & $-4.5$ ($-0.1$) & $-3.3$ ($-0.1$) & $-3.7$ ($-0.1$) & $-3.4$ ($-0.1$) & $-3.5$ ($-0.1$) & $-3.8$ ($-0.1$) & $-3.2$ ($-0.1$) & $-2.5$ ($-0.1$) \\
    {\ttfamily EOSHBQ2M135}     & $-4.6$ ($-0.2$) & $-3.4$ ($-0.2$) & $-3.9$ ($-0.2$) & $-3.5$ ($-0.2$) & $-3.6$ ($-0.2$) & $-4.0$ ($-0.2$) & $-3.3$ ($-0.2$) & $-2.6$ ($-0.2$) \\
    {\ttfamily EOSHQ2M135}      & $-4.8$ ($-0.4$) & $-3.5$ ($-0.4$) & $-4.0$ ($-0.4$) & $-3.7$ ($-0.3$) & $-3.7$ ($-0.3$) & $-4.1$ ($-0.4$) & $-3.4$ ($-0.3$) & $-2.7$ ($-0.3$) \\
    {\ttfamily EOS2HQ2M135}     & $-5.0$ ($-0.6$) & $-3.7$ ($-0.6$) & $-4.3$ ($-0.7$) & $-3.8$ ($-0.5$) & $-3.8$ ($-0.5$) & $-4.3$ ($-0.5$) & $-3.6$ ($-0.5$) & $-2.8$ ($-0.5$) \\
    \colrule
    {\ttfamily EOSBQ2M12}       & $-4.4$ ($-0.4$) & $-3.3$ ($-0.4$) & $-3.7$ ($-0.4$) & $-3.4$ ($-0.4$) & $-3.4$ ($-0.3$) & $-3.8$ ($-0.4$) & $-3.2$ ($-0.3$) & $-2.5$ ($-0.3$) \\
    {\ttfamily EOSHBQ2M12}      & $-4.5$ ($-0.5$) & $-3.4$ ($-0.5$) & $-3.8$ ($-0.5$) & $-3.5$ ($-0.4$) & $-3.5$ ($-0.4$) & $-3.9$ ($-0.5$) & $-3.2$ ($-0.4$) & $-2.6$ ($-0.4$) \\
    {\ttfamily EOSHQ2M12}       & $-4.5$ ($-0.5$) & $-3.4$ ($-0.5$) & $-3.9$ ($-0.5$) & $-3.5$ ($-0.5$) & $-3.5$ ($-0.5$) & $-3.9$ ($-0.5$) & $-3.3$ ($-0.5$) & $-2.6$ ($-0.4$) \\
    {\ttfamily EOS2HQ2M12}      & $-4.7$ ($-0.6$) & $-3.6$ ($-0.7$) & $-4.3$ ($-0.9$) & $-3.6$ ($-0.6$) & $-3.6$ ($-0.5$) & $-4.0$ ($-0.6$) & $-3.4$ ($-0.5$) & $-2.7$ ($-0.5$) \\
    \colrule
    {\ttfamily EOS15HQ5M135}    & $-9.6$ ($0.0 $) & $-6.5$ ($0.0 $) & $-6.7$ ($0.0 $) & $-7.0$ ($0.0 $) & $-7.2$ ($0.0 $) & $-8.1$ ($0.0 $) & $-6.5$ ($0.0 $) & $-4.8$ ($0.0 $) \\
    {\ttfamily EOS15HQ4M135}    & $-7.8$ ($0.0 $) & $-5.4$ ($0.0 $) & $-5.8$ ($0.0 $) & $-5.8$ ($0.0 $) & $-5.9$ ($0.0 $) & $-6.6$ ($0.0 $) & $-5.4$ ($0.0 $) & $-4.1$ ($0.0 $) \\
    {\ttfamily EOS15HQ3M135}    & $-6.6$ ($-0.5$) & $-4.8$ ($-0.5$) & $-5.3$ ($-0.5$) & $-5.0$ ($-0.5$) & $-5.1$ ($-0.5$) & $-5.7$ ($-0.5$) & $-4.7$ ($-0.4$) & $-3.7$ ($-0.4$) \\
    {\ttfamily EOS125HQ2M135}   & $-4.9$ ($-0.5$) & $-3.6$ ($-0.5$) & $-4.1$ ($-0.5$) & $-3.8$ ($-0.4$) & $-3.8$ ($-0.4$) & $-4.2$ ($-0.5$) & $-3.5$ ($-0.4$) & $-2.8$ ($-0.4$) \\
    {\ttfamily EOS15HQ2M135}    & $-4.9$ ($-0.5$) & $-3.7$ ($-0.5$) & $-4.2$ ($-0.6$) & $-3.8$ ($-0.5$) & $-3.8$ ($-0.5$) & $-4.3$ ($-0.5$) & $-3.6$ ($-0.5$) & $-2.8$ ($-0.5$) \\
    \colrule
    {\ttfamily EOSBlQ2M135}     & $-4.5$ ($-0.1$) & $-3.3$ ($-0.1$) & $-3.8$ ($-0.1$) & $-3.4$ ($-0.1$) & $-3.5$ ($-0.1$) & $-3.9$ ($-0.1$) & $-3.2$ ($-0.1$) & $-2.5$ ($-0.1$) \\
    {\ttfamily EOSHBlQ2M135}    & $-4.6$ ($-0.2$) & $-3.4$ ($-0.2$) & $-3.9$ ($-0.2$) & $-3.5$ ($-0.2$) & $-3.6$ ($-0.2$) & $-4.0$ ($-0.2$) & $-3.3$ ($-0.2$) & $-2.6$ ($-0.2$) \\
    {\ttfamily EOSHlQ2M135}     & $-4.7$ ($-0.4$) & $-3.5$ ($-0.3$) & $-4.0$ ($-0.4$) & $-3.7$ ($-0.3$) & $-3.7$ ($-0.3$) & $-4.1$ ($-0.3$) & $-3.4$ ($-0.3$) & $-2.7$ ($-0.3$) \\
    {\ttfamily EOS125HlQ2M135}  & $-4.9$ ($-0.5$) & $-3.6$ ($-0.4$) & $-4.1$ ($-0.5$) & $-3.8$ ($-0.4$) & $-3.8$ ($-0.4$) & $-4.2$ ($-0.4$) & $-3.5$ ($-0.4$) & $-2.8$ ($-0.4$) \\
    {\ttfamily EOS15HlQ2M135}   & $-4.9$ ($-0.5$) & $-3.7$ ($-0.5$) & $-4.2$ ($-0.5$) & $-3.8$ ($-0.5$) & $-3.8$ ($-0.5$) & $-4.2$ ($-0.5$) & $-3.5$ ($-0.5$) & $-2.8$ ($-0.4$) \\
    \colrule
    {\ttfamily EOSBsQ2M135}     & $-4.4$ ($0.0 $) & $-3.2$ ($0.0 $) & $-3.6$ ($0.0 $) & $-3.3$ ($0.0 $) & $-3.3$ ($0.0 $) & $-3.7$ ($0.0 $) & $-3.1$ ($0.0 $) & $-2.4$ ($0.0 $) \\
    {\ttfamily EOSHBsQ2M135}    & $-4.6$ ($-0.2$) & $-3.4$ ($-0.2$) & $-3.9$ ($-0.2$) & $-3.5$ ($-0.2$) & $-3.6$ ($-0.2$) & $-4.0$ ($-0.2$) & $-3.3$ ($-0.2$) & $-2.6$ ($-0.2$) \\
    {\ttfamily EOSHsQ2M135}     & $-4.8$ ($-0.4$) & $-3.6$ ($-0.4$) & $-4.0$ ($-0.4$) & $-3.7$ ($-0.4$) & $-3.7$ ($-0.4$) & $-4.1$ ($-0.4$) & $-3.4$ ($-0.4$) & $-2.7$ ($-0.4$) \\
    {\ttfamily EOS125HsQ2M135}  & $-4.9$ ($-0.5$) & $-3.7$ ($-0.5$) & $-4.2$ ($-0.5$) & $-3.8$ ($-0.5$) & $-3.8$ ($-0.5$) & $-4.2$ ($-0.5$) & $-3.5$ ($-0.5$) & $-2.8$ ($-0.4$) \\
    {\ttfamily EOS15HsQ2M135}   & $-5.0$ ($-0.6$) & $-3.7$ ($-0.6$) & $-4.3$ ($-0.6$) & $-3.8$ ($-0.5$) & $-3.9$ ($-0.5$) & $-4.3$ ($-0.5$) & $-3.6$ ($-0.5$) & $-2.8$ ($-0.5$) \\
    \colrule
    {\ttfamily EOSBssQ2M135}    & $-4.4$ ($0.0 $) & $-3.2$ ($0.0 $) & $-3.7$ ($0.0 $) & $-3.3$ ($0.0 $) & $-3.4$ ($0.0 $) & $-3.8$ ($0.0 $) & $-3.1$ ($0.0 $) & $-2.4$ ($0.0 $) \\
    {\ttfamily EOSHBssQ2M135}   & $-4.6$ ($-0.2$) & $-3.4$ ($-0.2$) & $-3.8$ ($-0.2$) & $-3.5$ ($-0.2$) & $-3.5$ ($-0.2$) & $-3.9$ ($-0.2$) & $-3.3$ ($-0.2$) & $-2.6$ ($-0.2$) \\
    {\ttfamily EOSHssQ2M135}    & $-4.8$ ($-0.6$) & $-3.6$ ($-0.6$) & $-4.1$ ($-0.7$) & $-3.7$ ($-0.5$) & $-3.8$ ($-0.5$) & $-4.2$ ($-0.5$) & $-3.5$ ($-0.5$) & $-2.8$ ($-0.5$) \\
    {\ttfamily EOS125HssQ2M135} & $-5.0$ ($-0.4$) & $-3.7$ ($-0.4$) & $-4.2$ ($-0.4$) & $-3.8$ ($-0.4$) & $-3.9$ ($-0.4$) & $-4.3$ ($-0.4$) & $-3.6$ ($-0.4$) & $-2.8$ ($-0.4$) \\
    {\ttfamily EOS15HssQ2M135}  & $-5.0$ ($-0.6$) & $-3.7$ ($-0.5$) & $-4.3$ ($-0.6$) & $-3.8$ ($-0.5$) & $-3.8$ ($-0.5$) & $-4.3$ ($-0.5$) & $-3.6$ ($-0.5$) & $-2.8$ ($-0.5$) \\
    \colrule
    \colrule
  \end{tabular}   
\end{table*}
%TTTTTTTTTTTTTTTTTTTTTTTTTTTTTTTTTTTTTTTTTTTTTTTTTTTTTTTTTTTTTTTTTTTTTTTTTTTTTT

We compute $\epsilon_\text{RPN}$ and $\epsilon_\text{BHBH}$ for all
binaries in Tables \ref{tab:NRruns2model} and \ref{tab:NRruns2tests};
in principle we could consider arbitrary binary configurations, but
(to be conservative) here we limit our calculations to the cases for
which we have evidence that our model works well.

We consider the following eight detectors: Advanced (Adv) LIGO, see
e.g.~Eq.\,(3.3) of \cite{Ajith:2009fz}; Advanced LIGO in the
zero-detuning, high-power configuration (AdvZDHP), as fitted in
Eq.\,(4.7) of \cite{Ajith:2011ec}; Advanced Virgo, Eq.\,(3.4) of
\cite{Ajith:2009fz}; the Einstein Telescope (ET) in broadband (B) and
xylophone (C) configuration, as found in the MATLAB files available at
\cite{ETnoiseMATLAB}; and finally, KAGRA in the variable broadband
(varBRSE), broadband optimized for NS-NS detection (maxBRSE), and
variable detuned mode (varDRSE). We provide fits to the three KAGRA
configurations in Appendix \ref{app:KAGRA} (to our knowledge, no such
fits have been published in the existing literature).

The results, collected in Table \ref{tab:SNRs}, may be summarized
as follows:
\begin{enumerate}[(1)]
\item The deviations between the PhenomC model and our model increase
  as one considers binaries with stronger tidal effects. This is
  expected: the stronger the tidal effects, the larger the deviations
  between a BH-NS waveform and a BH-BH waveform with the same
  constituent masses. However, for the purpose of SNR calculations the
  PhenomC model and our model are basically equivalent:
  $|\epsilon_\text{BHBH}|<0.01$ for (nonspinning) BH-NS mergers. This
  is due to the fact that SNR calculations for these binaries are
  largely dominated by the low-frequency inspiral contribution. This
  result implies that using NR based BH-BH binary models in (say) rate
  calculations is good enough also for BH-NS binaries, at least in the
  nonspinning case.
\item The numbers listed for $|\epsilon_\text{RPN}|$ show that
  NR-based modeling has an impact of at most $\sim 10\%$ in SNR
  calculations from BH-NS systems, as long as we truncate both signals
  at the Schwarzschild ISCO.
\item The deviations between different models are comparable for a
  given binary and different detectors (ET, KAGRA and Advanced
  LIGO/Virgo). Among the three configurations of KAGRA, the variable
  configuration in broadband mode systematically yields the smallest
  deviations (in absolute value). Among second-generation detectors,
  SNR calculations for Advanced LIGO are the most sensitive to the
  high-frequency behavior of BH-NS merger waveforms.
\een

%%%%%%%%%%%%%%%%%%%%%%%%%%%%%%%%%%%%%%%%%%%%%%%%%%%%%%%%%%%%%%%%%%%%%%%%%%%%%%
\section{Cutoff Frequencies}\label{sec:fCut}
%%%%%%%%%%%%%%%%%%%%%%%%%%%%%%%%%%%%%%%%%%%%%%%%%%%%%%%%%%%%%%%%%%%%%%%%%%%%%%
The amplitude of the GWs emitted by a coalescing compact binary dies
off at high frequency, once the newly formed object (be it a BH or a
NS) settles down to a stationary equilibrium configuration. In the
case of BH binaries, the GW amplitude drops at the frequency of the
dominant ($l=m=2$, $n=0$) QNM mode of the remnant BH. In the case of
BH-NS binaries, the cutoff, or shut down, frequency has received much
attention because, depending on the dynamical history of the system,
this frequency may originate from the tidal disruption of the NS. A
cutoff frequency in the GW amplitude of BH-NS binaries that is due to
the tidal disruption of the NS by the BH is dependent on the NS
EOS. This has suggested the idea of examining the cutoff frequency of
GWs emitted by mixed binaries in order to pin down the NS EOS (see
e.g.~\cite{Vallisneri:1999nq,Ferrari09,Ferrari:2009bw}).

The first studies of cutoff frequencies, $f_{\rm Cut}$, for BH-NS
binaries involved either semianalytical
\cite{Vallisneri:1999nq,Ferrari09,Ferrari:2009bw} or fully numerical
\cite{Taniguchi:2008a} (quasi)equilibrium approaches. More recent
estimates of $f_{\rm Cut}$ were determined via fully relativistic
numerical simulations of BH-NS mergers
\cite{Shibata:2009cn,Kyutoku:2010zd,Kyutoku:2011vz}. In these cases,
$f_{\rm Cut}$ was defined to be a parameter obtained from analytical
fits of the numerical GW data. This definition has the drawback of
being viable only if numerical data are available for the binary of
interest. Furthermore, the form of the fit to the GW data (and
therefore the definition of the cutoff frequency) had to be revised
when nonzero BH spins were considered.

In order to overcome these drawbacks, here we introduce a general
definition of $f_{\rm Cut}$ that is of immediate application to any
BH-NS GW spectrum, be it analytical or numerical. Our definition
allows for a straightforward comparison among GW spectra originating
from different models and/or calculations for the same binary, and for
consistent comparisons among binaries with different physical
parameters. The expression of the GW amplitude
$\tilde{A}_\text{Phen}(f)$ in Eq.\,(\ref{eq:PhenoMixed1}) shows that
at low frequencies, during the inspiral stage,
$\tilde{A}_\text{Phen}(m_0f)\sim (m_0f)^{-7/6}$. At high frequencies,
instead, we have $\tilde{A}_\text{Phen}(m_0f)\sim
(m_0f)^{-19/6}$. Furthermore (and as already noted in
\cite{Santamaria:2010yb} for the BH-BH case), the numerical data for
BH-NS binaries show a high-frequency falloff that is faster than
$(m_0f)^{-19/6}$: for example, in \cite{Kyutoku:2011vz} this falloff
was fitted by a function of the form $e^{-(f/f_0)^\sigma}/f$, where
$f_0$ and $\sigma$ are two positive, real parameters. These
considerations on the low- and high-frequency behavior imply that
$(m_0f)^2\tilde{h}(m_0f)$ must have a global maximum. Therefore we
first look for the frequency $f_{\rm Max}$ such that
$(m_0f)^2\tilde{h}(m_0f)$ has a maximum, and then we define $f_{\rm
  Cut}$ to be the frequency (greater than $f_{\rm Max}$) at which
\be
\label{eq:fCut}
em_0f_{\rm Cut}\tilde{h}(m_0f_{\rm Cut}) = m_0f_{\rm
  Max}\tilde{h}(m_0f_{\rm Max})\,.
\ee
This definition of the cutoff frequency is independent of the details
of the waveform, and it works for any $\tilde{h}(f)$ (given in either
analytical or numerical form). We would once more like to draw the
reader's attention to the conversion formula between dimensionless
frequencies of the form $m_0f$ and frequencies in Hz, which is given
in Eq.\,(\ref{eq:frequency-conversion}).

%FFFFFFFFFFFFFFFFFFFFFFFFFFFFFFFFFFFFFFFFFFFFFFFFFFFFFFFFFFFFFFFFFFFFFFFFFFFFF
\begin{figure*}[!tb]
\begin{tabular*}{\textwidth}{c@{\extracolsep{\fill}}c}
\includegraphics[scale=0.4,clip=true]{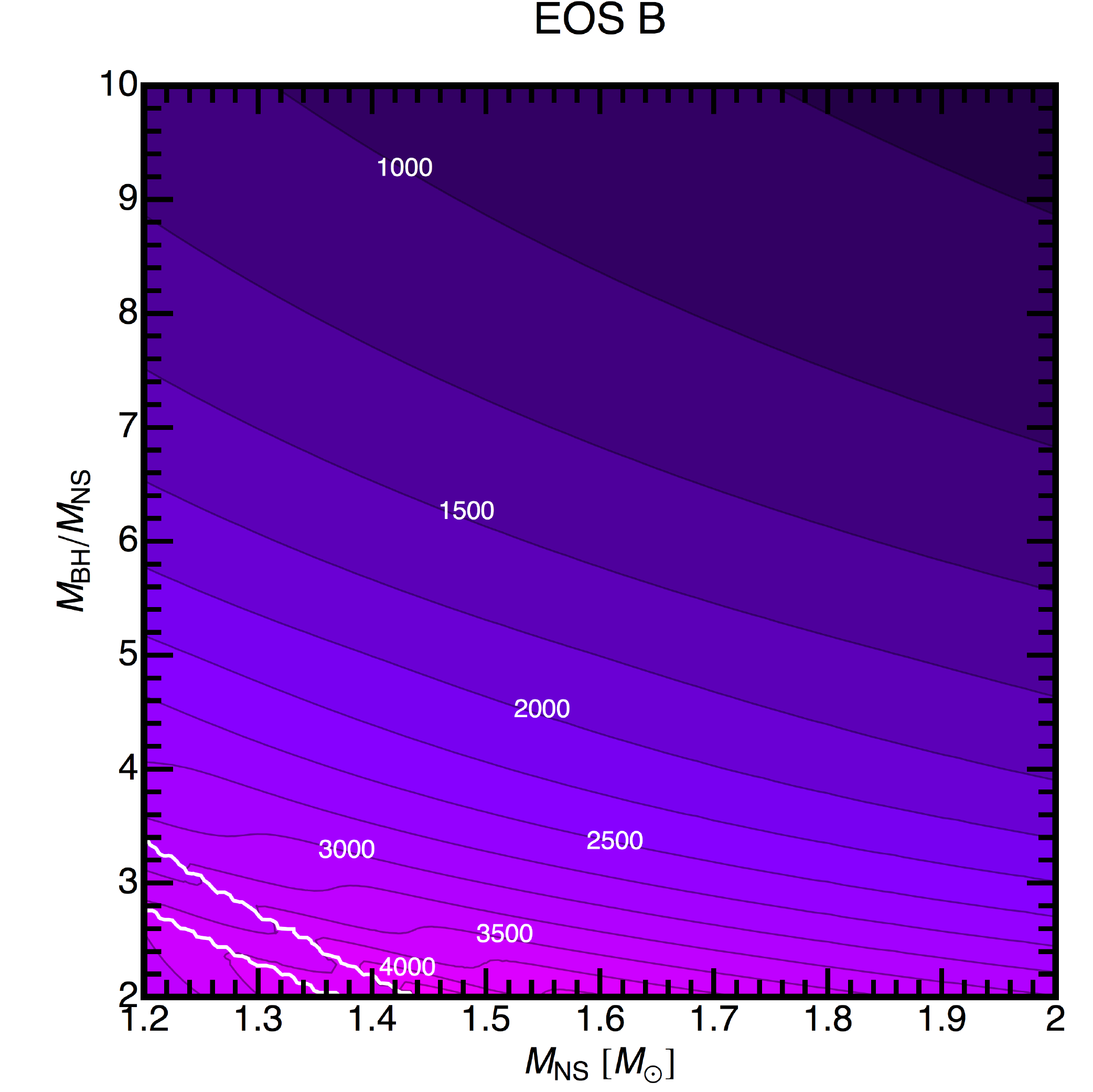}
&
\includegraphics[scale=0.4,clip=true]{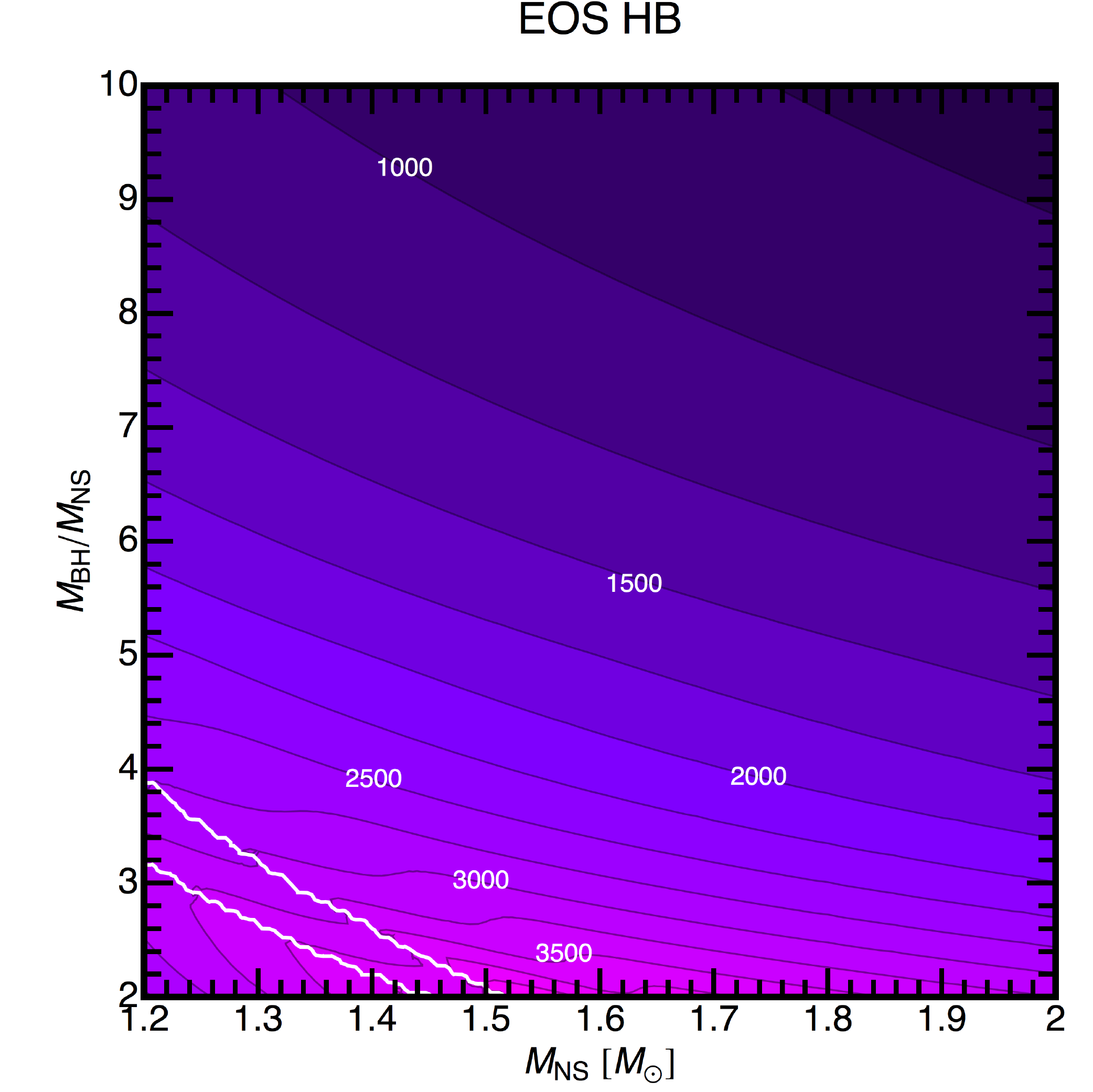}
\\
\includegraphics[scale=0.4,clip=true]{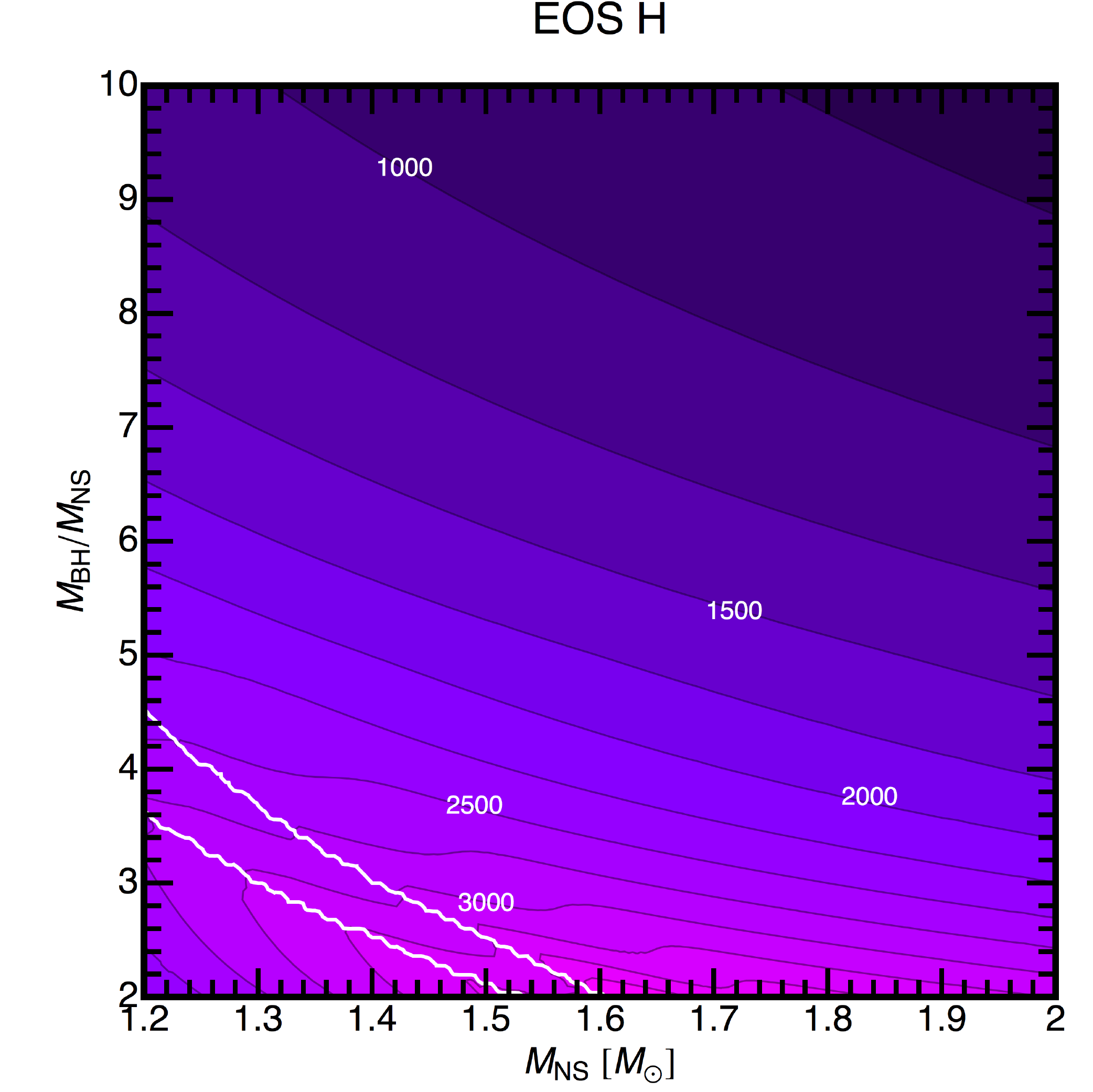}
&
\includegraphics[scale=0.4,clip=true]{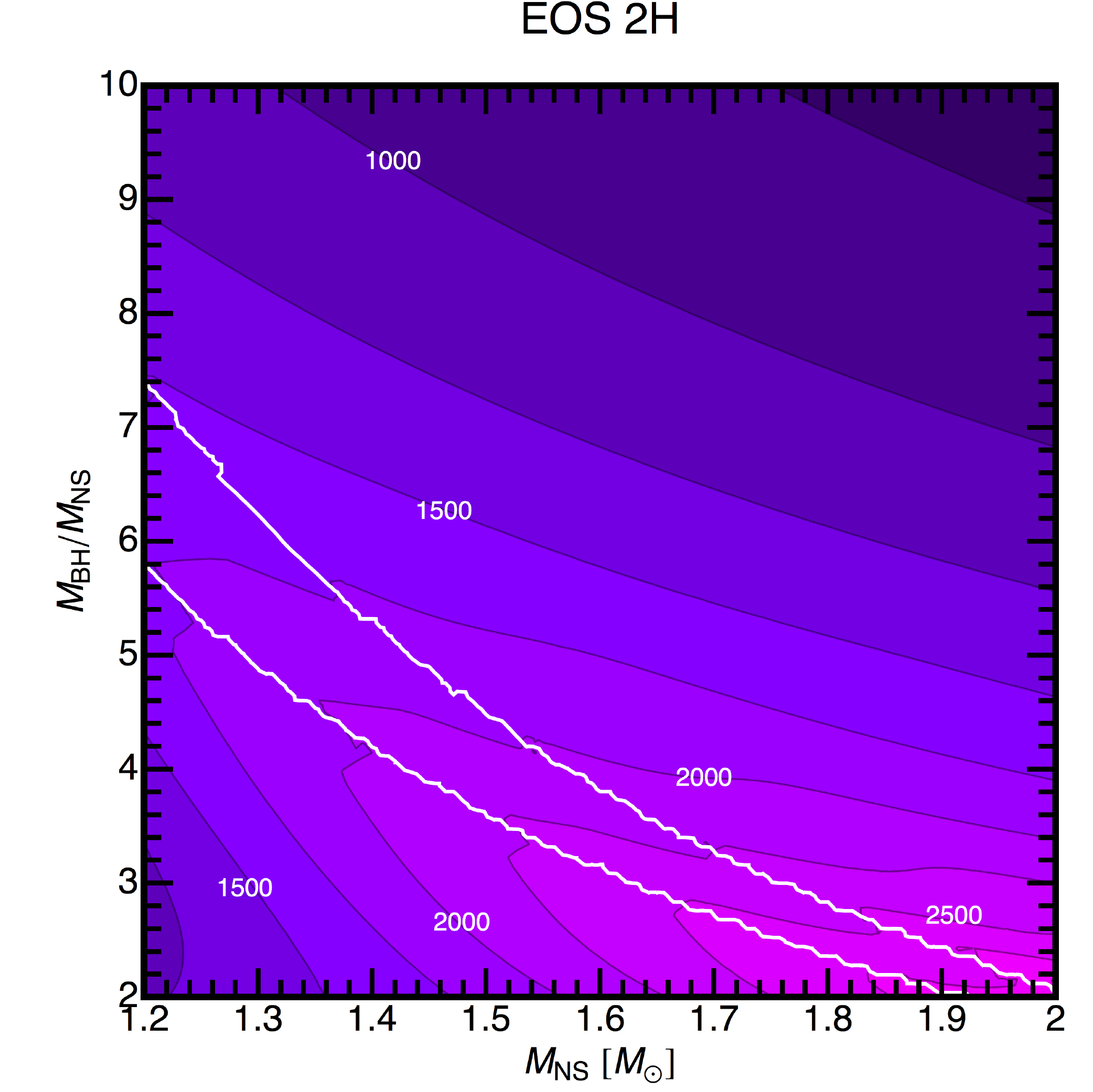}
\end{tabular*}
\caption{The cutoff frequency $f_{\rm Cut}$, as defined in
  Eq.\,(\ref{eq:fCut}), computed with our BH-NS GW amplitude model. We
  consider the EOSs B, HB, H, 2H, and report contour lines in Hz, with
  a spacing of $250\,$Hz. The two white lines in each panel divide the
  plane in three regions: a top-right one in which the BH-NS
  coalescences are nondisruptive, a bottom-left one in which they are
  disruptive, and a middle region in which mildly disruptive
  coalescences occur. This classification is discussed in
  Sec.~\ref{sec:classification}. \label{Fig:fCut}}
\end{figure*}
%FFFFFFFFFFFFFFFFFFFFFFFFFFFFFFFFFFFFFFFFFFFFFFFFFFFFFFFFFFFFFFFFFFFFFFFFFFFFF

Using our phenomenological model and the definition above, we computed
contour plots of the cutoff frequency in the
$(M_\text{NS},\,M_\text{BH}/M_\text{NS})$ plane. The results are shown
in Fig.~\ref{Fig:fCut}, where each panel refers to a different
piecewise polytropic EOS (B,\,HB,\,H,\,2H). The two white lines in
each panel divide the plane in three regions, following the
classification discussed in Sec.~\ref{sec:classification}.

In the top-right region the BH-NS coalescence is nondisruptive. This
region is by far the largest for the soft EOS B (i.e., when the NS
structure does not matter much) and it shrinks as we consider stiffer
and stiffer EOSs. The cutoff frequency in this region is essentially
the fundamental QNM of the remnant BH. In the bottom-left region the
merger is disruptive, and mildly disruptive coalescences occur in the
region comprised between the two white lines. This plot confirms the
conclusion of previous studies (e.g.~\cite{Pannarale:2011pk}),
i.e.~that the information from tidal disruptions is confined to high
frequencies, where second-generation detectors will not be very
sensitive for hypothetically typical events at a distance of $100$Mpc
or more: even for the stiffest EOS 2H, the observation of EOS effects
will require third-generation detectors such as ET, that are sensitive
at frequencies $\gtrsim 1\,$kHz.

What is most interesting (in our view) is that the calculations
presented in Fig.~\ref{Fig:fCut} could provide a basis to address the
inverse problem: given future observations of a tidal disruption
frequency, what can we say about the EOS prevailing in the interior?
We will address this question after working out a generalization of
the present calculations to the case of spinning BH-NS mergers.

%%%%%%%%%%%%%%%%%%%%%%%%%%%%%%%%%%%%%%%%%%%%%%%%%%%%%%%%%%%%%%%%%%%%%%%%%%%%%%
\section{Conclusions}
%%%%%%%%%%%%%%%%%%%%%%%%%%%%%%%%%%%%%%%%%%%%%%%%%%%%%%%%%%%%%%%%%%%%%%%%%%%%%%
In this paper we developed a phenomenological model for the
frequency-domain gravitational waveform amplitude of nonspinning BH-NS
mergers. The model was calibrated to general relativistic numerical
simulations using a piecewise polytropic neutron star EOS with a
$\Gamma_2=3.0$ core, and it encompasses the three possible outcomes of
the merger: no tidal disruption, mild, and strong tidal disruption. We
showed that the model is very accurate even in the most challenging
cases, namely when the core EOS has a very small polytropic exponent
and the binary mass ratio is small ($\Gamma_2=2.4$ and $Q=2$), so that
the frequency at the onset of tidal disruption and the ringdown
frequency of the final BH are very close to each other
($f_\text{RD}\simeq f_\text{tide}$). A \textsc{Mathematica} notebook
implementing the algorithm is publicly available online
\cite{Mathematica}.

We demonstrated that such an accurate modeling of the waveform
amplitude is probably unnecessary for SNR calculations and rate
estimates from nonspinning BH-NS binaries
(cf. \cite{Abadie:2010cf,Dominik:2012kk}), in the sense that BH-BH
phenomenological waveforms provide SNRs accurate within about a
percent for second- and third-generation GW interferometers. This may
not be true for mergers of {\em spinning} BH-NS binaries: by comparing
SNRs for GWs obtained with the PhenomC model of
\cite{Santamaria:2010yb} and with the model of \cite{Lackey:2013axa},
we found differences up to $\sim 10\%$. The most immediate (and
probably the most useful) application of the model will be to extract
information on the nuclear EOS from future high-frequency GW
observations, using e.g.~the high-frequency signal cutoff frequencies
that we provided in Fig.~\ref{Fig:fCut} (see
\cite{Vallisneri:1999nq,Ferrari09,Ferrari:2009bw} for previous studies
in this direction).

In the near future we plan to extend and improve the model as longer
and more accurate numerical waveforms become available. In particular,
we will improve the model in the underconstrained mild tidal
disruption regime, extend it to aligned, spinning binaries
\cite{Kyutoku:2011vz,Lackey:2013axa}, and possibly also to
precessing/inclined binaries
\cite{Foucart:2010eq,Foucart:2012vn}. Future work should also address
the development of a similar phenomenological model for the waveform
phasing, the extension to higher multipoles of the radiation (beyond
$l=m=2$), and possibly comparisons with the EOB formalism. All of
these extensions will rely critically on the accuracy of available
numerical simulations (see \cite{Bernuzzi:2011aq,Radice:2013hxh} for a
discussion in the context of NS-NS binaries).

%%%%%%%%%%%%%%%%%%%%%%%%%%%%%%%%%%%%%%%%%%%%%%%%%%%%%%%%%%%%%%%%%%%%%%%%%%%%%%
\section*{Acknowledgements}
%%%%%%%%%%%%%%%%%%%%%%%%%%%%%%%%%%%%%%%%%%%%%%%%%%%%%%%%%%%%%%%%%%%%%%%%%%%%%%
This work was supported in part by the DFG grant SFB/Transregio~7.
E.B.~is supported by NSF CAREER Grant No. PHY-1055103. K.K.~is
supported by a JSPS Postdoctoral Fellowship for Research Abroad.  We
are grateful to Masaki Ando for helpful correspondence on the KAGRA
power spectral density. F.P.~wishes to thank Frank Ohme for useful
discussions on the PhenomC model, and Luciano Rezzolla for reading the
manuscript. This work was supported by the Japanese Grant-in-Aid for
Scientific Research (21340051, 24244028), and by the Grant-in-Aid for
Scientific Research on Innovative Area (20105004). Part of this work
was completed during a long-term workshop on Gravitational Waves and
Numerical Relativity held at the Yukawa Institute for Theoretical
Physics of Kyoto University in 2013.
%%%%%%%%%%%%%%%%%%%%%%%%%%%%%%%%%%%%%%%%%%%%%%%%%%%%%%%%%%%%%%%%%%%%%%%%%%%%%%

\appendix

%%%%%%%%%%%%%%%%%%%%%%%%%%%%%%%%%%%%%%%%%%%%%%%%%%%%%%%%%%%%%%%%%%%%%%%%%%%%%%%
\section{Matching Numerical and Analytical
  Amplitudes}\label{app:factors}
%%%%%%%%%%%%%%%%%%%%%%%%%%%%%%%%%%%%%%%%%%%%%%%%%%%%%%%%%%%%%%%%%%%%%%%%%%%%%%%
When building our amplitude model [Eq.\,(\ref{eq:PhenoMixed1})], the
first thing to do is to ensure that the same convention is used for
the overall amplitude of the numerical gravitational waveform data and
in Eq.\,(\ref{eq:PhenomAmp}) for the BH-BH PhenomC GWs of
\cite{Santamaria:2010yb}, which are our starting point.

We begin by writing out the plus and cross polarization of the
quadrupole moment of the emitted gravitational radiation. These are
\beq
h_+ &=& \frac{1}{2}\left[ h_{22} ( _{-2}Y^{22} + \,_{-2}Y^{2-2*})\right. \nn\\
&+& \left. h_{22}^*( _{-2}Y^{22*} + \,_{-2}Y^{2-2})\right]\,,\\
&& \nn\\
h_\times &=& \frac{i}{2}\left[ h_{22} ( _{-2}Y^{22} - \,_{-2}Y^{2-2*})\right. \nn\\
&+& \left. h_{22}^*( _{-2}Y^{2-2} - \,_{-2}Y^{2-2})\right]\,,
\eeq
where the $_{-s}Y^{lm}$'s denote spin-weighted spherical harmonics. If
we pick an optimal observer, that is if we place the observer
``face-on'' by setting the angles $\theta=0$ and $\phi=0$ with respect
to the source, the harmonics become
\beq
_{-2}Y^{22} &=& \sqrt{\frac{5}{4\pi}}\,, \\
_{-2}Y^{2-2} &=& 0\,,
\eeq
and reduce the expressions for the two GW polarizations to
\beq
h_+ &=& \frac{1}{2}\sqrt{\frac{5}{4\pi}}\left( h_{22} + h_{22}^* \right)\,, \\
&& \nn\\
h_\times &=& \frac{i}{2}\sqrt{\frac{5}{4\pi}}\left( h_{22} -
  h_{22}^*\right)\,.
\eeq
These are the quantities we obtain from our numerical simulations. We
may also express $h_{+,\times}$ using the amplitude $A_{22}$ and the
phase $\Phi$ of $h_{22}=A_{22}e^{2i\Phi}$. This yields
\beq
h_+(t) &=& \sqrt{\frac{5}{4\pi}}|A_{22}(t)|\cos\Phi(t)\,, \\
&& \nn\\
h_\times(t) &=& \sqrt{\frac{5}{4\pi}}|A_{22}(t)|\sin\Phi(t)\,,
\eeq
where we explicitly wrote out the time dependence.

When transforming to the Fourier domain in the stationary phase
approximation,\footnote{This is done to obtain the
  $\tilde{A}_\text{PN}$ terms of Eq.\,(\ref{eq:PhenomAmp}) used in
  \cite{Santamaria:2010yb} and here.} special attention must be paid:
\beq
\tilde h_+(f) &=& \sqrt{\frac{5}{4\pi}}\frac{|A_{22}(t_\text{f})|}{2}\sqrt{\frac{\pi}{\dot{\omega}}}e^{i\Psi(f)}\,, \\
&& \nn\\
\tilde h_\times(f) &=&
\sqrt{\frac{5}{4\pi}}\frac{|A_{22}(t_\text{f})|}{2}\sqrt{\frac{\pi}{\dot{\omega}}}e^{i\Psi(f)}\,,
\eeq
where $\psi$ is the Fourier phase, and where $t_\text{f}$ is the
moment of time when the instantaneous frequency coincides with the
Fourier variable, i.e., $M\omega(t_\text{f})=2\pi f$, $\omega$ being
the time derivative of $\Phi(t)$. Notice the $1/2$ factor that appears
in these expressions. If one instead transforms directly
$h_{22}(t)=A_{22}(t)e^{2i\Phi(t)}$, as is done in
\cite{Santamaria:2010yb}, one gets
\beq
\tilde h_{22}(f) &=&
|A_{22}(t_\text{f})|\sqrt{\frac{\pi}{\dot{\omega}}}e^{i\Psi(f)}\,.
\eeq

We then consider the GW strain
\be
\label{GWstrain}
\sqrt{\frac{|\tilde h_+|^2+|\tilde h_\times|^2}{2}}\,,
\ee
which is the quantity calculated from our numerical data and plotted
throughout the paper. By using the previous equalities we see that
\beq
\sqrt{\frac{|\tilde h_+|^2+|\tilde h_\times|^2}{2}} &=&
\frac{1}{2}\sqrt{\frac{5}{4\pi}}|A_{22}(t_\text{f})|\sqrt{\frac{\pi}{\dot{\omega}}}\nn\\
&=&\frac{1}{2}\sqrt{\frac{5}{4\pi}}|\tilde h_{22}|
\eeq
This shows that the factor
\be
\frac{1}{2}\sqrt{\frac{5}{4\pi}}
\ee
must be used to translate between the PN frequency-domain expression
of \cite{Santamaria:2010yb} and the amplitude seen by an observer in
the direction of the rotational axis of the binary.

%%%%%%%%%%%%%%%%%%%%%%%%%%%%%%%%%%%%%%%%%%%%%%%%%%%%%%%%%%%%%%%%%%%%%%%%%%%%%%
\section{KAGRA Sensitivity Curve Fits}\label{app:KAGRA}
%%%%%%%%%%%%%%%%%%%%%%%%%%%%%%%%%%%%%%%%%%%%%%%%%%%%%%%%%%%%%%%%%%%%%%%%%%%%%%
In this Appendix we provide analytical fits to the estimated
sensitivity limits of KAGRA, available at the KAGRA webpage
\cite{KAGRAnoise}. We consider three different configurations. In the
first case, we fit the total noise data for the KAGRA variable
configuration in broadband mode (varBRSE); in the second case, we look
at the total noise curve of KAGRA in the broadband mode, optimized for
binary neutron star inspiral detection without detuning (maxBRSE); in
the third and last case, we examine the variable KAGRA configuration
in detuned mode (varDRSE).

The total noise data for the three KAGRA configurations may be fitted
with the following curve:
\begin{align}
\label{eq:10pfit}
S_h(f) &= s_0\left( a_{2}\bar{f}^2 +a_{1}\bar{f} +1+\right.\\
&\left. +\frac{b_{05}}{\sqrt{\bar{f}}}
+\frac{b_{1}}{\bar{f}}
+\frac{b_{2}}{\bar{f}^{2}}
+\frac{b_{3}}{\bar{f}^{3}}
+\frac{b_{4}}{\bar{f}^{4}}
+\frac{b_{5}}{\bar{f}^{5}}
+\frac{b_{16}}{\bar{f}^{16}}
\right)\,,\nn
\end{align}
where $\bar{f}=f/f_0$, $f_0$ being the frequency location of the
minimum of the quantum noise in the configuration one is
considering. This is $84.3335\,$Hz in broadband mode, i.e.~for varBRSE
and maxBRSE, and $83.3681\,$Hz in detuned mode, that is, for
varDRSE. The form of this fit follows automatically when considering
the individual contributions to the total noise and adding them
up. There are two power laws for the mirror noise, one for the seismic
noise, one for the suspension noise, and seven for the total quantum
noise. The mirror noise and the total quantum noise share a $\sim 1/f$
term so that there is a total of ten power laws and, hence, of ten
parameters to be fitted. The results of the fits are collected in
Table \ref{tab:KAGRAfits} for the three KAGRA configurations varBRSE,
maxBRSE, and varDRSE, and are shown in the three panels of
Fig.~\ref{Fig:KAGRAfits}, along with the original data.

%FFFFFFFFFFFFFFFFFFFFFFFFFFFFFFFFFFFFFFFFFFFFFFFFFFFFFFFFFFFFFFFFFFFFFFFFFFFFFF
\begin{figure}[htb]
\includegraphics[scale=0.35,clip=true]{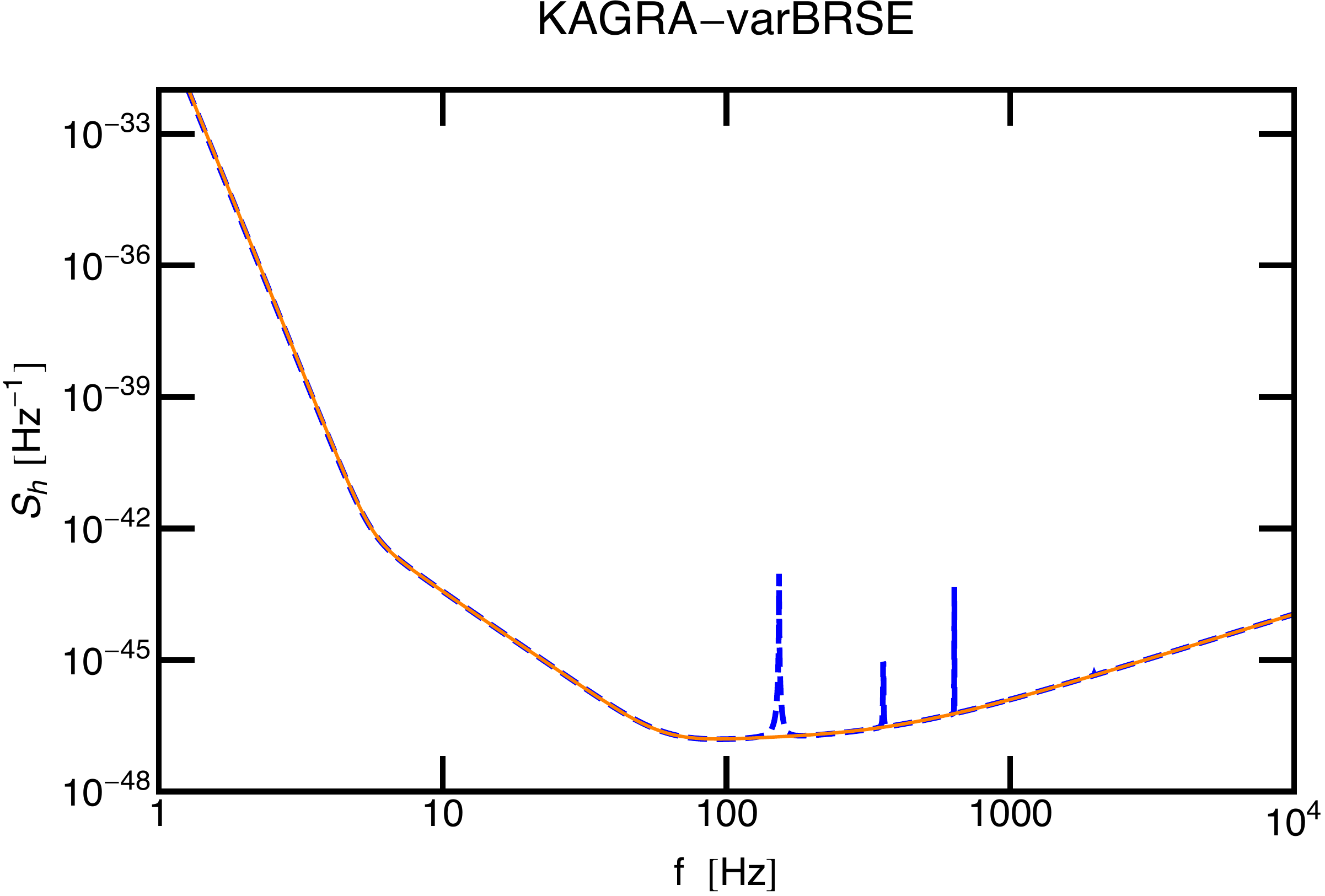}

\vspace{.5cm}

\includegraphics[scale=0.35,clip=true]{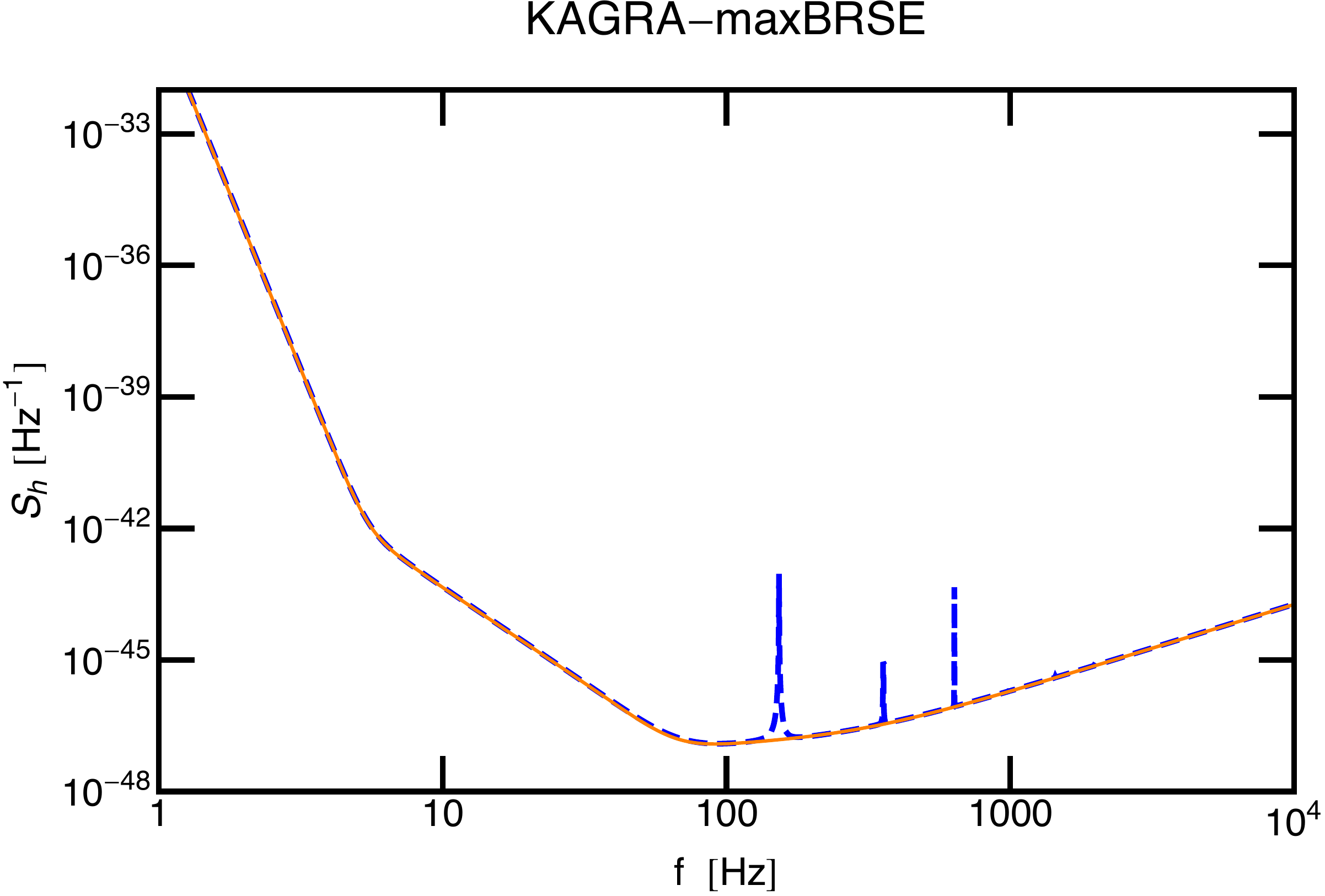}

\vspace{.5cm}

\includegraphics[scale=0.35,clip=true]{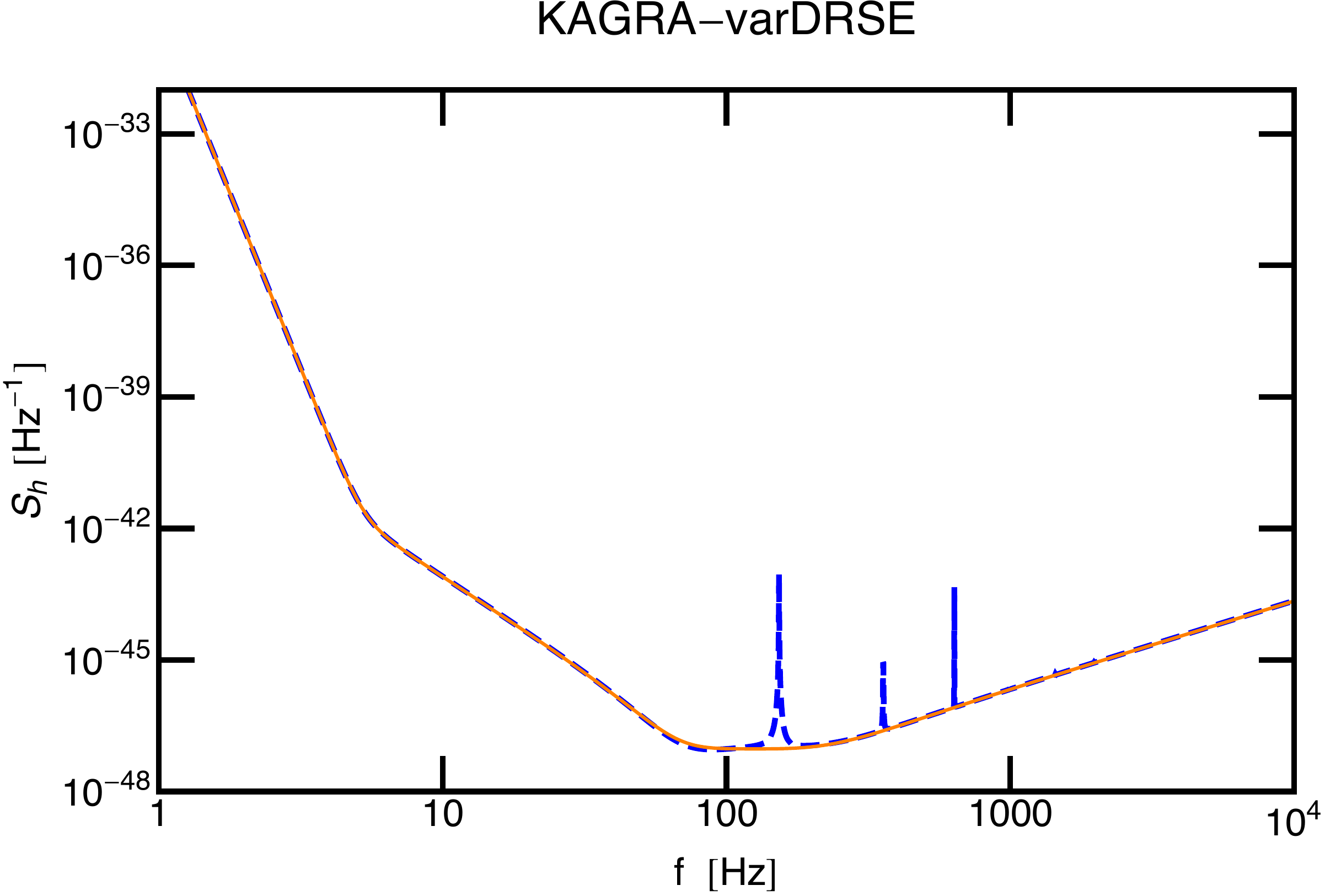}
\caption{Total noise data for KAGRA (dashed blue) and its best fit
  (continuous orange). The top panel refers to the variable KAGRA
  configuration in broadband mode (varBRSE); the middle panel refers
  to KAGRA in broadband mode and optimized for binary neutron star
  inspiral detection (maxBRSE); the bottom panel refers to the
  variable KAGRA configuration in detuned mode
  (varDRSE).\label{Fig:KAGRAfits}}
\end{figure}
%FFFFFFFFFFFFFFFFFFFFFFFFFFFFFFFFFFFFFFFFFFFFFFFFFFFFFFFFFFFFFFFFFFFFFFFFFFFFFF

\clearpage

\begin{widetext}
%TTTTTTTTTTTTTTTTTTTTTTTTTTTTTTTTTTTTTTTTTTTTTTTTTTTTTTTTTTTTTTTTTTTTTTTTTTTTTT
\begin{table*}
  \caption{Coefficients of the fit in Eq.\,(\ref{eq:10pfit}) for the
    total noise data of the three KAGRA configurations considered in
    this paper: varBRSE (KAGRA variable configuration in broadband
    mode); maxBRSE (KAGRA in broadband mode, optimized for binary neutron star
    inspiral detection); varDRSE (KAGRA variable configuration in
    detuned mode). The noise data for KAGRA is available at the KAGRA
    webpage \cite{KAGRAnoise}. \label{tab:KAGRAfits}}
  \begin{tabular}{|l||cc|ccccccccc|}
    \colrule
    \colrule
    KAGRA & $s_0$ [$10^{-47}\,\text{Hz}^{-1}]$ & $f_0$ [Hz] & $a_2$ & $a_1$ & $b_{05}$ & $b_1$ & $b_2$ & $b_3$ & $b_4$ & $b_5$ & $b_{16}$ [$10^{-15}$]\\
    \colrule
    varBRSE & $1.20522$ & $84.3335$ & $0.0653054$ &   $0.00563030$ & $0.535848$ & $0.109784$ & $-0.885726$ & $0.160197$ & $0.300831$ & $0.0350983$ & $5.97876$ \\
    maxBRSE & $1.25262$ & $84.3335$ & $0.108905$ & $-0.000260438$ &  $-1.27327$ & $2.74441$ & $-2.71327$ & $0.759074$ & $0.354601$ & $0.0389407$ & $5.41997$ \\
    varDRSE & $1.13778$ & $83.3681$ & $0.135200$ &   $-0.0194294$ &  $-9.81375$ & $17.3277$ & $-10.8376$ & $2.17689$ & $0.889623$ & $0.0510381$ & $7.17394$\\
    \colrule
    \colrule
  \end{tabular}
\end{table*}
%TTTTTTTTTTTTTTTTTTTTTTTTTTTTTTTTTTTTTTTTTTTTTTTTTTTTTTTTTTTTTTTTTTTTTTTTTTTTTT
\end{widetext}

\clearpage

\bibliography{PhenoMixedGWs}

\end{document}